\documentclass[fleqn,usenatbib]{mnras}

\usepackage[T1]{fontenc}

\DeclareRobustCommand{\DE}[3]{#2}
\let\DEthebibliography\thebibliography
\def\thebibliography{\DeclareRobustCommand{\DE}[3]{##3}\DEthebibliography}

\usepackage{graphicx}	% Including figure files
\usepackage{amsmath}	% Advanced maths commands
\usepackage{amssymb}	% Extra maths symbols
\usepackage{pdflscape} % landscape
\usepackage{tablefootnote}

\usepackage{newtxtext,newtxmath}

   \title[BLOeM: Pipeline Properties of OB stars]{Binarity at LOw Metallicity (BLOeM): Pipeline-Determined Physical Properties of OB Stars\thanks{Based on observations collected at the European Southern Observatory
under ESO program ID 112.25W2}}

   \author[Bestenlehner et al.]{J.M. Bestenlehner$^{1, 2}$,
          Paul A. Crowther$^{1}$\thanks{Corresponding author; (paul.crowther@sheffield.ac.uk)},
          V.~A. Bronner$^{3, 4}$,
          S. Sim\'{o}n-D\'{i}az$^{5, 6}$,
          D. J. Lennon$^{5, 6}$, \newauthor
          J. Bodensteiner$^{7, 8}$, 
          N. Langer$^{9}$,      
          P. Marchant$^{10, 11}$,
          H. Sana$^{10}$,
       F.~R.~N. Schneider$^{3, 12}$,
          T. Shenar$^{13}$
          \\
$^{1}$ Astrophysics Research Cluster, School of Mathematical and Physical Sciences, University of Sheffield, Hicks Building, Hounsfield Road, Sheffield S3 7RH, UK \\
$^{2}$ School of Chemical, Materials and Biological Engineering, University of Sheffield, Sir Robert Hadfield Building, Mappin Street, Sheffield S1 3JD, UK\\
$^{3}$ Heidelberg Institute for Theoretical Studies, Schloss-Wolfsbrunnenweg 35, 69118 Heidelberg, Germany\\
$^{4}$ Department of Physics \& Astronomy, Universit\"{a}t Heidelberg, Im Nuenheimer Feld 226, 69120 Heidelberg, Germany\\
$^{5}$ Instituto de Astrof\'isica de Canarias, Calle vi\'{a} L\'{a}ctea, s/n, 38205, La Laguna, Santa Cruz de Tenerife, Spain\\
$^{6}$ Departamento de Astrof\'isica, Universidad de La Laguna, E-38205 La Laguna, Tenerife, Spain\\
$^{7}$ European Southern Observatory, Karl-Schwarzschild-Strasse 2, 85748 Garching bei M\"{u}nchen, Germany\\
$^{8}$ Anton Pannekoek Institute of Astronomy, University of Amsterdam, Postbus 94249, 1090 GE Amsterdam, Netherlands\\
$^{9}$ Argelander-Institut f\"{u}r Astronomie, Universit\"{a}t Bonn, Auf dem H\"{u}gel 71, 53121 Bonn, Germany\\
$^{10}$ Institute of Astronomy, KU Leuven, Celestijnenlaan 200D, B-3001 Leuven, Belgium\\
$^{11}$ Sterrenkundig Observatorium, Universiteit Gent, Krijgslaan 281 S9, B-9000 Gent, Belgium\\
$^{12}$ Zentrum f\"{u}r Astronomie der Universit\"{a}t Heidelberg, Astronomisches Rechen-Institut, M\"{o}nchhofstr. 12-14, 69120, Heidelberg, Germany\\
$^{13}$ School of Physics and Astronomy, Tel Aviv University, Tel Aviv 6997801, Israel
       } 
       
   \date{Accepted 2025 May 23. Received 2025 May 20; in original form 2025 March 7}

   \pubyear{2025}

\begin{document}
\label{firstpage}
\pagerange{\pageref{firstpage}--\pageref{lastpage}}
\maketitle

  \begin{abstract}
We aim to determine the physical properties of OB stars from the multi-epoch VLT/FLAMES BLOeM spectroscopic survey of the Small Magellanic Cloud.  We apply a pipeline designed to analyse large spectroscopic samples of OB stars to the co-added, initial 9 epochs of the BLOeM survey, utilising grids of synthetic model spectra computed with the stellar atmosphere code {\sc fastwind}. 69 OB stars are excluded from the analysis owing to disk emission or significant contamination by secondaries in SB2 binaries. We determine physical properties of 778 OB stars, including $T_{\rm eff}$, $\log g$, $\log L/L_{\odot}$ and $\varv_{\rm e} \sin i$. There appears to be a bimodality in $\varv_{\rm e} \sin i$ of single O stars, while $\varv_{\rm e} \sin i$ distributions of OB stars are strikingly different for single (median 78 km\,s$^{-1}$) and binary (median 200 km\,s$^{-1}$) systems. Inferred temperatures are broadly in agreement with literature results for stars in common, plus results from a grid-based automization tool for a subset of O and early B stars, although uncertainties are larger for surface gravities. Rotational velocities are broadly in line with
an independent tool applied to the same subset. We recover the anticipated lower mass cutoff at 8 $M_{\odot}$ from the survey design using a Bayesian inference method coupled with SMC metallicity evolutionary models, with median masses of 12.6 $M_{\odot}$ (19.8 $M_{\odot}$) for B-type (O-type) stars. Spectroscopic masses exceed evolutionary masses, albeit with large uncertainties in surface gravities. We also provide an updated catalogue of O stars in the SMC since half of the 159 BLOeM O stars are newly classified as O-type stars.   
   \end{abstract}
   % 250 words 
   
\begin{keywords}  
stars: atmospheres -- stars: early-type –– stars: massive -- stars: fundamental parameters -- stars: rotation
\end{keywords}
   
%%%%%%%%%%%%%%%%% BODY OF PAPER %%%%%%%%%%%%%%%%%

\section{Introduction}
\label{intro}

Massive stars ($M_{\rm init} \geq 8 M_{\odot}$), despite their rarity, are major contributors to the radiative, chemical, and mechanical feedback of star-forming galaxies, owing to their high temperatures, production of $\alpha$-elements, and powerful stellar winds \citep{Geen+2023}. They are responsible for core-collapse supernovae \citep{Smartt2015}, gamma-ray bursts \citep{Gehrels+2009} and compact objects responsible for gravitational waves \citep{Abbott+2016}, especially at low metallicity.

Massive stars in the Milky Way are overwhelmingly found in close binaries \citep{Sana+2012}, affecting the evolution of the system  \citep{deMink+2014}, and consequently the lifetime, feedback and ultimate fate of each component. Large spectroscopic surveys of massive stars in the Large Magellanic Cloud (LMC), with a present-day metallicity of 1/2 Z$_{\odot}$, also reveal a high close binary fraction amongst massive stars \citep{Sana+2013}. 

The proximity of the Small Magellanic Cloud (SMC), with a present-day metallicity of 1/5 Z$_{\odot}$ \citep{RussellDopita1990}, provides our best view of individual metal-poor massive stars. Binary at LOw Metallicity \citep[BLOeM,][]{Shenar+2024} involves a multi-epoch spectroscopic survey of 929 massive stars in the SMC using the Fibre Large Array Multi Element Spectrograph \citep[FLAMES,][]{FLAMES} at the Very Large Telescope (VLT). The selection criteria for BLOeM targets focused on bright, blue sources from the Gaia DR3 catalogue \citep[see figure 2 of][]{Shenar+2024}, to ensure targets were representative of massive stars in the SMC. The use of a fibre-fed instrument (FLAMES) hindered sampling of crowded environments, such as the NGC~346 star-forming region \citep{Massey+1989, Dufton+2019, Rickard+2022}. Early results also favour a high close binary fraction of O and B-type stars \citep{BLOeM_O, BLOeM_B}.

\begin{figure}
%\sidecaption
\centering
\includegraphics[width=\columnwidth]{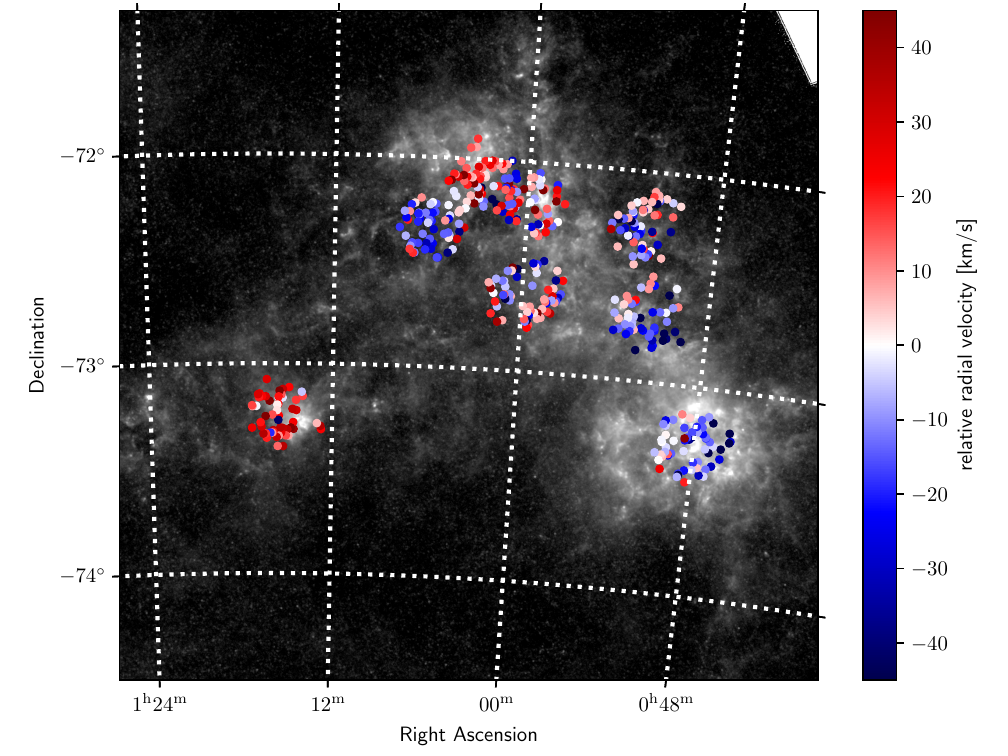}
  \caption{Radial velocities of single BLOeM OB stars -- according to initial 9 epoch dataset -- relative to 159 km\,s$^{-1}$ average of sample, overlaid on a {\it Herschel} SPIRE 350$\mu$m map of the SMC \citep{Meixner+2013}. Higher radial velocities for OB stars
  in the wing (south east) has previously been reported by \citet{EvansHowarth2008}.}
  \label{rv_ra-dec}
\end{figure}

Multiple systems in tight orbits range from double-lined (SB2) spectroscopic binaries in which both components contribute significantly at optical wavelengths, to single-lined (SB1) systems in which one component dominates, owing to a faint stellar or compact companion. Techniques used to analyse SB2 systems include spectral disentangling \citep{Mahy+2020}, which can also be used for SB1 systems to detect or rule out faint stellar companions \citep{Shenar+2022}. In all cases, it is necessary to determine
stellar parameters for OB stars, which is generally resource intensive. Spectral analysis of metal poor B stars is especially challenging since metal lines, which serve as primary temperature diagnostics \citep[e.g.][]{BeckerButler1990}, are much weaker than for Milky Way counterparts \citep{Walborn1983}.

In contrast to late-type stars, spectroscopic studies of hot, luminous stars usually involve one of two approaches. Coarse physical parameters can be estimated from spectral type-temperature calibrations, as was undertaken by \citet{Shenar+2024} for the BLOeM sample. Alternatively, detailed analysis of individual stars can be undertaken, owing to the large parameter space involved and requirement to use sophisticated non-LTE model atmospheres. Studies of very large samples typically involve a grid-based star-by-star approach \citep{Holgado+2018, Castro+2018, Ramachandran+2019}.
Here, we exploit a new pipeline for the efficient analysis of very large samples of optical OB spectra \citep{Bestenlehner2024}. This study of the entire BLOeM OB sample will be complemented by bespoke studies of sub-samples, and upcoming studies focused on specific quantities such as rotational velocities (Berlanas et al. in prep).

We present BLOeM datasets in Section~\ref{obs} and briefly describe the pipeline used to analyse OB stars in Section~\ref{analysis}. We present our derived physical parameters in Section~\ref{properties}, including comparisons with previous results. Section~\ref{rotation} discusses rotational velocities, while
Section~\ref{gbat} presents tailored analyses of a subset of BLOeM stars using the grid-based {\it interactive} tool {\sc iacob-gbat} \citep{SimonDiaz+2011} for comparison with pipeline results. Spectroscopic masses are compared to evolutionary mass determinations in Section~\ref{masses}, followed by a consideration of the BLOeM O star sample within the
context of the global SMC population in Section~\ref{SMC}. Finally, brief conclusions are drawn in Section~\ref{conclusions}. Appendices include pipeline results, comparisons with previous studies and an updated catalogue of O stars in the SMC, since there have been many discoveries since the census of \citet{Bonanos+2010}.

%%%%%%%%%%%%%%%%%%%%%%%%%%%%%%%%%%%%%%%%%%%%%%%%%%%%%%%%%%%%%%%%%%%%%%%%%%%%%%%%%%%%%%%%%%%%%%%%%%%%%%

\begin{table}
\caption{Breakdown of 847 OB stars identified in the BLOeM survey \citep{Shenar+2024} by spectral type and single versus multiple, according to analysis of the initial 9 epoch dataset \citep{BLOeM_O, BLOeM_B, BLOeM_Bsuper, BLOeM_OeBe, BLOeM_BAF}. Sources excluded from the present study (69 sources) include a subset of SB2 binaries, OBe stars plus a few OB stars contaminated by strong nebular emission. Miscellaneous targets excluded from analysis are B[e] supergiants (BLOeM 2-116, 3-012, 4-055), sources with B+A composite appearance (BLOeM 3-006, 8-009, 8-056) and two B9 supergiants (BLOeM 5-036, 5-086) for which fits were unsatisfactory.
% or {\sc PoWR} \citep{Grafener+2002, Sander+2015}. 
}
% Mass-loss rates are presented as $\dot{M}/\sqrt{f_{v}}$ to reflect differences in adopted or derived wind clumping factors.
\label{BLOeM}
\begin{center}
\begin{tabular}
{
l@{\hspace{2mm}}
r@{\hspace{2mm}}
r@{\hspace{2mm}}
r@{\hspace{2mm}}
r@{\hspace{2mm}}
r@{\hspace{2mm}}
r
}

    \hline\hline
    Spectral & \multicolumn{2}{c}{-- Included --} & \multicolumn{3}{c}{-- Excluded --} & Total\\
    Type     &  Single & Multiple  &  Single & Multiple &  Misc.    & \\ 
\hline
O-type &   71 &  66 & 14 & 8 & 0 & 159\\ 
B-type &  380 & 261 & 32 & 7 & 8 & 688 \\
\hline
Total  & 451  & 327 & 46 & 15 & 8 & 847\\
\hline
\end{tabular}
\end{center}
\end{table}

\section{BLOeM observations}\label{obs}

The BLOeM survey (PI: Shenar, Co-PI: Bodensteiner) involves 25 epoch spectroscopy of 929 massive stars with FLAMES at the VLT, using the LR02 setup ($\lambda\lambda$3950-4550\AA, $R$=6,200) between October 2023 and late 2025. Targets were drawn from a {\it Gaia} catalogue of bright, blue stars, which peaks at $G \sim$ 14.6 mag, and has a limiting magnitude of $G$ = 16.5 mag, as shown in figure~2 of \citet{Shenar+2024}.
The use of 8 FLAMES fields allowed a reasonable fraction of the SMC to be considered, albeit with limited sampling of young, luminous stars in rich star-forming regions \citep[e.g.][]{Evans+2006, Dufton+2019}. The data reduction process is described in \citet{Shenar+2024}.  

For the present study the first 9 epochs (Oct 2023 to Dec 2023) are considered, with individual spectroscopic datasets obtained by co-adding two normalized back-to-back 615 sec exposures. Average radial velocities, $\varv_{\rm rad}$, and dispersions, $\sigma(\varv_{\rm rad})$ are obtained for all OB stars and presented in Table~\ref{table:targets} with the exception of stars exhibiting unusual spectral features (e.g. B[e] supergiants).

\begin{figure*}
\centering
  \includegraphics[width=2\columnwidth]{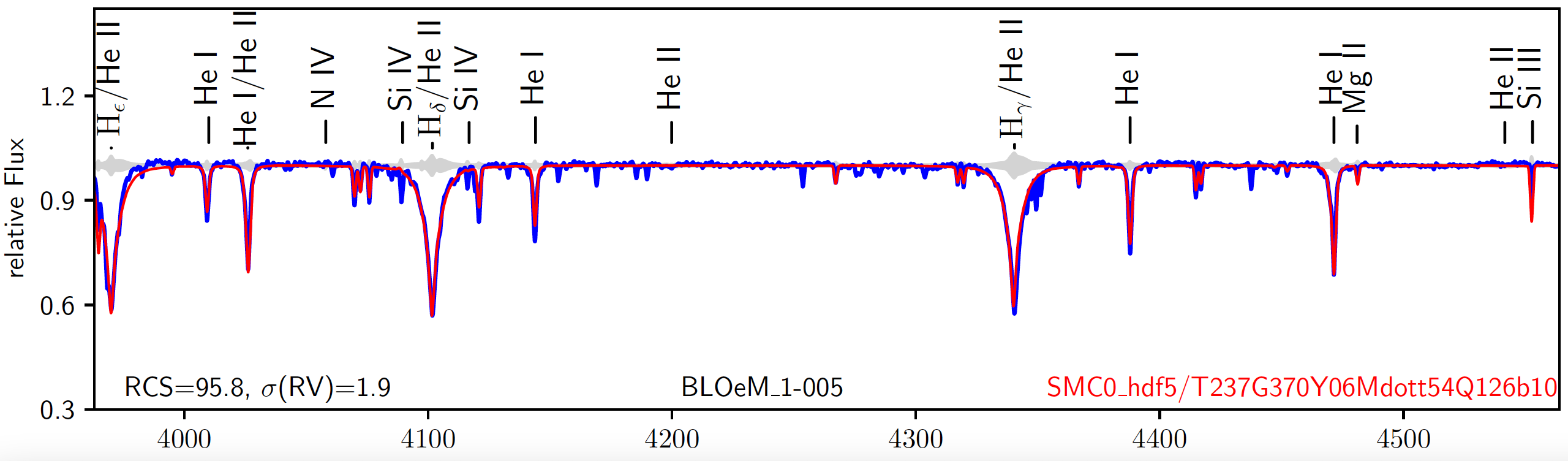}
  \includegraphics[width=2\columnwidth]{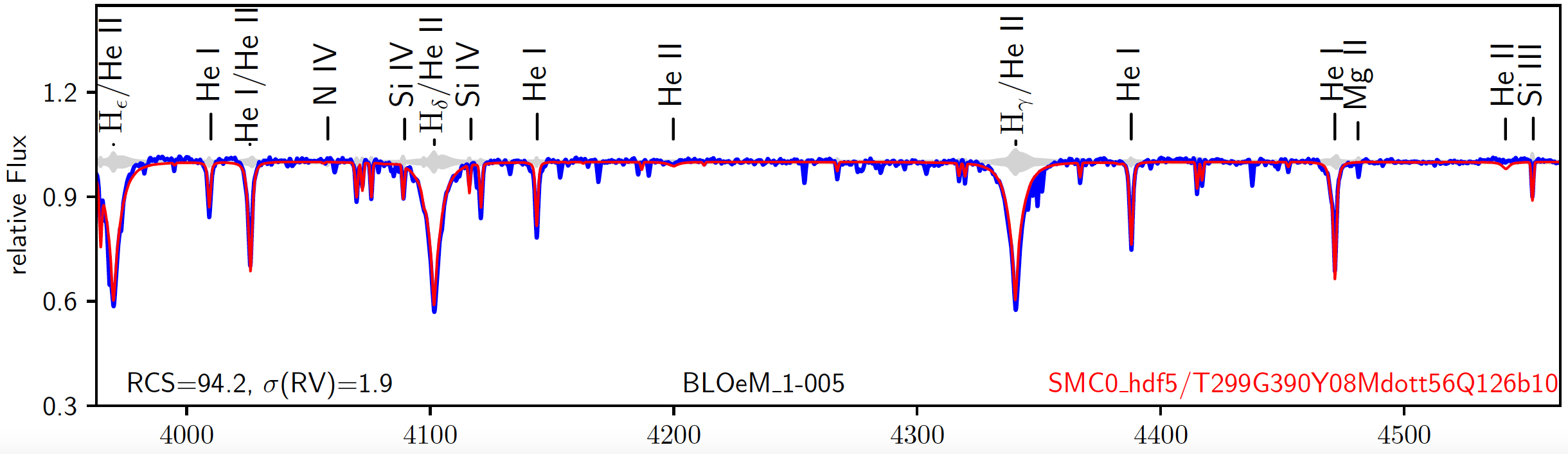}
  \caption{Comparison between the pipeline fits (red) obtained for BLOeM 1-005 (B1\,II, blue) for the unweighted solution (upper panel, $T_{\rm eff} = 23.6^{+0.7}_{-0.8}$ kK, $\log g$/(cm s$^{-2}$) = $3.64^{+0.15}_{-0.16}$) versus the solution with additional weight given to Si\,{\sc iv} $\lambda$4089 (lower panel, $T_{\rm eff} = 29.9\pm1.2$~kK, $\log g$/(cm s$^{-2}$) = $3.93^{+0.34}_{-0.17}$). It is apparent that both solutions reproduce H\,{\sc i} and He\,{\sc i} lines plus Si\,{\sc iii} $\lambda$4553, with the higher temperature solution matching Si\,{\sc iv} $\lambda\lambda$4089-4116 and the lower temperature solution reproducing Mg\,{\sc ii} $\lambda$4481. The grey shaded area is the square root of the diagonal elements of the model-error uncertainty matrix calculated by the pipeline. RCS refers to the reduced $\chi^{2}$ and $\sigma$(RV) refers to the dispersion in radial velocities.} 
  \label{BLOeM_1-005}
\end{figure*}

The primary purpose of multi-epoch spectroscopy is to investigate the multiplicity of massive stars at low metallicity. Binarity is assessed via peak-to-peak radial velocities of $\geq$20 km\,s$^{-1}$ at the 4$\sigma$ significance level, with the initial nine epoch dataset split into five studies, focused on O stars \citep{BLOeM_O}, OBe stars \citep{BLOeM_OeBe}, non-supergiant early B stars \citep{BLOeM_B}, early B supergiants \citep{BLOeM_Bsuper} and cooler supergiants \citep{BLOeM_BAF}. Short period spectroscopic binaries (some of which may be higher order systems) from these studies are indicated in Table~\ref{table:targets}, and include supergiants for which variability arises either from a companion (SB1) or intrinsic line profile variability (lpv). The true multiplicity fraction of BLOeM stars is doubtless higher, such that stars categorised as `single' are preliminary, with definitive results awaiting analysis of the complete 25 epoch dataset.

\citet{Shenar+2024} also describes cross-correlation and co-addition of individual normalized observations to improve S/N for classification and quantitative analysis. This
is the primary dataset used in the present study. The LR02 setup includes the majority of diagnostics necessary for quantitative studies of OB stars, including multiple He\,{\sc i-ii} lines for the determination of temperatures for O and early B stars, plus N\,{\sc iv} $\lambda$4058 for early O stars. Si\,{\sc iv} $\lambda\lambda$4089--4116, Si\,{\sc iii} $\lambda$4553, Si\,{\sc ii} $\lambda\lambda$4128--31 and Mg\,{\sc ii} $\lambda$4481 are available for B stars lacking He\,{\sc ii} diagnostics, together with multiple He\,{\sc i} lines. H$\gamma$ and H$\delta$ permit surface gravities to be determined, noting H$\epsilon$ lies at the edge of the LR02 spectral coverage. H$\alpha$ and He\,{\sc ii} $\lambda$4686 are excluded, so it is not possible to determine wind properties from the current BLOeM observations.

The grid used in our spectroscopic pipeline is suitable for the determination of physical parameters of OB stars, so 81 AF supergiants are excluded. Their physical parameters are considered by \citet{BLOeM_BAF}. In addition, the subset of SB2 systems in which both components are prominent in the co-added datasets are also excluded, as are OB stars in which the Balmer (and sometimes He\,{\sc i}) lines exhibit strong emission components, i.e. OBe stars and OB stars within regions of strong nebulosity \citep[e.g. NGC~346,][]{Evans+2006}. We also exclude B[e] supergiants from our analysis. 

In total we present analyses of 778 OB stars, representing 84\% of the BLOeM sample of 929 stars, or 92\% of the 847 OB stars. Confirmed or suspected spectroscopic binaries (SB1, SB2, SB3) are indicated in Table~\ref{table:targets} and represent 42\% (329 stars) of the total sample studied. A breakdown of OB statistics from BLOeM \citep{Shenar+2024} and the present study is provided in Table~\ref{BLOeM}.

% AzV 177
% AzV 377
% AzV 388
% AzV 476: O4 primary (Pauli+2022)

\begin{figure*}
\centering
  \includegraphics[width=2\columnwidth]{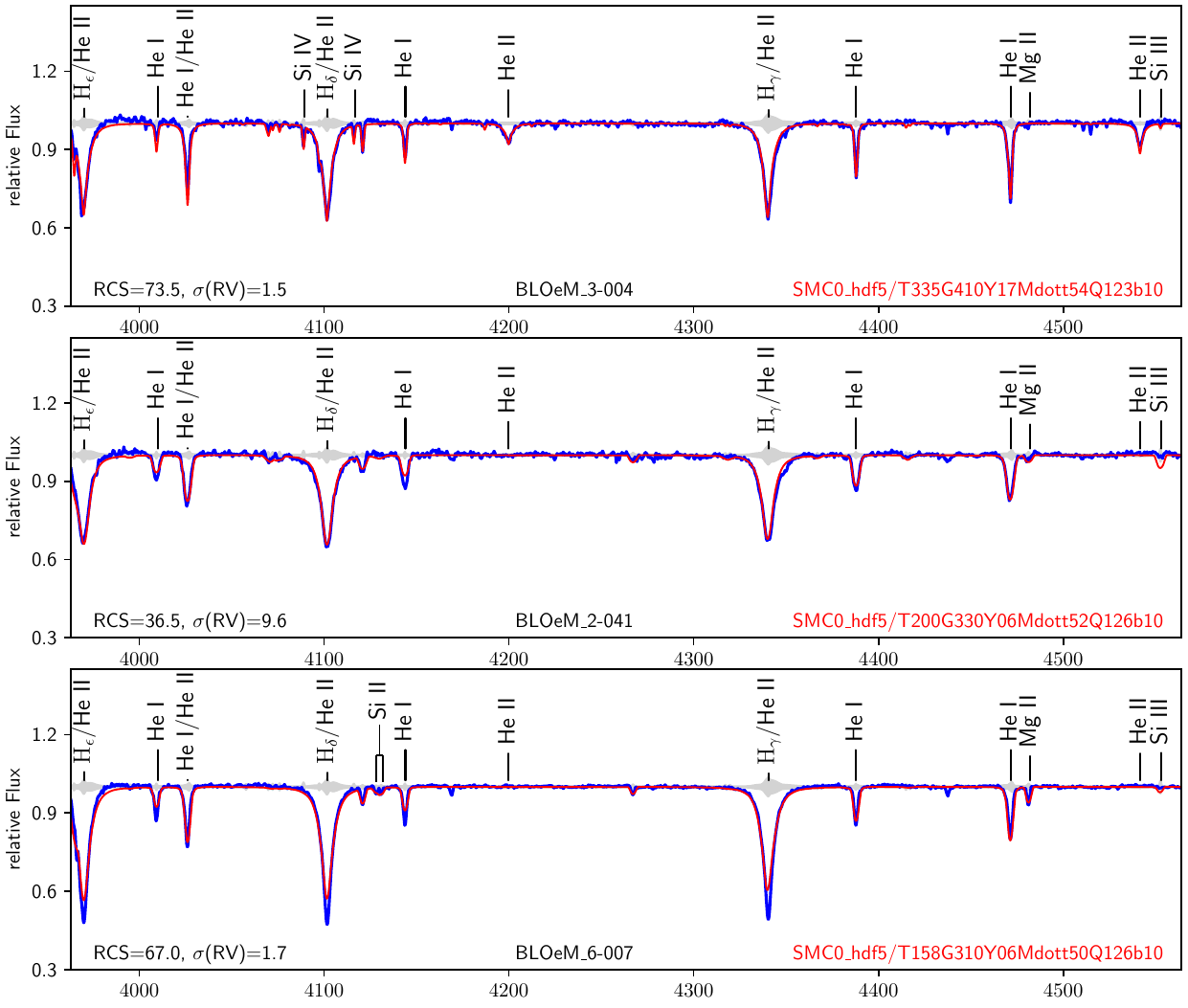}
  \caption{Comparison between the pipeline fits (red) obtained for visually faint OB stars, from top to bottom: BLOeM 3-004 (O9.7\,IV:) for which $T_{\rm eff} = 33.7^{+1.5}_{-2.3}$~kK, $\log g$/cm s$^{-2}$ = $4.12^{+0.34}_{-0.43}$, BLOeM 2-041 (B2:\,II), for which $T_{\rm eff} = 20.1^{+4.7}_{-2.7}$~kK, $\log g$/cm s$^{-2}$ = $3.30^{+0.74}_{-0.40}$ and BLOeM 6-007 (B5\,II), for which $T_{\rm eff} = 15.9\pm 0.8$~kK, $\log g$/cm s$^{-2}$ = $3.07^{+0.17}_{-0.29}$. The grey shaded area is the square root of the diagonal elements of the model-error uncertainty matrix calculated by the pipeline. RCS refers to the reduced $\chi^{2}$ and $\sigma$(RV) refers to the dispersion in radial velocities.}
  \label{BLOeM_3-004_BLOeM_2-041_BLOeM_6-007}
\end{figure*}

\section{Spectroscopic pipeline}\label{analysis}

For our spectroscopic analysis pipeline we employ grids of synthetic model spectra computed with v10.6 of
the non-LTE atmosphere code {\sc fastwind} \citep{Puls+2005, Rivero-Gonzalez+2012} including  H, He, C, N, O, Si and Mg as explicit elements at the SMC metallicity ($0.2Z_{\odot}$). 
% respectively with initial, semi-processed and fully processed CNO composition. 
Grids covered the following parameter space $\log T_{\rm eff}$ (K) over [4.0, 4.775] in 0.025 dex steps, corresponding to 10kK $\leq T_{\rm eff} \leq$ 60kK, $\log g$ (cm\,s$^{-2}$) over [1.5, 4.5] in 0.2 dex steps, and Helium abundances in mass-fraction $Y$ over [0.15, 0.55] in 0.05 steps. Convergence difficulties were experienced at the lowest temperatures ($T_{\rm eff} \leq$ 15kK) impacting on fits to late B supergiants.

Although the FLAMES LR02 setup excludes typical wind diagnostics, the wind-strength parameter $\log Q$ was retained as a variable, ranging from --11.4 to --15.0 in 0.3 dex steps, where $Q = \dot{M} (R_{\ast} v_{\infty})^{-3/2}$ with units $M_{\odot}$\,yr$^{-1}$, $R_{\odot}$ and km\,s$^{-1}$. 
%In addition, we varied nitrogen abundances for high temperature grids from $\log T_{\rm eff}$/K = 4.6 to 4.775, ranging from initial N-abundance \citep{Vink+2023} to fully CNO processed, as the ionisation balance between nitrogen ions becomes the main temperature diagnostic.
A smooth wind with volume filling factor $f_{\rm v} = 1 $ and $\beta = 1$ velocity law was assumed and the micro-turbulent velocity was set to $\varv_{\rm mic}=10$ km\,s$^{-1}$ in the model grids.

Typical macro-turbulent velocities for OB stars are in the range between a few km\,s$^{-1}$ to several tens of km\,s$^{-1}$, although can reach  higher values \citep{SimonDiaz+2017}. The velocity resolution of the LR02 FLAMES dataset is 48~km\,s$^{-1}$. We convolved our synthetic grid with a fixed $\varv_{\rm mac} = 20$ km\,s$^{-1}$ and assumed any additional broadening is due to rotation, with projected rotational velocities of $\varv_{\rm e} \sin i = [0, 10, 20, 35, 50, 75, 100, 150, 200, 250, 300, 350, 400, 450, 500]$ km\,s$^{-1}$.
% For the terminal wind velocities, $v_{\infty}$, in our grid we adopted the empirical calibration of Magellanic Cloud OB stars by \citet{Hawcroft+2023}, namely
%\begin{equation}
%    \varv_{\infty} = \left[ (92 \pm 3) T_{\rm eff}/{\rm kK)} - (1040 \pm 100) \right] Z/Z_{
%    \odot}^{(0.22\pm0.03)} 
%\end{equation}\label{Hawcroft}
%in km\,s$^{-1}$ for LMC ($T_{\rm eff} \geq$ 15 kK, $Z = 0.5 Z_{\odot}$) and SMC ($T_{\rm eff} \geq$ 20 kK, $Z = 0.2 Z_{\odot}$) OB stars. These results involved Sobolev with Exact  Integral (SEI) fits to ULLYSES \citep{ULLYSES} observations supplemented with literature results, and was extrapolated to lower temperature with a minimum $\varv_{\infty}$ = 250 km\,s$^{-1}$. Individual wind velocities are used when determining mass-loss rates from wind densities, $Q$, as discussed in Section~\ref{vely}.

A complete description of the pipeline\footnote{\url{https://github.com/jbestenlehner/mdi\_analysis\_pipeline}} is provided in \citet{Bestenlehner2024}. In brief, we used the full FLAMES spectral range including the observational error spectrum by utilising a $\chi^2$ minimisation Ansatz: 
\begin{equation}
    \chi^2 = (\vec{d} - \mathrm{R}\vec{s})^{\mathrm T}\mathrm{N}^{-1}(\vec{d} - \mathrm{R}\vec{s})
\end{equation}
with $\vec{d}$ the observed and $\vec{s}$ the synthetic spectra, $\mathrm{R}$ the instrumental responds matrix and observational, diagonal error matrix $\mathrm{N}$. As model uncertainties should be budgeted into the parameter determination, we `de-idealised' the model spectrum $\vec{s}$ according to \citet{Bestenlehner2024}.

Our sample is fairly heterogeneous, ranging from early O dwarfs to late B supergiants, albeit with a large number of early B stars. Therefore, the model-error is averaged over the entire parameter space of our sample. This impacted the overall performance of the pipeline, because a meaningful model-error should ideally be based on a sample of similar objects \citep[c.f. the discussion in][]{Bestenlehner2024}.
%\subsection{Synthetic spectra grids}

%The velocity resolutions for the XShooter UVB and VIS arm are $\sim 45$ and $\sim 25$km/s. 

The combined BLOeM datasets are cross-correlated with synthetic spectral templates to determine a mean radial velocity ($\varv_{\rm rad}$), and then corrected for this shift before being sampled on the wavelength grid of the synthetic spectra. Fig.~\ref{rv_ra-dec} shows radial velocities of single OB stars with respect to the +159 km\,s$^{-1}$ mean value of the BLOeM sample. For comparison, \citet{Hilditch+2005} obtained mean systemic velocities of +196 km\,s$^{-1}$ for  OB eclipsing binaries in the SMC while \citet{EvansHowarth2008} obtained a mean of +172.0 km\,s$^{-1}$ for the 2dFS sample, and highlighted differences between the bar (+167.4 km\,s$^{-1}$) and the wing (+189.5 km\,s$^{-1}$) which are also apparent in Fig.~\ref{rv_ra-dec}.

% 2 sigma outliers: SMC AzV 234 (89), ELS 26 (276), AzV 267 (256), AzV 296 (260), 372 (265)

% 2 sigma outliers: LMC Sk-67 22 (472), Sk-65 47 (423), Sk-67 191 (388), BI 237 (394), Sk-66 171 (421)

Hydrogen lines are the most prominent spectroscopic features in the blue spectra of OB stars and dominate the $\chi^2$, with He lines sometimes as weak as metal lines. Firstly, we initialize a wavelength array with 0.1\AA\ spacing around the spectral lines in our {\sc fastwind} LINES-list. Secondly, we increased the number of wavelength points by a factor of 5 beyond $\pm5$\AA\ of the central wavelength of the Balmer lines, because $\log g$ is based on the pressure-broadened wings. Thirdly, we increased the number of wavelength points by a factor of 25 within $\pm 1$\AA\ of the central wavelength of the Helium and metal lines.

Our default approach is not to increase the weighting of any specific spectral features for those samples involving a broad range of spectral types, such as BLOeM. However, weak Si\,{\sc iv} $\lambda\lambda$4089, 4116 features were poorly reproduced for a large subset of early B stars, leading to an unphysical gap in solutions close to $T_{\rm eff} \sim$ 25 kK. 

Increased weight for both Si\,{\sc iv} lines improved temperatures to the detriment of surface gravities (both lie within the wing of H$\delta$) so we ultimately elected to adopt an increased weighting  of solely Si\,{\sc iv} $\lambda$4089. The higher weighing of $\lambda$4089 generally led to improved fits, without adversely affecting surface gravities. This was achieved by incorporating more data points around this line (4088.85$\pm$0.25\AA). 

O\,{\sc ii} $\lambda$4089.29 \citep{Wenaker1990} was not included in the {\sc fastwind} line list for spectral line synthesis, but contributes to the Si\,{\sc iv} $\lambda$4089 feature in early B stars \citep[see][]{HardorpScholz1970, BeckerButler1988, Kilian+1991, deBurgos+2023b}. However, the pipeline is designed to handle model deficiencies such as missing spectral lines or inaccurate physics \citep[see][Sect.~2]{Bestenlehner2024}. 

Test calculations incorporating O\,{\sc ii} $\lambda$4089\footnote{O\,{\sc ii} oscillator strengths were obtained from the Vienna Atomic Line Database (VALD), which compare closely to R-Matrix calculations from \citet{BeckerButler1988}.} have been undertaken for {\sc fastwind} models at $\log g$/(cm\,s$^{-2}$) = 3.3 for $T_{\rm eff}$ = 30~kK, 25~kK and 20~kK, indicating that O\,{\sc ii} $\lambda$4089 is a minor, major and primary contributor to the blend, respectively.  At $T_{\rm eff}$ = 25~kK the addition of O\,{\sc ii} would significantly boost the strength of the $\lambda$4089 feature, and so would impact on the favoured solution. 
At $T_{\rm eff}$ = 30~kK several other high ionization lines (e.g. He\,{\sc ii}) are present, so the contribution from O\,{\sc ii} is not anticipated to adversely impact the favoured solution. At $T_{\rm eff}$ = 20~kK, the blend is weak, with primarily Si\,{\sc iii} and Mg\,{\sc ii} observed, so again the solution is not anticipated to be impacted by the omission of O\,{\sc ii} $\lambda$4089.

% although its central wavelength is sufficiently offset from Si\,{\sc iv} to not lie within the spectral range for which higher weighting is applied. 

We have also considered an alternate increased weighting of Si\,{\sc iv} $\lambda$4116, the weaker component of the doublet, but ultimately favoured $\lambda$4089 owing to its greater strength in early B stars. To reiterate, many spectral lines contributed to the pipeline fit (including Si\,{\sc iv} $\lambda$4116), in contrast to usual practice which focus {\it solely} on Si lines in early B stars \citep[e.g.][]{Dufton+2018}, albeit with additional weighting to Si\,{\sc iv} $\lambda$4089 that produced more robust solutions.

By way of example, Fig.~\ref{BLOeM_1-005} illustrates unweighted (upper panel) and weighted (lower panel) solutions (red) for BLOeM 1-005 (B1\,II, blue) for which $T_{\rm eff}$ = 23.6$^{+0.7}_{-0.6}$ kK, $\log g$/(cm s$^{-2}$) = $3.64^{+0.15}_{-0.16}$ and $T_{\rm eff} = 29.9\pm1.2$~kK, $\log g$/(cm s$^{-2}$) = $3.93^{+0.34}_{-0.17}$ are obtained, respectively. 
The unweighted solution reproduces most features (including Mg\,{\sc ii} $\lambda$4481) aside for Si\,{\sc iv} $\lambda$4089--4116, with Si\,{\sc iii} $\lambda$4553 somewhat too strong. In contrast, the weighted solution addresses the mismatch to the Si\,{\sc iv} $\lambda\lambda$4089--4116 doublet, and improves the match to Si\,{\sc iii} $\lambda$4553, albeit at the expense of Mg\,{\sc ii} $\lambda$4481. BLOeM 1-005 is representative of OB stars analysed in this study, since its {\it Gaia} $G$-band brightness ($G$ = 14.6 mag) corresponds to the photometric peak of the BLOeM sample.

The stellar atmosphere grid is non-rectilinear since a subset of models did not converge or failed to compute due to unphysical parameter space (e.g. Eddington limit). Before determining the uncertainties we fill the gaps in the probability distribution function (PDF) with zero-values, so that the PDF becomes a 4D($T_{\rm eff}-\log g - \log Q - Y$) rectilinear grid. The 4D grid was then interpolated to artificially increase the grid resolution using the multidimensional interpolation function {\sc scipy.interpolate.interpn} with cubic-spline method to obtain more accurate parameters and less grid-specific uncertainties.

\begin{figure}
\centering
  \includegraphics[width=\columnwidth]{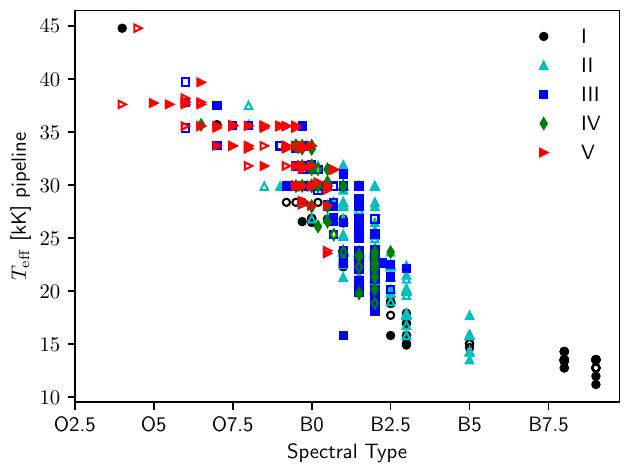}
  \caption{Pipeline effective temperatures, $T_{\rm eff}$ for BLOeM OB stars using spectral types from \citet{Shenar+2024}. Single stars according to analysis of the initial 9 epochs of BLOeM \citep{BLOeM_O, BLOeM_B, BLOeM_Bsuper, BLOeM_OeBe, BLOeM_BAF},  are open symbols, multiples are filled symbols.
  %Evolutionary masses of post-main sequence stars are determined from inspection of SMC tracks from \citet{Schootemeijer+2019}.
  }
  \label{SpT_T}
\end{figure}

\begin{figure}
\centering
  \includegraphics[width=\columnwidth]{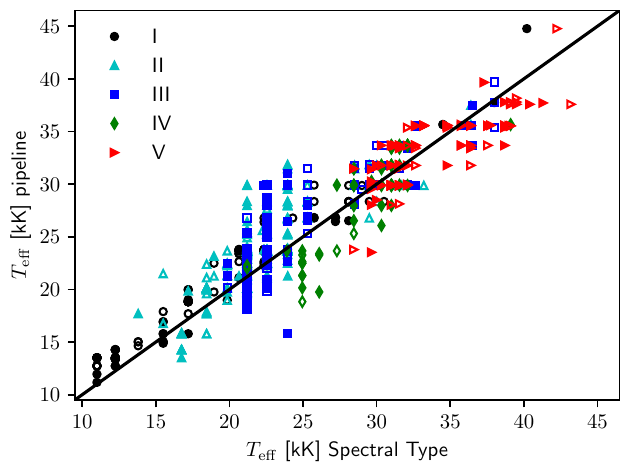}
  \caption{Comparison between adopted $T_{\rm eff}$ of BLOeM OB stars from SMC calibrations \citep{Shenar+2024} and pipeline-derived, $T_{\rm eff}$. Single stars according to analysis of the initial 9 epochs of BLOeM \citep{BLOeM_O, BLOeM_B, BLOeM_Bsuper, BLOeM_OeBe, BLOeM_BAF} are open symbols, multiples are filled symbols.
  %Evolutionary masses of post-main sequence stars are determined from inspection of SMC tracks from \citet{Schootemeijer+2019}.
  }
  \label{calib}
\end{figure}

\begin{figure}
\begin{center}
  \includegraphics[width=0.85\columnwidth,bb=48 169 523 640]{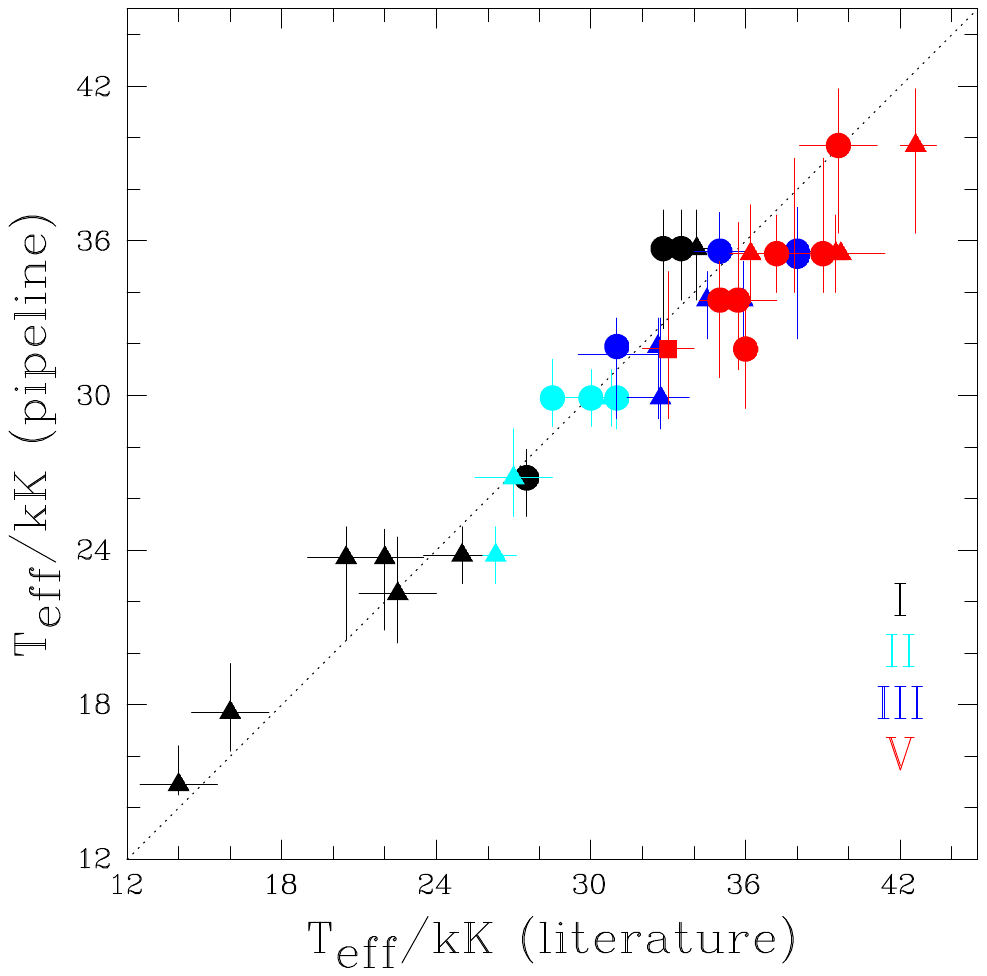}
  \caption{Comparison between $T_{\rm eff}$ for BLOeM OB stars from literature studies (circles: {\sc cmfgen}, triangles: {\sc fastwind}, squares: {\sc tlusty}) and the current pipeline, colour coded by luminosity class. References are provided in the Appendix in Tables~\ref{O-lit}-\ref{B-lit}.}
  \label{OB_Teff}
  \end{center}
\end{figure}
We used the following standard deviations in 4D; $1\sigma$: 0.0902, $2\sigma$: 0.5940 and $3\sigma$: 0.9389, following \citet{Wang+2015}. CNO abundances and $\varv_{\rm e} \sin i$ were not included as they mainly improve the fit to the nitrogen lines and the line broadening, but also a 6D grid interpolation becomes computationally very expensive. In a few instances the 4D grid leads to multiple
minima in which local minima with the lowest $\chi^{2}$ solutions  preferred. 
In the few instances for which $\Delta T/T_{\rm eff} >$ 10\%, there are no significant differences between the fits obtained.

% 1-052, 3-016, 3-025, 6-090, 7-114

% We estimated N-abundances in the 2D$-T_{\rm eff}-N$ PDF for stars hotter than $\log T_{\rm eff}$/K $\geq 4.6$, because the ionisation balance of the nitrogen lines becomes the main temperature diagnostic, and projected rotational velocities in 1D-PDF, because line broadening is largely independent of the stellar parameters.

In order to determine bolometric luminosities we adopted a distance modulus of 18.98 mag \citep{Graczyk+2020} for the SMC, and used optical \citep{GAIA-EDR3} and near-IR photometry for the determination of interstellar reddening. Note that K$_{\rm s}$-band
photometry presented in Table~A2 of \citet{Shenar+2024} is a mixture of 2MASS \citep{2MASS} and aperture photometry from VMC \citep{Cioni+2011} rather than PSF photometry of the latter survey. For the present study K$_{\rm s}$-band photometry are utilised, either from VMC PSF photometry or 2MASS Point Source Catalog (PSC) if $m_{K_{\rm s}} < 13.2$ mag (see Table~\ref{table:targets}).

Individual reddening parameters $R_{5495}$ and $E_{4405 - 5495}$ were obtained by fitting individual photometric fluxes to the model spectral energy distribution (SED) employing the reddening law of \citet{MaizApellaniz+2014}. $R_{\rm V} = 3.0$ for the SMC bar has been determined by \citet{Gordon+2024}. Inferred interstellar extinctions are modest, with an average of $A_{5495} \simeq A_{\rm V} = 0.39 \pm 0.14$ mag, as expected for {\it Gaia} colour selected targets towards SMC sightlines, with individual values included in Table~\ref{table:targets}.

% Neither 2MASS nor VMC PSF photometry is available for BLOeM 7-086, so no bolometric luminosity is provided for this star.

\begin{figure*}
%\centering
  \includegraphics[width=1.5\columnwidth]{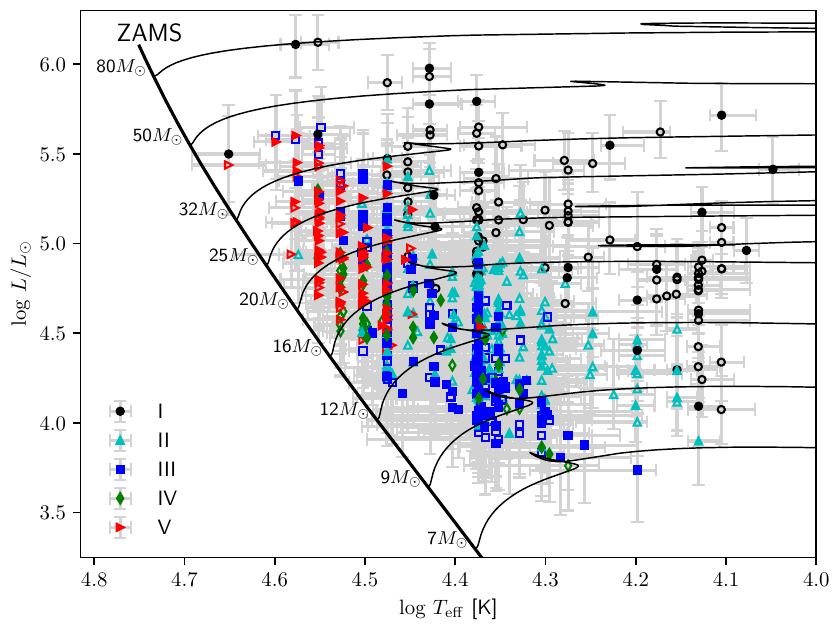}
  \caption{Hertzsprung-Russell diagram  of the BLOeM OB sample (colour coded by luminosity class). Open symbols are single according to analysis of the initial 9 epochs of BLOeM \citep{BLOeM_O, BLOeM_B, BLOeM_Bsuper, BLOeM_OeBe, BLOeM_BAF}, filled symbols are multiple.  Evolutionary tracks for SMC massive stars are from %\citet{Brott+2011}. 
  %Evolutionary masses of post-main sequence stars are determined from inspection of SMC tracks from 
  \citet{Schootemeijer+2019} for non-rotating stars ($\alpha_{\rm SC} = 10, \alpha_{\rm OV} = 0.33$).
  }
  \label{HRD}
\end{figure*}

\section{Physical properties of BLOeM OB stars}\label{properties}

Table~\ref{table:targets} presents inferred physical parameters for 778 OB stars from BLOeM. For completeness we include radial velocities (and dispersions) of all OB stars. Online material includes spectral fits for each star (model in red, observations in blue) at 
\href{10.5281/zenodo.15526149}. 69 SB2 systems, OBe stars, OB stars with strong nebular emission and B[e] supergiants are excluded from our analysis.

By way of example, Figure~\ref{BLOeM_3-004_BLOeM_2-041_BLOeM_6-007} presents the solution (model in red) for several visually faint OB stars, from top to bottom: BLOeM 3-004 (O9.7\,IV:, $G$ = 16.0 mag), BLOeM 2-041 (B2:\,II, $G$ = 16.2 mag) and BLOeM 6-007 (B5\,II, $G$ = 15.0 mag). The overall fit quality to H\,{\sc i}, He\,{\sc i-ii}, Si\,{\sc iv} $\lambda$4088--4116, Si\,{\sc ii} $\lambda$4128--31 and Mg\,{\sc ii} $\lambda$4481 lines is satisfactory, although  Si\,{\sc iii} $\lambda$4553 is over predicted in BLOeM 2-041, and the cores of strong He\,{\sc i} and Balmer lines are under predicted in BLOeM 6-007.

% Two realisations of the pipeline were run in order to assess the robustness of the results obtained from FLAMES/LR02 datasets. For the overwhelming majority of stars, inferred temperatures agreed to within 1\% for the overwhelming majority of stars, but 8 stars (1\% of the total) differed by 5\%. For these 8 stars, some of which were affected by strong nebular emission, both fits were inspected and the favoured solution adopted. Uncertainties provided in Table~\ref{table:targets} follow from our pipeline, though exclude systematic differences between different codes and approaches set out in \citet{Sander+2024}. 

% 4-057, 4-070, 5-004, 6-016, 6-032 6-037, 7-073, 8-018

\subsection{Stellar temperatures}\label{temp}

Figure~\ref{SpT_T} compares BLOeM spectral types with pipeline-derived effective temperatures. Overall there is a clear correlation between spectral type and inferred temperature, although there is a large (unrealistic) spread in temperatures for stars close to B1. This spread is highlighted in Fig.~\ref{calib}, which compares $T_{\rm eff}$ adopted from calibrations in \citet{Shenar+2024} with pipeline values. This issue arises despite the increased weighting to Si {\sc iv} $\lambda$4089, with lower temperatures obtained if Si\,{\sc iv} is not reproduced (recall Fig.~\ref{BLOeM_1-005}).

A subset of the BLOeM stars have been subject to earlier quantitative spectral analysis efforts, primarily those in common with the ULLYSES/XShootU sample \citep{ULLYSES, Vink+2023}. We compare our derived temperatures to detailed literature results in the Appendix in Table~\ref{O-lit} (\ref{B-lit}) for O-type (B-type) stars. Previous studies utilised UV and optical spectroscopic datasets, plus either {\sc cmfgen} \citep{Hillier-Miller1998}, {\sc fastwind} \citep{Puls+2005, Rivero-Gonzalez+2012} or {\sc tlusty}  \citep{HubenyLanz1995}. 

Overall pipeline-derived temperatures agree reasonably well with detailed studies within the uncertainties, as illustrated for OB stars in Fig.~\ref{OB_Teff}, although large uncertainties are obtained in some instances (e.g. BLOeM 4-020, B0\,Ib-Iab). For the BLOeM subset of late B stars, \citet{BLOeM_BAF} have estimated temperatures from comparison with {\sc cmfgen} models. Pipeline temperatures are systematically warmer for B5 and B8 subtypes by 1.0 kK, and 0.9 kK, respectively, increasing to 2.4 kK for B9 supergiants, arising from {\sc fastwind} model convergence difficulties at the lowest temperatures (He\,{\sc i} lines are generally overestimated). 

In addition to previously detailed spectroscopic studies for BLOeM OB stars, \citet{Castro+2018} have also determined temperatures of a large sample of SMC field OB stars from the RIOTS4 survey \citep{Lamb+2016} using a grid of {\sc fastwind} models. \citet{Castro+2018} relied solely on H and He diagnostics, so their temperatures will be less robust for B stars in which He\,{\sc ii} is not observed. 25 OB stars are in common between the present study and \citet{Castro+2018}, listed in the Appendix (Table~\ref{RIOTS4}), with $\log T_{\rm eff}$(pipeline) - $\log T_{\rm eff}$(Castro) = +0.04$\pm$0.10 dex. 

\citet{Bestenlehner+2025} have also applied the pipeline described in Section~\ref{analysis} to XShootU datasets \citep{Vink+2023}. 30 OB stars are in common between the present study and \citet{Bestenlehner+2025}, with parameters compared in the Appendix (Table~\ref{OB-BLOeM-XShootU}). Our derived temperatures agree well with the XShootU pipeline analysis, with $\log T_{\rm eff}$(BLOeM) - $\log T_{\rm eff}$(XShootU) = +0.00$\pm$0.02 dex, indicating that the lack of wind spectral diagnostics does not adversely impact stellar temperatures. We will revisit effective temperatures in Section~\ref{gbat}.

\subsection{Stellar luminosities}

Figure~\ref{HRD} presents pipeline results for OB stars in a Hertzsprung-Russell (HR) diagram, superimposed upon non-rotating SMC metallicity evolutionary tracks from \citet{Schootemeijer+2019}, for which semiconvection and overshooting parameters follow \citet{Brott+2011}. This represents a more robust HR diagram than that presented in \citet{Shenar+2024} which was based upon spectral type calibrations. 

The lack of O stars close to the theoretical
zero age main sequence (ZAMS) is striking, in common with previous  Milky Way \citep{Holgado+2020}, LMC \citep{Sabin-Sanjulian+2017, Ramachandran+2018b} and SMC \citep{Castro+2018, Ramachandran+2019, Schootemeijer+2021} analyses of large samples of OB stars. O stars {\it are} observed close to the ZAMS in young, rich star clusters such as NGC~3603 in thee Milky Way \citep{Melena+2008} and R136 in the LMC \citep{Crowther+2016, Brands+2022}. No close counterparts to R136 exist in the SMC, with the extended star-forming region NGC~346 also deficient in luminous ZAMS stars \citep{Rickard+2022}, although compact clusters whose O stars are located close to the ZAMS have been observed \citep{HeydariMalayeri+1999a, HeydariMalayeri+1999b, Martins+2004}.

Aside from the deficit of ZAMS stars and those close to $T_{\rm eff} \sim$ 26kK ($\log T_{\rm eff}$/K $\sim 4.4$, recall Sect.~\ref{temp}) it is apparent that a large fraction of the BLOeM OB stars lie close to the terminal age main-sequence (TAMS), although the precise TAMS is not well established from evolutionary models. One would expect very few post-MS  for standard single star evolution, since evolution is predicted to be rapid toward cool supergiants. Mid to late B supergiants are unambiguously post-MS stars \citep[see also][]{deBurgos+2025}, whereas the situation for early B (super)giants is less clear (B dwarfs are too faint given the BLOeM selection criteria). From a comparison with evolutionary predictions set out in Section~\ref{masses}, 57 stars from the total sample of 778 are unambiguously in a post-MS evolutionary phase, providing the
TAMS from \citet{Brott+2011} is correct.

A major advantage of BLOeM over the majority of previous spectroscopic studies of the Magellanic Clouds is the multi-epoch nature of the survey. Figure~\ref{HRD2} provides separate HR diagrams for single (upper panel) and multiple (lower panel) systems,  together with \citet{Brott+2011} tracks, potentially highlighting binary interaction products \citep[see e.g.][]{Menon+2024}.

Stellar luminosities of individual BLOeM stars are provided in Table~\ref{table:targets}. The average stellar luminosity of O-type (B-type) stars in our sample is $\log (L/L_{\odot}$) =  5.10$\pm$0.31 (4.58$\pm$0.38). Table~\ref{O-lit} (\ref{B-lit}) in the Appendix includes comparisons between pipeline-derived stellar luminosities of O-type (B-type) stars and those from the wider literature, for which agreement is overall satisfactory (mostly within 0.1 dex). For the 25 stars in common with \citet{Castro+2018}, $\log L/L_{\odot}$(pipeline) - $\log L/L_{\odot}$(Castro) = +0.12$\pm$0.22~dex (Appendix, Table~\ref{RIOTS4}). For the 30 OB stars in common with the XShootU pipeline study of \citet{Bestenlehner+2025}, $\log L/L_{\odot}$(pipeline) - $\log L/L_{\odot}$(XShootU) =  +0.11$\pm$0.18.

\subsection{Surface gravities}\label{gravities}

\begin{figure}
%\centering
  \includegraphics[width=\columnwidth]{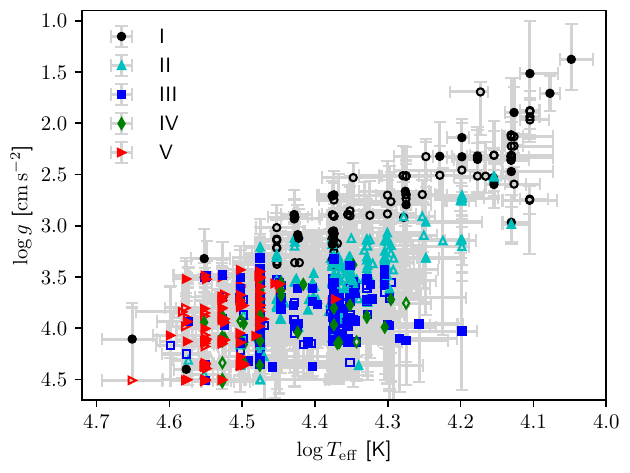}
  \caption{Comparison between effective temperatures, $T_{\rm eff}$, and surface gravities, $\log g$, of BLOeM OB stars (Kiel diagram). Open symbols are single stars according to the initial 9 epochs of BLOeM \citep{BLOeM_O, BLOeM_B, BLOeM_Bsuper, BLOeM_OeBe, BLOeM_BAF}, filled symbols are multiple.
  %Evolutionary masses of post-main sequence stars are determined from inspection of SMC tracks from \citet{Schootemeijer+2019}.
  }
  \label{logT-logg}
\end{figure}

Fig.~\ref{logT-logg} shows a Kiel diagram for the analysed OB stars, with surface gravities ranging from the vicinity of $\log g \sim 4$ for O-type dwarfs, to $\log g \sim 1.5$ for late B supergiants. The average surface gravity of O-type (B-type) stars in our sample is $\log g$/(cm s$^{-2}$) = 3.78 $\pm$ 0.44 (3.59 $\pm$ 0.53). Overall statistics are dominated by early B (super)giants \citep[recall figure~8 from][]{Shenar+2024}.

\begin{figure}
%\centering
\begin{center}
  \includegraphics[width=0.85\columnwidth, bb=48 169 523 640]{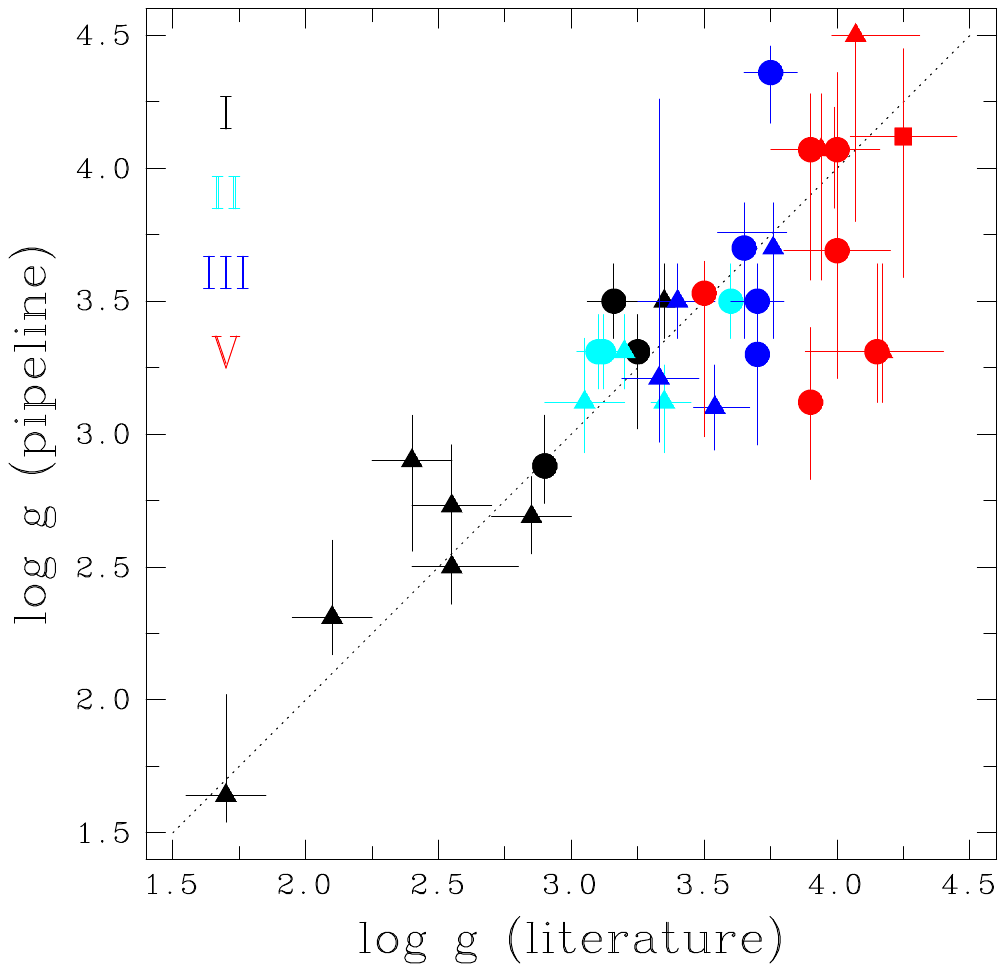}
  \caption{Comparison between $\log g$ for BLOeM OB stars from literature studies (circles: {\sc cmfgen}, triangles: {\sc fastwind}, squares: {\sc tlusty}) and the current pipeline, colour coded by luminosity class. References are provided in the Appendix in Tables~\ref{O-lit}-\ref{B-lit}.}
  \label{OB_logg}
\end{center}
\end{figure}

Table~\ref{O-lit} (\ref{B-lit}) in the Appendix compares pipeline gravities of O-type (B-type) stars to literature values. Overall agreement is satisfactory. However, significantly lower gravities are inferred from the pipeline for some dwarfs and giants (e.g. BLOeM 7-072, O8\,Vnn), as illustrated in Fig.~\ref{OB_logg} for OB stars. We will revisit surface gravities in Section~\ref{gbat}.

Both H$\gamma$ and H$\delta$ possess metallic lines in their damping wings, only some of which are explicitly included in {\sc fastwind} synthetic spectra (e.g. O\,{\sc ii} $\lambda\lambda$4345-51 in Fig.~\ref{BLOeM_1-005}). For the 30 OB stars in common with the XShootU pipeline study of \citet{Bestenlehner+2025}, $\log g$(BLOeM) - $\log g$(XShootU) = 0.06$\pm$0.38.

Spectroscopically derived surface gravities must be corrected for the effect of centrifugal forces, as highlighted by \citet{Herrero+1992}. Gravities corrected for centrifugal forces, denoted $g_{c}$, are obtained from 
\[
g_{c} = g + (\varv_{\rm e} \sin i)^2 /R_{\ast}\]
using radii via the Stefan--Boltzmann relation, and $\varv_{\rm e} \sin i$ discussed in Section~\ref{rotation}. These are included in Table~\ref{table:targets}. In most instances corrections are modest, but can exceed 0.1 dex for rapid rotators e.g. $\log g_{c} - \log g$ = 0.40 dex for BLOeM 6-090 (B2\,III) with $\varv_{\rm e} \sin i \sim$ 400 km\,s$^{-1}$.

\subsection{Elemental abundances}

Helium is our primary focus regarding elemental abundances in OB stars.  The baseline He abundance from H\,{\sc ii} regions \citep{RussellDopita1990} is $N$(He)/$N$(H) = 0.09 by number or $Y \sim$ 25\% by mass, whereas our grid permits lower helium mass fractions to avoid a truncated PDF. Although He weak stars are known, these results should be viewed with caution. High He mass fractions for a significant subset of OB supergiants are more plausible, some of which infer $Y$ = 40--50\%, with several main sequence stars favouring $Y$ = 55\%, the upper limit of the grid \citep[see also]
[]{MartinezSebastian+2025}. We revisit the significance of He mass fractions for O and early B stars in Sect.~\ref{gbat}.
% Figure~\ref{Y} compares projected rotational velocities to He mass fractions.
% There is no apparent relationship between He enrichment and $\varv \sin i$, % suggesting that processes other than rotation dominate surface He enrichment. 

%Fig.~\ref{age-He} compares evolutionary ages (in Myr) to surface He abundances (in mass fraction). O stars have ages of up to 5 Myr, whereas B stars span X Myr $\leq \tau \leq Y$ Myr. 

%Fig.~\ref{Y} reveals no clear relationship between helium enrichment and projected rotational %velocity, suggesting that processes other than rotation dominate surface helium enrichment.

%\begin{figure}
%\centering
%  \includegraphics[width=\columnwidth]{figs/run1_vs_run2_4oct2024.png}
%  \caption{Comparison between temperatures obtained for OB stars from two independent calculations of the pipeline, $T_{1}$ vs $T_{2}$. Temperatures agree to within 1\% for 89\% of the sample, but 63 outliers differ by $\Delta T/T_{1} \geq$5\%.}
%  \label{run1_2}
%\end{figure}

\begin{figure}
\begin{center}
  \includegraphics[width=0.9\columnwidth]{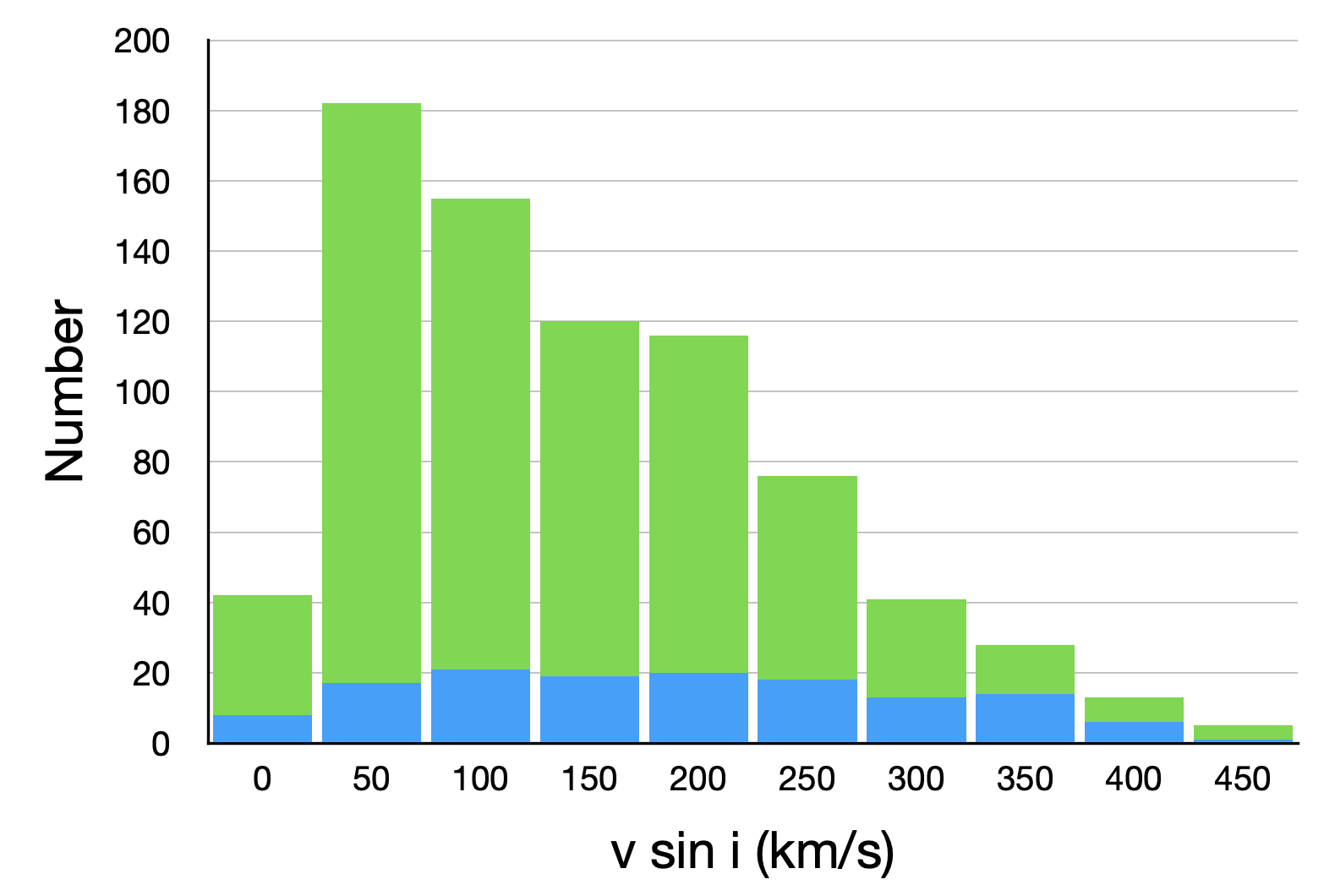}
   \includegraphics[width=0.9\columnwidth]{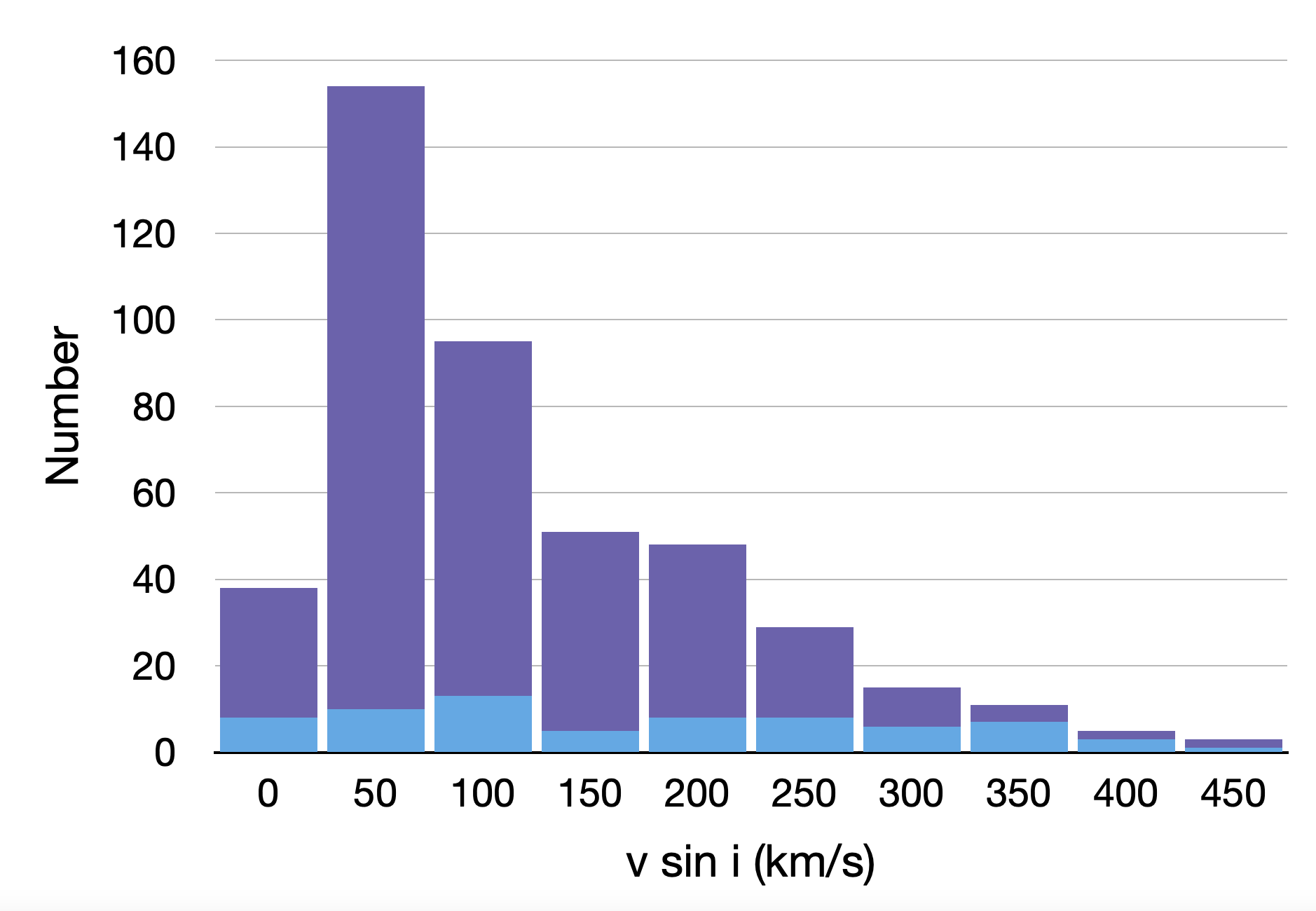}
   \includegraphics[width=0.9\columnwidth]{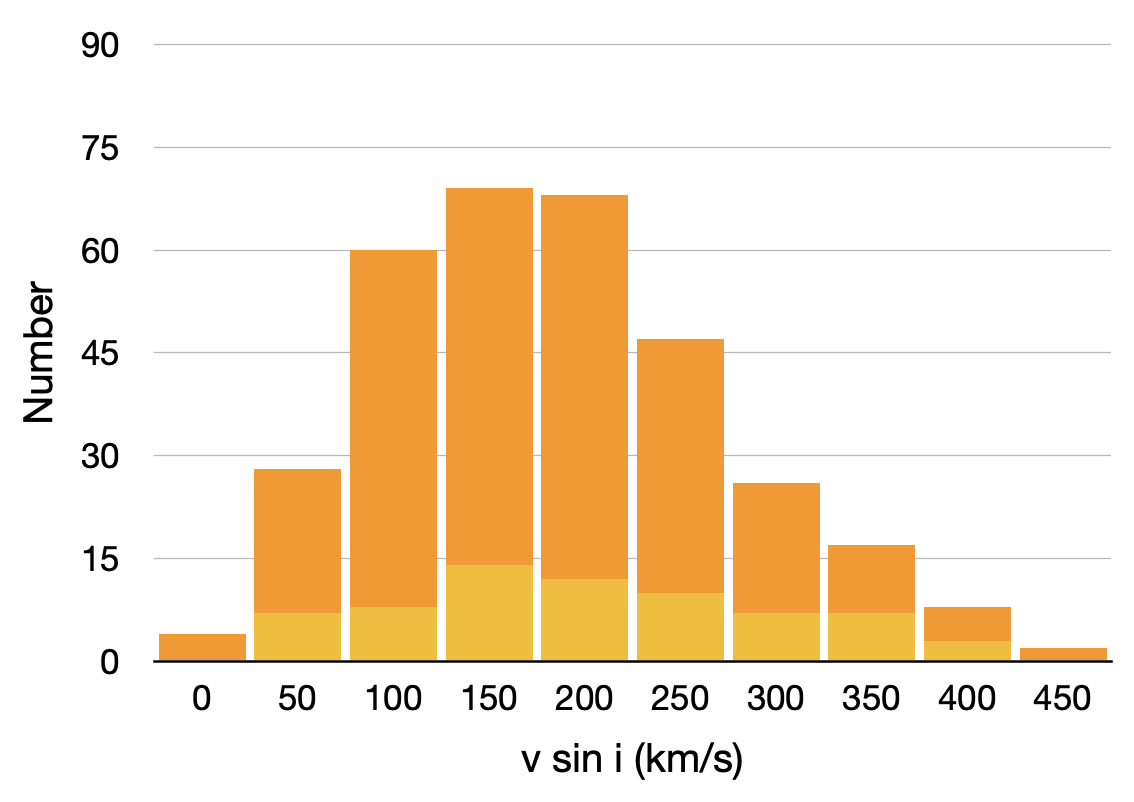}
  \caption{Histogram of projected rotational velocities $\varv_{\rm e} \sin i$ (km\,s$^{-1}$) of all O (blue) and B stars (green) in the top panel, sorted into 50 km\,s$^{-1}$ bins (e.g. 50 km\,s$^{-1}$ refers to 50$\pm$25 km\,s$^{-1}$), aside from the 0 bin which refers to 0--25 km\,s$^{-1}$; Central panel: As above for single O (pale blue) and B (purple) stars according to the initial 9 epochs of the BLOeM survey; Lower panel: As above for multiple O (yellow) and B (orange) stars.
   %Evolutionary masses of post-main sequence stars are determined from inspection of SMC tracks from \citet{Schootemeijer+2019}.
  }
  \label{vsini}
  \end{center}
\end{figure}

\begin{figure*}
%\centering
% \sidecaption
  \includegraphics[width=12cm]{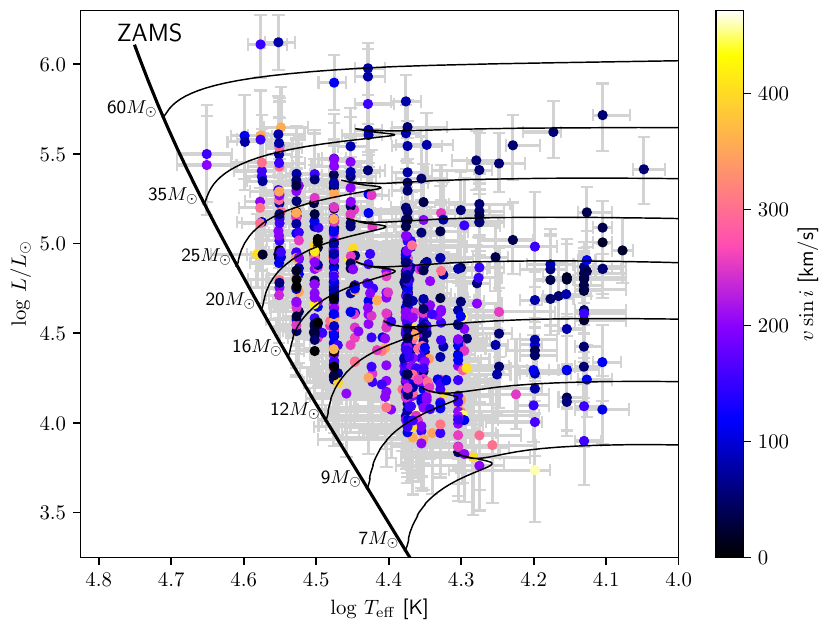}
  \caption{Hertzsprung-Russell diagram  of BLOeM sample (colour coded by $\varv_{\rm e} \sin i$), together with evolutionary tracks for non-rotating SMC massive stars from \citet{Brott+2011}.}
  \label{HRD_vsini_Brott}
\end{figure*}

\section{Rotational velocities}\label{rotation}

\subsection{Pipeline results}

The distribution of projected rotational velocities for BLOeM O (blue) and B (green) stars is presented in Fig.~\ref{vsini} (top panel). Median values are $\varv_{\rm e} \sin i$ = 200 km\,s$^{-1}$ (113 km\,s$^{-1}$) for O-type (B-type) stars, including 8\% (25\%) of fast rotators with $\varv_{\rm e} \sin i > 275$ km\,s$^{-1}$. 
Recalling Section~\ref{analysis}, the synthetic grid was convolved with a fixed $\varv_{\rm mac}$ = 20 km\,s$^{-1}$, with any additional broadening assumed to be attributed to rotation. Consequently, pipeline results will likely overestimate the true $\varv_{\rm e} \sin i$ in many instances, and instrumental broadening hinders
reliable $\varv_{\rm e} \sin i$ for slow rotators.
Table~\ref{summary} provides an overview of rotational velocities obtained for our sample. Table~\ref{O-lit} (\ref{B-lit}) in the Appendix compares pipeline-derived rotational velocities of O-type (B-type) stars to literature results. Rotational velocities from our pipeline are similar to, or somewhat larger than, literature results.

Since close binary evolution can strongly modify rotational velocities \citep{deMink+2014}, Fig.~\ref{vsini} also shows histograms of rotational velocities for (apparently) single stars (middle panel) and spectroscopic binaries (lower panel), revealing strikingly different distributions. Median values for single (binary) stars are $\varv_{\rm e} \sin i$ = 78 km\,s$^{-1}$ (200 km\,s$^{-1}$). The histogram for single stars suggests a bimodality in rotational velocities for O stars, reminiscent of single early B stars from the VLT FLAMES Tarantula Survey \citep[VFTS,][]{Dufton+2013}. 

This bimodality is not apparent for single B-type stars, although giants make up the overwhelming majority of BLOeM B stars (O stars are primarily dwarfs). The histogram for multiple systems reveals that high $\varv_{\rm e} \sin i$ bins are overrepresented with respect to single stars. Nevertheless, further study is warranted since our sample includes a subset of known SB2's, which are likely to artificially boost inferred rotational velocities of binary systems. In addition, many OBe stars -- usually found to be rapid rotators -- are also excluded.

Figure~\ref{HRD_vsini_Brott} shows the Hertzsprung-Russell diagram for BLOeM OB stars, now colour coded by $\varv_{\rm e} \sin i$, and using the non-rotating SMC metallicity tracks from \citet{Brott+2011}. Higher temperature OB stars ($\log T_{\rm eff}$/K $\geq 4.3$) exhibit a broad range of projected rotational velocities, whereas cooler B supergiants predominantly possess modest $\varv_{\rm e} \sin i$ values. There is also a dearth of slow rotators at intermediate temperatures ($\log (T_{\rm eff}$/K) $\sim 4.4$),  suggestive of a physical origin.  Figure~\ref{HRD_vsini_2} in the Appendix separates the HR diagram into single (upper panel) and multiple (lower panel) systems, also colour coded by $\varv_{\rm e} \sin i$, with evolutionary models from \citet{Schootemeijer+2019}.

\begin{table*}
\caption{Summary of median masses ($M_{\rm evol}$), ages ($\tau$), and projected rotational velocities ($\varv_{e} \sin i$) for 778 BLOeM OB stars analysed in this study, separated into single and multiple systems according to the initial 9 epoch BLOeM dataset \citep{BLOeM_O, BLOeM_B, BLOeM_Bsuper, BLOeM_OeBe, BLOeM_BAF}, and into main sequence (MS) versus post-MS according to rotating models from \citet{Brott+2011}.
% or {\sc PoWR} \citep{Grafener+2002, Sander+2015}. 
}
% Mass-loss rates are presented as $\dot{M}/\sqrt{f_{v}}$ to reflect differences in adopted or derived wind clumping factors.
\label{summary}
\begin{center}
\begin{tabular}
{
l@{\hspace{2mm}}
r@{\hspace{2mm}}r@{\hspace{2mm}}r@{\hspace{2mm}}r@{\hspace{5mm}}
r@{\hspace{2mm}}r@{\hspace{2mm}}r@{\hspace{2mm}}r@{\hspace{5mm}}
r@{\hspace{2mm}}r@{\hspace{2mm}}r@{\hspace{2mm}}r|@{\hspace{10mm}}
l@{\hspace{2mm}}
r@{\hspace{2mm}}r@{\hspace{2mm}}r@{\hspace{2mm}}r
}
    \hline\hline
    Spectral & 
    \multicolumn{4}{c}{All} & 
    \multicolumn{4}{c}{Single} &
    \multicolumn{4}{c}{Multiple} &
    Evol. & 
    \multicolumn{4}{c}{All} \\
    Type    &  
    N & $M_{\rm evol}$ & $\tau$ & $\varv_{e} \sin i$ &
    N & $M_{\rm evol}$ & $\tau$ & $\varv_{e} \sin i$ &
    N & $M_{\rm evol}$ & $\tau$ & $\varv_{e} \sin i$ &
    Phase       &
    N & $M_{\rm evol}$ & $\tau$ & $\varv_{e} \sin i$ \\
    
             &
      & $M_{\odot}$    & Myr    & km\,s$^{-1}$ &
      & $M_{\odot}$    & Myr    & km\,s$^{-1}$ &
      & $M_{\odot}$    & Myr    & km\,s$^{-1}$ &
             &
      & $M_{\odot}$    & Myr    & km\,s$^{-1}$       \\
\hline
O & 137 & 19.8 &  4.9 & 200 &  69 & 19.9 &  4.8 & 153 &  68 & 19.8 &  5.1 & 201& MS & 721 & 12.8 & 9.6 & 153 \\ 
B & 641 & 12.6 & 10.8 & 113 & 380 & 12.7 & 10.9 &  78 & 261 & 12.5 & 10.6 & 156& Post-MS & 57 & 14.2 & 11.3 & 55\\
\hline
\end{tabular}
\end{center}
\end{table*}

\begin{figure}
\begin{center}
  \includegraphics[width=0.9\columnwidth]{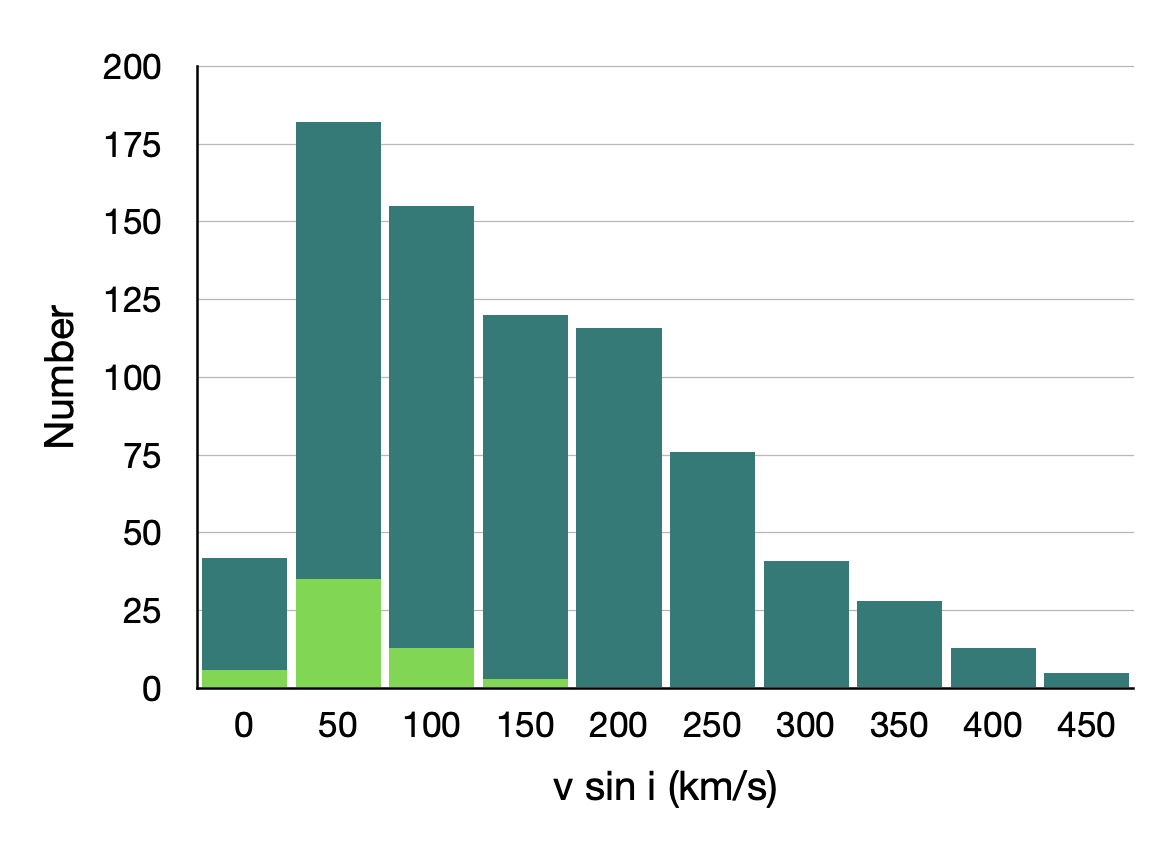}
  \caption{Histogram of projected rotational velocities $\varv_{\rm e} \sin i$ (km\,s$^{-1}$) of main sequence (dark green) and post-main sequence (pale green) OB stars, according to \citet{Brott+2011} rotating models, sorted into 50 km\,s$^{-1}$ bins aside for the 0 bin (e.g. 50 km\,s$^{-1}$ refers to 50$\pm$25 km\,s$^{-1}$).
   %Evolutionary masses of post-main sequence stars are determined from inspection of SMC tracks from \citet{Schootemeijer+2019}.
  }
  \label{vsini_MS}
  \end{center}
\end{figure}

Figure~\ref{vsini_MS} presents a histogram of projected rotational velocities, separated into main sequence (dark green) and post-main sequence (pale green) OB stars -- according to \citet{Brott+2011} rotating models discussed in Section~\ref{masses} -- illustrating that overall statistics are dominated by the former. The median $\varv_{\rm e} \sin i$ of MS (post-MS) stars is 154 (55) km\,s$^{-1}$. \citet{Vink+2010} have previously discussed low rotational velocities of cool B supergiants in the Milky Way and LMC.

\citet{PennyGies2009} have previously estimated rotational velocities of 55 bright SMC O-type stars and B supergiants from high resolution {\it FUSE} spectroscopy, for which the $\varv_{\rm e} \sin i$ distributions of both `unevolved' (IV-V) and `evolved' (II-I) stars peak below 100 km\,s$^{-1}$, in common with Fig.~\ref{vsini}.

\citet{Dufton+2019} have previously investigated the rotational velocities of large populations of massive stars in the NGC~346 star forming region of the SMC. They primarily focused on single B stars for which a median $\varv_{\rm e} \sin i$ = 136 km\,s$^{-1}$ was obtained, somewhat higher than our results for single B stars in the field (78 km\,s$^{-1}$). \citet{Dufton+2019} compare cumulative velocity distributions of single B stars in NGC~346 with other environments in their figure 6, which reveals a high velocity tail. $\sim$10\% of their single B stars exceed 300 km\,s$^{-1}$, somewhat higher than the BLOeM sample of single B stars (4\% exceed 300 km\,s$^{-1}$.

\citet{Ramirez-Agudelo+2015} have previously investigated the rotational velocities of VFTS O stars in the LMC, finding that primaries in binaries closely resembled those of single stars. However, wind-induced spin-down will be stronger in the LMC than the SMC, so perhaps the O star birth spin distribution is bimodal, but not retained at high metallicity due to spin-down.

\subsection{{Pipeline versus} {\sc iacob-broad} results: $\varv_{e} \sin i$}\label{broad}

To assess the reliability of pipeline-derived $\varv_{\rm e} \sin i$, we applied the widely used tool {\sc iacob-broad} \citep{SimonDiaz-Herrero2014} to a representative subset of the OB sample, namely BLOeM identifications with labels X-XX0. Of these, 77 stars are included in our study, recalling AF supergiants and some OB stars were excluded (SB2, OBe, strong nebulosity). 

Owing to the limited spectral range of BLOeM we focus primarily on He\,{\sc i} $\lambda$4387. Rotational velocities can be obtained either via a Fourier Transform (FT) or Goodness-of-Fit (GOF) approach. In principle, the GOF method is preferred, since it also allows the determination of macroturbulence, $\varv_{\rm mac}$. However, this relies on suitable metal lines being available. Si\,{\sc iii} $\lambda$4553 is a suitable alternative diagnostic for the majority of the BLOeM sample, although this line is absent in O stars and late B supergiants. 

We select the FT approach  for comparison with pipeline results for O (blue triangles) and B (green squares) stars in Fig.~\ref{iacob_broad}, although results from both FT and GOF methods are provided in the Appendix in Table~\ref{iacob}. 
Pipeline-derived  $\varv_{\rm e} \sin i$ typically exceed direct measurements, owing to the `quantized' broadening values involved plus macroturbulent broadening, $\varv_{\rm mac}$ may be significantly higher than the 20 km\,s$^{-1}$ adopted. By way of example, we have applied {\sc iacob-broad} to the Si\,{\sc iii} $\lambda$4553 profile in BLOeM 1-020 (B0\,III), the results of which are presented in Figure~\ref{BLOeM_1-020-SiIII4553}. Neglecting other sources of broadening, the GOF
value of $\varv_{\rm e} \sin i$ = 121 km\,s$^{-1}$ (shown in green) is in close agreement to $\varv_{\rm e} \sin i = 113 \pm 19$ km\,s$^{-1}$ determined from the pipeline, with $\varv_{\rm e} \sin i$ = 89 km\,s$^{-1}$ obtained with a non-zero $\varv_{\rm mac}$ (shown in blue). In many instances -- such as BLOeM 1-020 -- there may be an important $\varv_{\rm mac}$ contribution, such that the pipeline would naturally overestimate $\varv_{\rm e} \sin i$. In addition, potential stellar companions may also cause GOF results for strong He\,{\sc i} lines to exceed those of weak He\,{\sc i} and metal lines, noting that BLOeM 1-020 is a SB1 according to \citet{BLOeM_B}. Definitive results await an upcoming dedicated study of rotational velocities of BLOeM OB stars (S.~Berlanas et al. in prep).

%\begin{figure}
%\centering
%  \includegraphics[width=7cm,angle=-90]{figs/Pipeline_Y.pdf}
%  \caption{Comparison between Helium mass fractions and projected rotational
%  velocities for BLOeM OB stars.}
%  \label{Y}
%\end{figure}

% We focus on nitrogen in early O stars, whose baseline value is $\log$ (N/H) + 12 = 6.66 in the SMC \citep{Vink+2023} by number. Modest enrichment is obtained for some O dwarfs, with nitrogen fully processed in a few instances (e.g. BLOeM X-YYY).

% The overwhelming majority of evolutionary models \citep{Brott+2011} predict no surface He or N enrichment for the optimised {\sc bonnsai} parameters ($T_{\rm eff}, \log L/L_{\odot}, v_{\rm e} \sin i$). These are in clear tension with the preferred He and N mass fractions, indicating deficiencies in current evolutionary models for single stars or other processes at play (e.g. close binary evolution).

%\begin{figure}
%\centering
%  \includegraphics[width=9 cm]{figs/logT_logg.pdf}
%  \caption{Kiel diagram for pipeline results (LMC: blue circles, SMC: green triangles). Low  gravities at high temperatures are excluded in order to avoid exceeding the Eddington limit.}
%  \label{logT_logg}
%\end{figure}

\section{Pipeline versus {\sc iacob-gbat} analysis: Temperatures, gravities, abundances}\label{gbat}

Pipeline results were compared to literature temperatures, gravities and luminosities in Section~\ref{analysis}. Literature results were usually obtained from datasets covering a significantly broader spectroscopic range, extending to the ultraviolet in some instances \citep[e.g.][]{Hillier+2003, Martins+2024}. Consequently, here we undertake a star-by-star quantitative analysis of a representative subset of the BLOeM OB stars, based on the dataset outlined in Section~\ref{obs}.

To perform the quantitative spectroscopic analysis, we focus on the same subset as that discussed above in relation to {\sc iacob-broad} rotational velocities, 
although physical parameters could not be determined for stars lacking He\,{\sc ii} lines -- classified as B1 or later. For the remainder,
spectroscopic parameters ($T_{\rm eff}, \log g$, $Y$) are derived using {\sc iacob-gbat}  \citep{SimonDiaz+2011, Sabin-Sanjulian+2014, Holgado+2018}, together with a grid of {\sc fastwind} models, ensuring consistent observational and stellar atmospheres to the pipeline. A velocity law with $\beta = 1$ was adopted and the wind density parameter was set to $\log Q = -13.5$. Results from the {\sc iacob-gbat} analysis are presented in the Appendix (Table~\ref{iacob}). Helium abundances are shown by number ratio, $y$ = $N$(He)/$N$(H), where $y=$0.085 corresponds to a mass fraction of $Y$=0.25, the baseline He content in the SMC adopted by \citet{Brott+2011}.

\begin{figure}
%\centering
\begin{center}
\includegraphics[width=0.85\columnwidth]{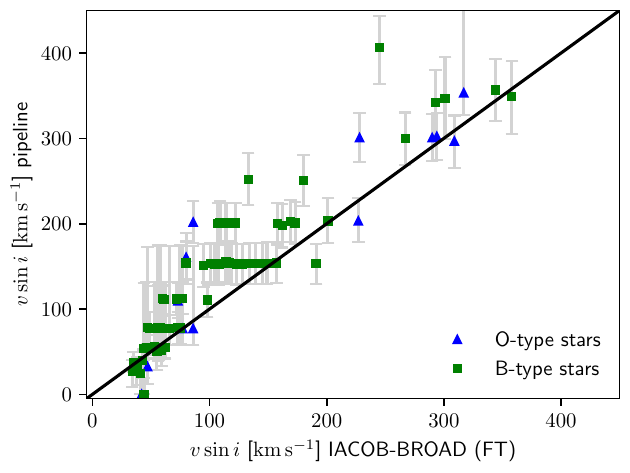}
  \caption{Comparison between $\varv_{\rm e} \sin i$ for a subset of O (blue triangles) and B (green squares) BLOeM stars from {\sc iacob-broad} \citep[][]{SimonDiaz-Herrero2014} and the spectroscopic pipeline.
  %Evolutionary masses of post-main sequence stars are determined from inspection of SMC tracks from \citet{Schootemeijer+2019}.
  }
  \label{iacob_broad}
  \end{center}
\end{figure}

% \begin{figure}
% %\centering
% \begin{center}
% %  \includegraphics[width=0.85\columnwidth,bb=48 169 523 650]{figs/iacob_broad.pdf}
% \includegraphics[width=\columnwidth,bb=26 304 563 788]{figs/BLOeM_1-020_HEI4387.pdf}
%  \caption{\bf {\sc iacob-broad} \citep[][]{SimonDiaz-Herrero2014} Fourier Transform (FT) and Goodness-of-Fit (GOF) results for
%   He\,{\sc i} $\lambda$4387 in BLOeM 1-020 (B0\,III).
%  %Evolutionary masses of post-main sequence stars are determined from inspection of SMC tracks from \citet{Schootemeijer+2019}.
%  }
%  \label{BLOeM_1-020-HeI4387}
%  \end{center}
% \end{figure}

\begin{figure}
%\centering
\begin{center}
\includegraphics[width=\columnwidth,bb=26 304 563 788]{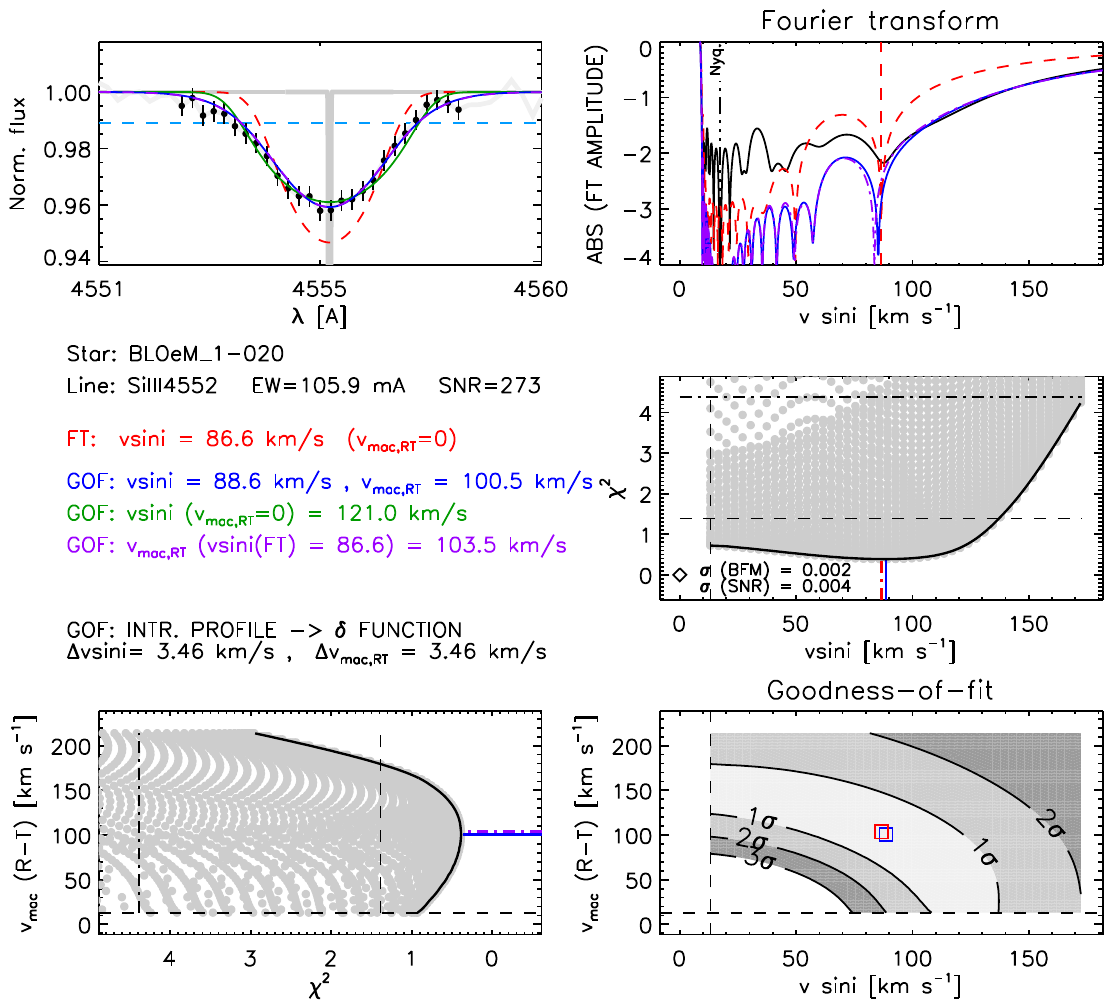}
  \caption{{\sc iacob-broad} \citep[][]{SimonDiaz-Herrero2014} Fourier Transform (FT) and Goodness-of-Fit (GOF) results for
  Si\,{\sc iii} $\lambda$4553 in BLOeM 1-020 (B0\,III).
  %Evolutionary masses of post-main sequence stars are determined from inspection of SMC tracks from \citet{Schootemeijer+2019}.
  }
  \label{BLOeM_1-020-SiIII4553}
  \end{center}
\end{figure}

\begin{figure*}
%\centering
\begin{center}
\includegraphics[width=1.5\columnwidth,angle=180,bb=20 23 778 458]{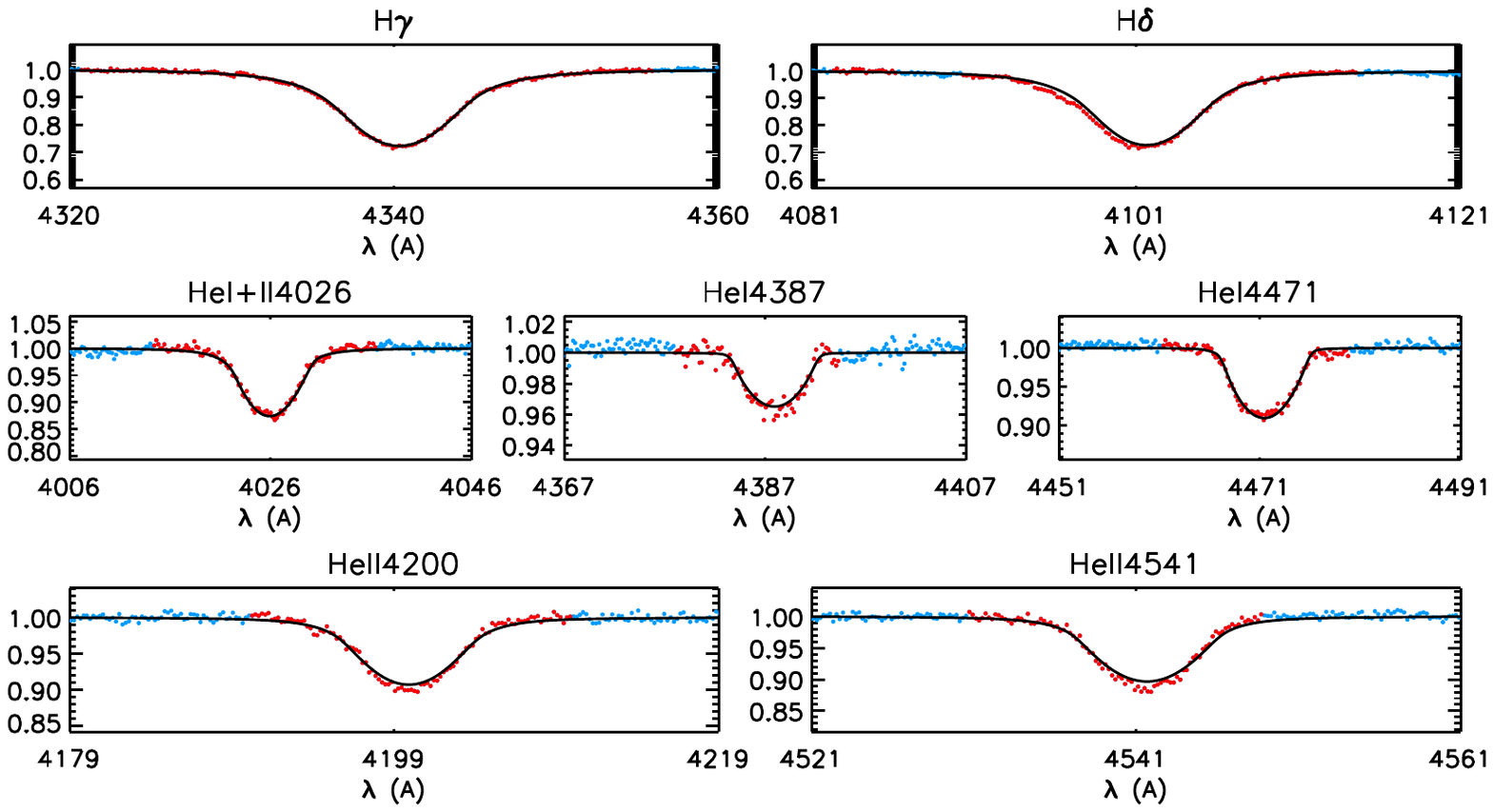}
 \caption{{\sc iacob-gbat} hydrogen and helium spectral line fits (black lines) to BLOeM 8--030 (O6.5\,Vn), in which selected regions (excluded) are indicated in red (blue). Physical parameters are  $T_{\rm eff}$ = 38.2$\pm$0.8 kK, $\log g = 3.82 \pm 0.08$ and $y = 0.130\pm0.023$, with $\varv_{e} \sin i$ = 290 km\,s$^{-1}$ (from {\sc iacob-broad}). 
  }
  \label{BLOeM_8-030-fits}
  \end{center}
\end{figure*}

\begin{figure*}
%\centering
\begin{center}
\includegraphics[width=1.5\columnwidth,angle=180,bb=20 23 778 458]{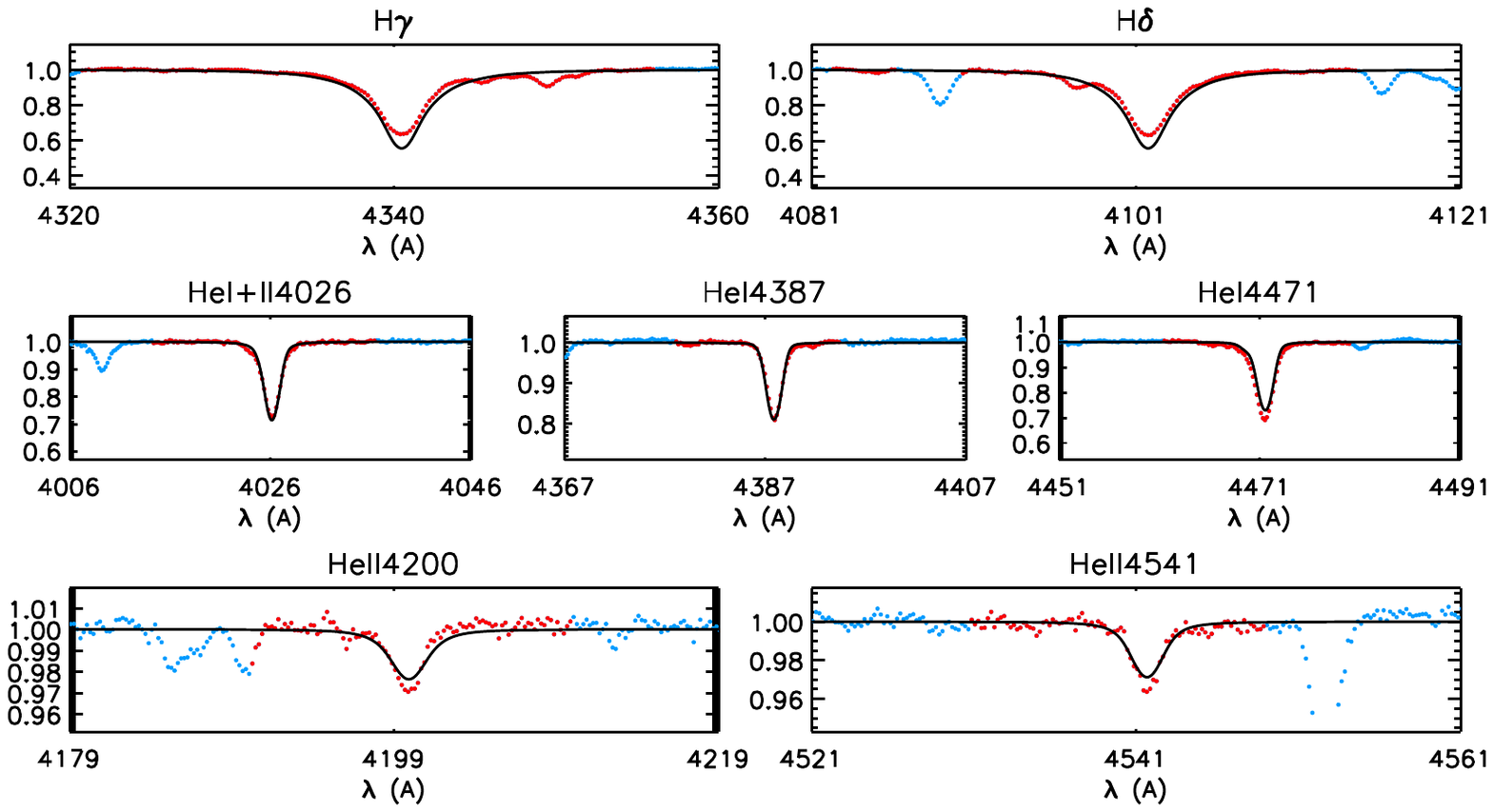}
  \caption{{\sc iacob-gbat} hydrogen and helium spectral line fits (black lines) to BLOeM 3--090 (B0.2\,Ia), in which selected regions (excluded) are indicated in red (blue). Physical parameters are  $T_{\rm eff}$ = 28.0$\pm$1.1 kK, $\log g = 3.19 \pm 0.21$ and $y < 0.06^{+2.3}$, with $\varv_{e} \sin i$ = 74 km\,s$^{-1}$ (from {\sc iacob-broad}).  
  }
  \label{BLOeM_3-090-fits}
  \end{center}
\end{figure*}

Figures~\ref{BLOeM_8-030-fits}--\ref{BLOeM_3-090-fits} present line profile fits to BLOeM 8--030 (O6.5\,Vn) and 3--090 (B0.2\,Ia) obtained with {\sc iacob-gbat}. Spectral regions selected for fitting are shown in red, with regions excluded shown in blue. Overall fit quality is excellent, allowing temperatures, surface gravities and helium abundances (limits for BLOeM 3--090) to be determined in these cases.

\begin{figure}
%\centering
\begin{center}
\includegraphics[width=0.85\columnwidth]{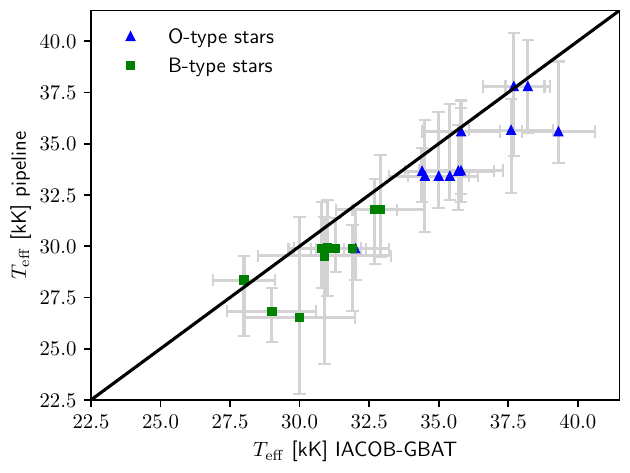}
\includegraphics[width=0.85\columnwidth]{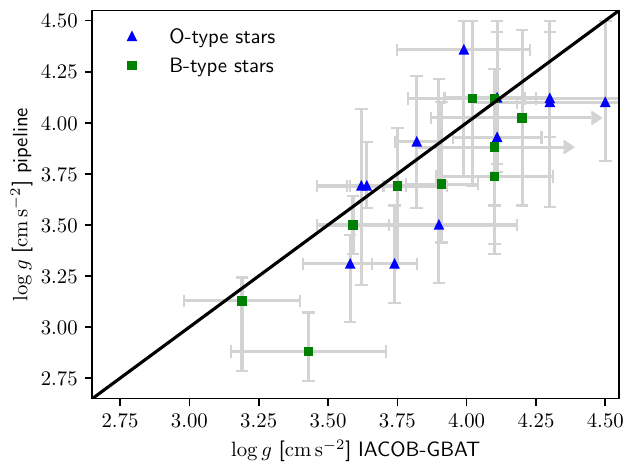}
\includegraphics[width=0.85\columnwidth]{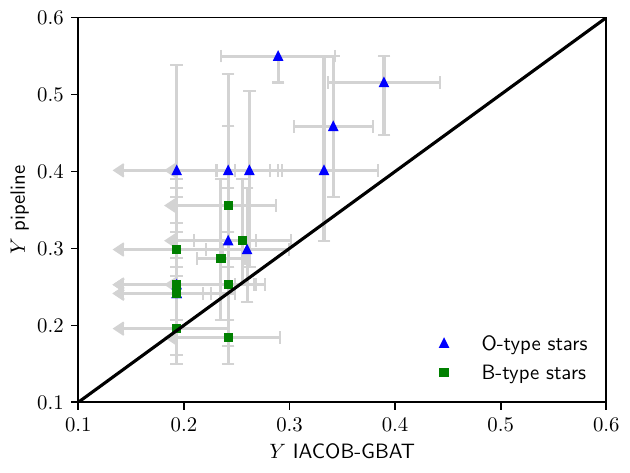}
  \caption{Comparison between {\sc iacob-gbat} \citep{SimonDiaz+2011} and pipeline effective temperatures for a subset of O (blue triangles) and B (green squares) BLOeM stars (top panel). Middle and lower panels: As above for $\log g$ and helium mass fraction, $Y$, respectively. $Y$=0.25 is the SMC baseline according to \citet{Brott+2011}.
  %Evolutionary masses of post-main sequence stars are determined from inspection of SMC tracks from \citet{Schootemeijer+2019}.
  }
\label{iacob_gbat}
\end{center}
\end{figure}

Figure~\ref{iacob_gbat} compares {\sc iacob-gbat} results for $T_{\rm eff}$, $\log g$ and helium mass fraction $Y$ to those from the spectroscopic pipeline. Pipeline effective temperatures are $1.5\pm1$ kK lower for O and early B stars - albeit consistent within formal uncertainties. Pipeline surface gravities for O and early B stars are also somewhat lower 
than {\sc iacob-gbat} ($0.1\pm0.2$ dex), albeit with considerable scatter and sizeable uncertainties.

Interactive fitting has the advantage of permitting specific regions in the wings of Balmer lines to be fit using {\sc iacob-gbat} (e.g. excluding O\,{\sc ii} $\lambda\lambda$4345-51 from H$\gamma$), whereas the entire spectrum is incorporated into the spectroscopic pipeline. Finally, uniformly higher He abundances are inferred from the spectroscopic pipeline for O stars, with
better consistency achieved for early B stars, albeit with considerable uncertainties in both approaches. 

In summary the comparison between results from the spectroscopic pipeline and {\sc iacob-gbat}/{\sc iacob-broad} is relatively satisfactory, though caution should be advised regarding pipeline-derived surface gravities and especially He abundances.

\section{Stellar masses and ages}\label{masses}

  Individual spectroscopic masses, $M_{\rm spec}$, inferred from surface gravities and radii, are presented in Table~\ref{table:targets}. The median value of all O-type (B-type) stars is 23.0 $M_{\odot}$ (16.4 $M_{\odot}$). Spectroscopic masses are highly sensitive to surface gravities, which are uncertain owing to the limited spectral range of the BLOeM dataset, and may also be influenced by convective turbulence \citep[e.g.][]{Cantiello+2009}. Alternatively, initial masses, $M_{\rm init}$, current masses, $M_{\rm evol}$ and ages, $\tau$, may be determined from comparisons to evolutionary models, assuming they have evolved as single stars (which may not be the case for many of the present sample).

\begin{figure}
\begin{center}
  \includegraphics[width=\columnwidth]{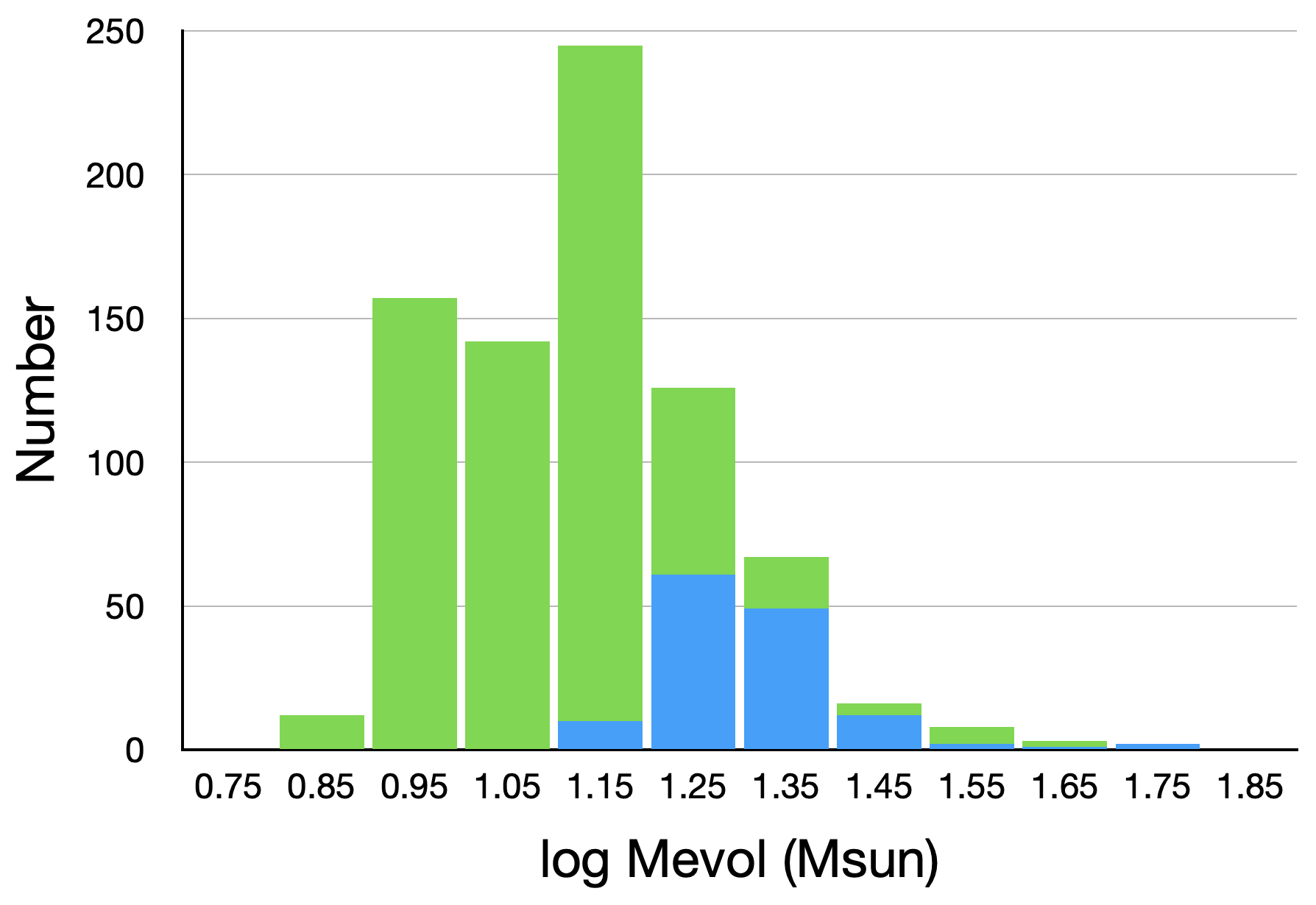}
  \caption{Histogram of (logarithmic) current masses ($M_{\odot}$) of BLOeM O (blue) and B (green) stars, with O stars dominant for $\log M_{\rm evol}/M_{\odot} \geq 1.35 \pm 0.05$ and B stars dominant for $\log M_{\rm evol}/M_{\odot} \leq 1.15 \pm 0.05$. Masses are based on \citet{Brott+2011} rotating evolutionary models, plus \citet{Hastings+2021} evolutionary models for two luminous O supergiants.
   %Evolutionary masses of post-main sequence stars are determined from inspection of SMC tracks from \citet{Schootemeijer+2019}.
  }
  \label{BLOeM_Minit}
  \end{center}
\end{figure}

\begin{figure*}
%\sidecaption
\includegraphics[width=1.5\columnwidth]{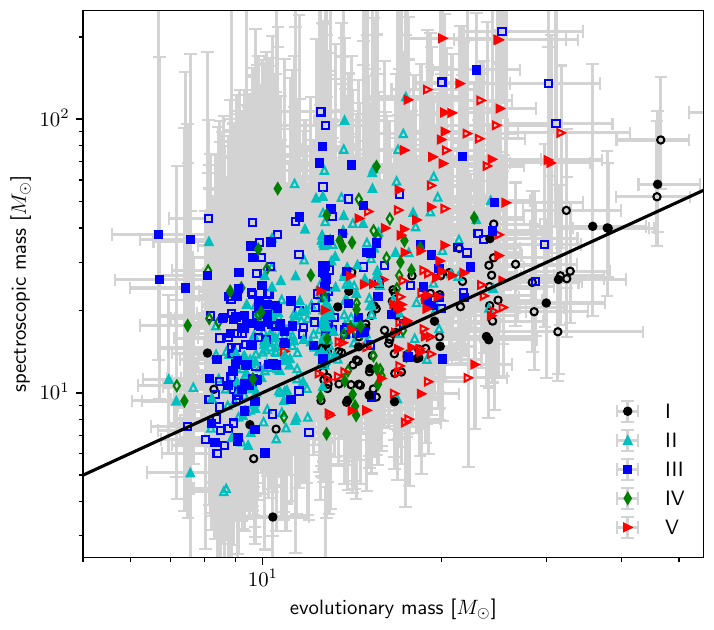}
  \caption{Comparison between (current) evolutionary masses and spectroscopic masses of BLOeM OB stars, based on \citet{Brott+2011} rotating models, plus \citet{Hastings+2021} evolutionary models for two luminous O supergiants above the upper mass limit of the \citet{Brott+2011} models (BLOeM 3-042 and 4-058), colour coded by luminosity class (filled symbols are binaries).}
  \label{Msp_vs_Mev}
\end{figure*}

\subsection{Evolutionary masses}

For core H burning main-sequence (MS) stars, these were obtained from a Bayesian inference method (V.~Bronner et al. in prep), coupled to SMC metallicity evolutionary models. This is similar to {\sc bonnsai}\footnote{The BONNSAI web-service is available at {\url www.astro.uni-bonn.de/stars/bonnsai} } \citep{Schneider+2014} albeit with updated techniques. Our primary evolutionary models involved the rotating grid from \citet{Brott+2011}, using spectroscopic temperatures, luminosities and $v_{\rm e} \sin i$ as input observables. Recalling Section~\ref{gravities}, we exclude spectroscopic gravities from the input observables. The only prior adopted was a Salpeter Initial Mass Function (IMF), with uniform priors for initial rotational velocities and ages. We have investigated
the effect of different rotational velocity priors on the results, and obtain
differences of 1--2\% using the empirical results from \citet{Dufton+2019}, a gaussian prior based on \citet{Hunter+2008} or Fig.~\ref{vsini_MS} rescaled by 4/$\pi$.

Evolutionary masses presented are current values, $M_{\rm evol}$, with initial masses usually only a few percent higher due to the modest mass-lost during the MS evolution at SMC metallicity.  Since the upper mass limit of the SMC grid from \citet{Brott+2011} was 60 $M_{\odot}$, it was necessary to use a non-rotating SMC grid \citep[upper limit 100 $M_{\odot}$,][]{Hastings+2021} for two luminous O-type supergiants close to this limit, namely BLOeM 3-042 (Sk~18) and BLOeM 4-058 (Sk~80), with evolutionary masses of 60$^{+14}_{-12}$ $M_{\odot}$ and  $61^{+15}_{-9} M_{\odot}$, respectively.
 
 For evolved post-MS stars, the determination of masses is more problematic since evolutionary models exhibit more variety than during the MS. However, the luminosity at the end of the contraction phase following the TAMS provides a credible mass estimate. Post-MS stars were identified as being located more than 2$\sigma$ from the theoretical TAMS, again following Bronner et al. (in prep) adopting the \citet{Brott+2011} rotating evolutionary models. Three sources located within 2$\sigma$ from the TAMS failed the posterior predictive check (BLOeM 1-111, 2-093, 3-001) so these were also considered to be post-MS stars. 

Masses obtained for post-MS stars may differ from the true value, since additional mass-loss may occur during the cool supergiant phase. Individual evolutionary masses, $M_{\rm evol}$, are included in Table~\ref{table:targets},
and assume pre-red loop evolution. SMC stars in this mass range are predicted to lose up to 5\% of their TAMS mass prior to core He depletion \citep{Hastings+2021}.
For comparison we also obtained parameters with the grid of non-rotating models from \citet{Schootemeijer+2019} using identical
semiconvection ($\alpha_{\rm SC}$ = 10) and overshooting ($\alpha_{\rm OV}$ = 0.33) parameters to \citet{Brott+2011}.

We present a histogram of initial (logarithmic) masses of BLOeM OB stars in Fig.~\ref{BLOeM_Minit}, separated into O (blue) and B (green) subtypes. O stars dominate the sample above 20 $M_{\odot}$ whereas B stars dominate below 16 $M_{\odot}$. The median evolutionary mass of all O-type (B-type) stars is 19.8 $M_{\odot}$ (12.6 $M_{\odot}$). Table~\ref{summary} provides an overview of evolutionary masses obtained for our sample, separated into single and binary O and B stars. Subdivided into BLOeM fields \citep[figure 1 from][]{Shenar+2024}, median OB masses range from 10.6 $M_{\odot}$ (Field 8) to 15.2 $M_{\odot}$ (Field 3). We shall revisit OB populations across different BLOeM fields in Sect.~\ref{ages}.

The target selection criteria for the BLOeM survey focused on stars with initial masses in excess of 8 $M_{\odot}$ \citep{Shenar+2024}. Indeed, Fig.~\ref{BLOeM_Minit} reveals a sharp cutoff to masses at $\log M_{\rm init}/M_{\odot} = 0.9$ or 8 $M_{\odot}$. A key goal of BLOeM is to determine the slope of the IMF of massive stars in the SMC. We defer a determination of the IMF to a future study in this series once all epochs have been collected (late 2025). The complete multi-epoch dataset will permit a more robust census of single stars to be established, together with a careful analysis of binaries from which  individual component masses will be determined.

\begin{figure}
\begin{center}
  \includegraphics[width=\columnwidth]{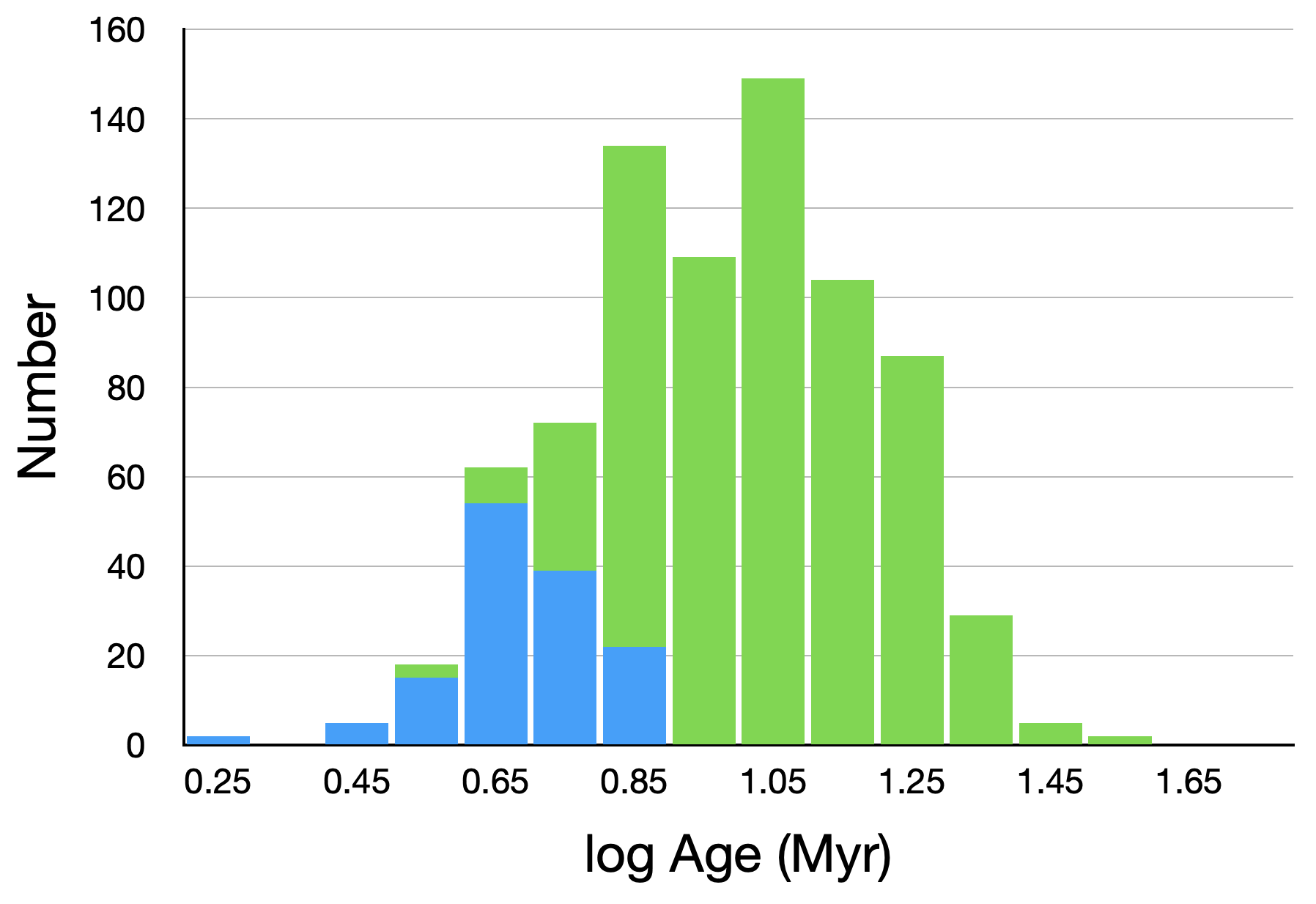}
  \caption{Histogram of (logarithmic) ages (in Myr) of BLOeM O (blue) and B (green) stars, based on \citet{Brott+2011} rotating evolutionary models, plus \citet{Hastings+2021} evolutionary models for two luminous O supergiants. 
   %Evolutionary masses of post-main sequence stars are determined from inspection of SMC tracks from \citet{Schootemeijer+2019}.
  }
  \label{BLOeM_Age}
  \end{center}
\end{figure}

\subsection{Spectroscopic versus evolutionary masses}

  Fig.~\ref{Msp_vs_Mev} compares spectroscopic and (current) evolutionary masses of OB stars from the BLOeM survey (filled symbols are known binaries) based on \citet{Brott+2011} rotating models. Overall, $M_{\rm spec} \geq M_{\rm evol}$, with the possible exception of supergiants (black symbols). Comparisons are hindered by large uncertainties in $\log g_{c}$, plus some of the stars are likely products of binary interaction for which evolutionary masses will be in error. In contrast, the original
   mass discrepancy between spectroscopic and evolutionary values for OB stars identified by \citet{Herrero+1992} involved $M_{\rm evol} \geq M_{\rm spec}$. 
  
  \citet{Schneider+2018} failed to identify a statistically significant mass discrepancy amongst OB stars from the VFTS survey of 30 Doradus in the LMC \citep{Evans+2011}, and no major discrepancy was identified by \citet{Bestenlehner+2025} for pipeline results of higher luminosity LMC and SMC OB stars from the XShootU survey. For completeness, Figure~\ref{Msp_vs_Mev_Abel} compares spectroscopic masses to evolutionary masses obtained with non-rotating SMC models from \citet{Schootemeijer+2019}, which also reveals  $M_{\rm spec} \geq M_{\rm evol}$.

  For the BLOeM sample, the discrepancy may arise as a result of the limited spectral window available (recall Fig.\ref{iacob_gbat}) or the focus on non-supergiant B stars. Indeed, \citet{Schneider+2018} found $M_{\rm spec} \geq M_{\rm evol}$ for B dwarfs within the VFTS sample. Regardless, various explanations for the discrepancy have been proposed. Recall that spectroscopic gravities are sensitive to turbulent velocities, for which a fixed value of 20 km\,s$^{-1}$ is adopted in our study. 2D simulations suggest significantly higher turbulent broadening \citep{Debnath2024}, albeit dependent on metallicity \citep{Cantiello+2009}. 
  
\subsection{Stellar ages}\label{ages}

Stellar ages following the same approach as that described above for evolutionary masses, and are included in Table~\ref{table:targets}. Since \citet{Brott+2011} evolutionary models were adopted, inferred MS lifetimes are believed to be underestimated by $\sim$15\% \citep[][see fig 5.2]{Marchant2017} with respect to MESA models \citep{Paxton+2011, Paxton+2015}. Fig.~\ref{BLOeM_Age} presents a histogram of ages of O (green) and B (blue) subtypes, with median stellar ages of 4.9 Myr and 10.8 Myr, respectively, reflecting the shorter lifetimes of higher mass stars. The youngest O stars have ages of $\sim$3 Myr (e.g. BLOeM 4-058) whereas the oldest B stars reach 30 Myr (e.g. BLOeM 6-062). 

\begin{figure}
%\sidecaption
\includegraphics[width=\columnwidth]{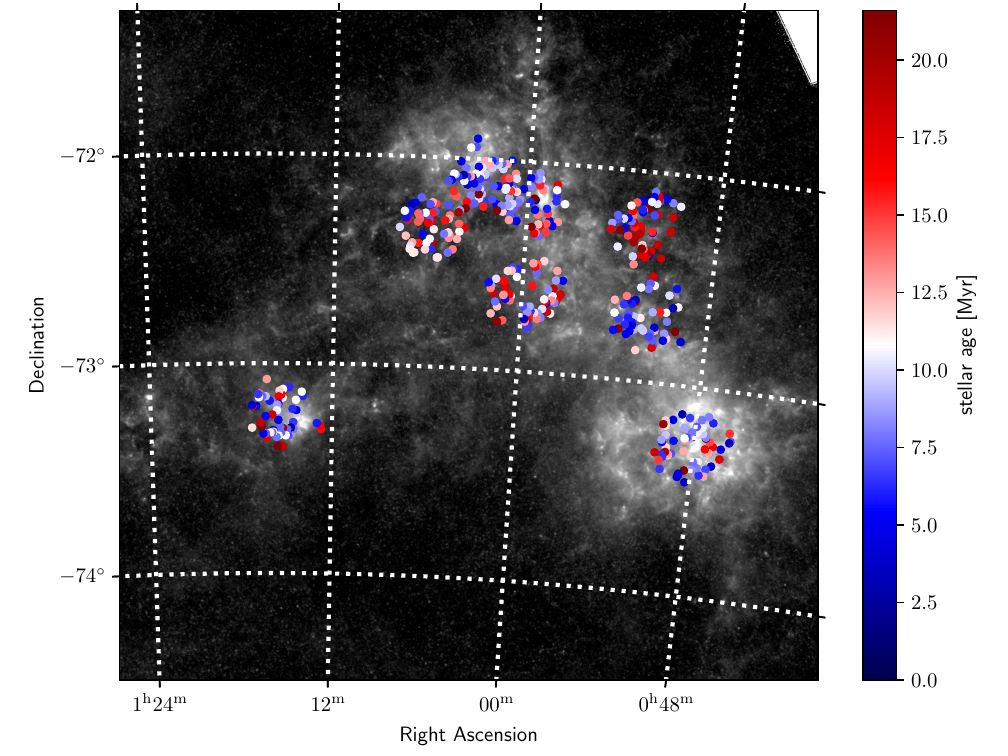}
  \caption{Ages of BLOeM OB stars, overlaid on a {\it Herschel} SPIRE 350$\mu$m map of the SMC \citep{Meixner+2013}. Field 8 (upper right) hosts OB stars with the highest  median age (13.1 Myr) with the remainder in the range 7.7--11 Myr.}
  \label{age_ra-dec}
\end{figure}

Fig.~\ref{age_ra-dec} overlays ages of OB stars on a {\it Herschel} SPIRE 350$\mu$m dust map of the SMC \citep{Meixner+2013}. Subdivided into BLOeM fields \citep[figure~1 from][]{Shenar+2024}, median OB ages range from 7.7 Myr (Field 1) to 13.1 Myr (Field 8). Table~\ref{summary} provides an overview of evolutionary ages obtained for our sample.

Of course, a large fraction of BLOeM OB targets comprise binary systems, so inferred masses (ages) represent upper (lower) limits to the primary component. In addition, mass exchange during close binary evolution can rejuvenate mass gainers, giving the false appearance of youth, so detailed masses and ages await analysis of the complete time series BLOeM dataset.

 % The $L/M$ ratio is affected by the adopted core overshooting parameter, and there is some evidence that a reduced overshooting  parameter may be more appropriate at lower masses \citep{Schootemeijer+2019}. Adopting the mass dependent $\alpha_{\rm OV}$ from  \citet{Hastings+2021}, an alternative mass-TAMS luminosity relation is obtained

   %\begin{equation} 
   % \log M_{\rm init} = 0.8816 -0.2593 \log L_{\rm TAMS} + 0.0671 (\log L_{\rm TAMS})^2. 
   % \end{equation}

% where masses and luminosities are in solar units.

%  Using {\sc bonnsai} inferred masses from main-sequence models of \citet{Brott+2011} and the mass-dependent
%  $\alpha_{\rm OV}$ models of \citet{Hastings+2021} for post-main sequence stars, we present a histogram of initial masses of BLOeM OB stars in Fig.~\ref{BLOeM_Minit}. The median evolutionary mass of the BLOeM B-type (O-type) stars is $12.5 M_{\odot} (25.4 M_{\odot}$). 

  % log M_i =  0.6288 - 0.1764 * log L + 0.0602 * (log L)^2
  % log age(Myr) = 4.702 - 0.9651 * log L + 0.0471 * (log L)^2

 \section{BLOeM in the context of the global SMC O star population}\label{SMC}

BLOeM was designed to sample representative O and early B stars in the SMC, with 929 science targets
drawn from a master {\it Gaia} catalogue of 5576 stars representing 1/6 of the global population \citep{Shenar+2024}. \citet{Bonanos+2010} have previously provided a catalogue of 5324 massive stars in the SMC comprising literature spectral types. This included 277 O-type stars, plus the 12 known Wolf-Rayet stars in the SMC (5 of which also host O stars). 

At face value this suggests that the BLOeM survey -- including 159 O stars -- comprises over half of the known O stars within the SMC. However, nearly 50\% of the O stars from BLOeM were newly classified as such, either representing the first spectral classification or a revision from the previous literature. We have therefore compiled an updated catalogue of spectroscopically confirmed O stars in the SMC, adapted from I.D.~Howarth (priv. comm.), to incorporate newly identified O stars from BLOeM plus additions from e.g. 2dFS 2dFS \citep{Evans+2004-2dF}, RIOTS4 \citep{Lamb+2016} and  \citet{Dufton+2019}. This is presented in RA order in Table~\ref{census}.

% I.D.~Howarth (priv. comm.) has previously compiled a list of spectroscopically confirmed O stars in the SMC, incorporating additions to the \citet{Bonanos+2010} compilation, such as RIOTS4 \citep{Lamb+2016}. 

O type classifications solely based from UV spectroscopy are excluded \citep[e.g.][]{Prinja1987, SmithNeubigBruhweiler1997} from the present compilation. However, we do include the embedded ionizing source of the compact H\,{\sc ii} region N88A \citep{HeydariMalayeri+1999b, Testor+2010}, owing to its high ionizing photon production rate, although this itself may comprise multiple O star components.  A number of stars have been classified as either O9.5 or B0, so the updated catalogue of SMC O stars provided in Table~\ref{census} includes alternate classifications. 75 BLOeM sources are newly identified as O stars which brings the current total of systems to 449, so BLOeM comprises 1/3 of the known O star population of the SMC, of which $\sim$10\% lie within the NGC~346 star-forming
region. The current total will doubtless be incomplete, with the upcoming VISTA/4MOST spectroscopic survey 1001MC \citep{1001MC} set to provide definitive numbers.

There is a well known deficiency of luminous early O stars in the SMC \citep{Schootemeijer+2021}, so it is unsurprising that the earliest O-type stars within the sample are BLOeM 2-079 (O4:\,V+early B) and BLOeM 3-049 (O4\,I(n)). At present, there are six known O2--3 stars in the SMC, NGC~346 MPG 355 \citep{Walborn+2004}, NGC~346 MPG 435 \citep{Dufton+2019, RickardPauli2023}, NGC~346 ELS 7 \citep{Bestenlehner+2025}, Sk~183 \citep{Evans+2012, Ramachandran+2019}, AzV 14 \citep{Pauli+2023}, and AzV 435 \citep{Massey+2005}, plus several O3.5 stars \citep{Bestenlehner+2025}.

Individual BLOeM stars for which $\log (Q_{0}/{\rm s}^{-1}) \geq 49.0$ are listed in Table~\ref{Q_0}, which also includes their ionizing output in the neutral He continuum, $Q_{1}$, and the ratio of these rates. Collectively these 17 sources provide $Q_{0} = 3.2 \times 10^{50}$ s$^{-1}$, over 40\% of the cumulative $Q_{0} = 7.5 \times 10^{50}$ s$^{-1}$ Lyman continuum ionizing output of the 778 BLOeM OB stars. For context, this represents $\sim$20\% of the global H$\alpha$-derived $Q_{0} = 3.4 \times 10^{51}$ s$^{-1}$ ionizing output of the SMC \citep{Kennicutt+2008}. Since BLOeM samples 1/3 of the known SMC O population one might have anticipated a greater fraction. However, the earliest O stars and Wolf-Rayet stars -- neither populations included in BLOeM -- are anticipated to dominate the ionizing output of individual H\,{\sc ii} regions or more generally the galaxy as a whole \citep{Doran+2013, Ramachandran+2019}.

\begin{table}
\caption{Lyman continuum ionizing photon rates of BLOeM OB stars exceeding $Q_{0} = 10^{49}$ s$^{-1}$, including neutral He continuum ionizing photon rates ($Q_{1}$), and their ratio $\log Q_{1}/Q_{0}$.
% or {\sc PoWR} \citep{Grafener+2002, Sander+2015}. 
}
% Mass-loss rates are presented as $\dot{M}/\sqrt{f_{v}}$ to reflect differences in adopted or derived wind clumping factors.
\label{Q_0}
\begin{center}
\begin{tabular}
{
c@{\hspace{2mm}}
r@{\hspace{2mm}}
r@{\hspace{2mm}}
l@{\hspace{2mm}}
c@{\hspace{2mm}}
c@{\hspace{2mm}}
c
}

    \hline\hline
    BLOeM & Sk & AzV & Spectral & $\log Q_{0}$ & $\log Q_{1}$ & $\log Q_{1}/Q_{0}$\\
          &    &     & Type     & s$^{-1}$     & s$^{-1}$       &  \\
\hline
4-058 &   80 & 232       & O7\,Iaf$^{+}$    & 49.73 & 48.75 & --0.98 \\
3-042 &   18 & 26        & O6\,I(f)+O7.5    & 49.70 & 48.83 & --0.87 \\ 
1-072 &      &           & O5\,V(n)+O6.5(n) & 49.33 & 48.49 & --0.84 \\
2-016 &      & 80        & O6\,III:nn(f)p   & 49.33 & 48.42 & --0.90 \\ 
2-020 &      & 83        & O7\,Iaf$^{+}$    & 49.28 & 48.36 & --0.92 \\
3-081 &      &           & O6\,III:         & 49.28 & 48.52 & --0.75 \\
3-049 &      &           & O4\,I(n)         & 49.27 & 48.66 & --0.61 \\
7-069 &  84 & 243        & O6.5\,V          & 49.22 & 48.47 & --0.75 \\
6-033  &    &            & O4.5\,V:         & 49.19 & 48.60 & --0.59 \\
1-102  &    & 345a       & O6\,III(n)       & 49.17 & 48.30 & --0.87 \\
2-035  &     & 95        & O7.5\,III((f))   & 49.10 & 48.13 & --0.97 \\
2-075  &     & 133       & O6\,Vn((f))      & 49.08 & 48.14 & --0.94 \\
6-105  &     &           & O6\,V:n          & 49.08 & 48.20 & --0.88 \\
3-051  &      &          & O5.5:\,V         & 49.07 & 48.31 & --0.75 \\
2-098  &       &         & O6.5\,V((f))     & 49.00 & 48.15 & --0.86 \\
2-007 &  35    & 70      & O9.5\,II-I       & 49.00 & 46.53 & --2.47 \\
\hline
\end{tabular}
\end{center}
\end{table}

\section{Conclusions}\label{conclusions}

Previous quantitative studies have included large samples of OB stars in the Milky Way \citep{Castro+2014, deBurgos+2023b, Holgado+2020, Holgado+2022} and Magellanic Clouds \citep{Sabin-Sanjulian+2017, Ramirez-Agudelo+2017, Castro+2018, Ramachandran+2019, Bestenlehner+2025}. Nevertheless, the present study -- involving a large and representative sample of hot, massive stars in the SMC -- is unprecedented in its scale, owing to the use of a dedicated spectroscopic pipeline \citep{Bestenlehner2024} applied to large grids of synthetic spectra computed with {\sc fastwind} \citep{Puls+2005, Rivero-Gonzalez+2012}. 

We limit our analysis to those OB stars unaffected by strong disk emission, so OBe, sgB[e] stars are excluded, together with instances of strong nebular emission and/or significant contamination from secondaries in SB2 systems. Our study therefore focuses on a total of 778 stars, or 92\% of the total OB sample from BLOeM.

Stellar temperatures are generally in line with previous determinations for SMC OB stars, except that the pipeline fails to reproduce Si\,{\sc iv} $\lambda$4089 in some instances, so
underestimates the temperatures of some early B stars. Nevertheless,
stellar temperatures (Fig.~\ref{OB_Teff}) and surface gravities (Fig.~\ref{OB_logg}) 
are generally in satisfactory agreement with previous detailed studies based on extensive UV and optical spectroscopy. 

Temperatures are also in good agreement with pipeline analysis of BLOeM stars in common with XShootU \citep{Bestenlehner+2025} plus {\sc iacob-gbat} bespoke results for a subset of BLOeM O and early B stars (Fig.~\ref{iacob_gbat}). There is greater scatter for surface gravity comparisons, and He abundance comparisons with {\sc iacob-gbat} suggesting the pipeline overestimates He abundances. Both may arise from the limited spectral range of the current BLOeM dataset.

We establish median BLOeM O (B) masses of 19.8 (12.6) $M_{\odot}$ with a few O supergiants exceeding 50 $M_{\odot}$ (e.g. BLOeM 4-058 a.k.a. Sk~80), and a significant fraction close to the theoretical TAMS according to rotating models of \citet{Brott+2011}. Evolution is expected to be rapid between the TAMS and cool supergiant phase for single stars, so the presence of such stars is difficult to explain without considering binary evolution unless the theoretical TAMS extend to cooler temperatures. A comparison between spectroscopic and evolutionary masses (Fig.~\ref{Msp_vs_Mev}) reveals systematically higher values for the former, with the potential exception of OB supergiants. 

The pipeline analysis also provides estimates of rotational velocities, $v_{\rm e} \sin i$, with known binaries (mostly SB1) possessing relatively high rotational velocities, and  an apparent bimodality amongst single O stars (Fig.~\ref{vsini}) which resembles that of single B stars in the Tarantula region of the LMC identified by \citet{Dufton+2013}. Definitive results await an upcoming dedicated study (S.~Berlanas et al. in prep), although pipeline results are broadly in line with {\sc iacob-broad} Fourier Transform results from He\,{\sc i} $\lambda$4387 for a subset of BLOeM OB stars (Fig.~\ref{iacob_broad}).

Future studies will utilise the entire 25 epoch BLOeM dataset, permitting the identification of additional binaries, derive orbital properties for known SB1 and SB2 systems, individual fits for disentangled spectra, allowing searches for compact companions, and determine the IMF of single stars and binaries.

%Links to \href{https://drive.google.com/file/d/1JKfqFAcCYnXUoYcBmlx1yDD3mmq5x-36/view?usp=sharing}{LMC} and \href{https://drive.google.com/file/d/12Sl8rWVCZg1q81-c_qrLaO6rr_tDRMY9/view?usp=sharing}{SMC} fits.

%%%%%%%%%%%%%%%%%%%%%%%%%%%%%%%%%%%%%%%%%%%%%%%%%%%%%%%%%%%%%%%%%%%%%%%%%%%%%%%%%%%%%%%%%%%%%%%%%%%%%%%%%%%%%%%%%%%%%
%%%%%%%%%%%%%%%%%%%%%%%%%%%%%%%%%%%%%%%%%%%%%%%%%%%%%%%%%%%%%%%%%%%%%%%%%%%%%%%%%%%%%%%%%%%%%%%%%%%%%%%%%%%%%%%%%%%%%

\section*{Acknowledgements}
Based on observations collected at the European Southern Observatory under program id 112.25W2.
JMB and PAC acknowledge financial support from the Science and Technology Facilities Council via research grant ST/V000853/1 (P.I. Vik Dhillon). This work has received funding from the European Research Council (ERC) under the European Union's Horizon 2020 research and innovation programme (grant agreements 945806/TEL-STARS, 101164755/METAL, 101165213/Star-Grasp). VAB and FRNS acknowledge support from the Klaus Tschira Foundation, and the Deutsche Forschungsgemeinschaft (DFG, German Research Foundation) under Germany's Excellence Strategy EXC 2181/1-390900948 (the Heidelberg STRUCTURES Excellence Cluster). SS-D acknowledges support from the Spanish Ministry of Science and Innovation and Universities (MICIU) through the Spanish State Research Agency (AEI) through grants PID2021-122397NB-C21 and the Severo Ochoa Program 2020-2023 (CEX2019-000920-S). PM acknowledges support from the FWO senior fellowship number 12ZY523N. TS is supported by the Israel Science Foundation (ISF) under grant number 2434/24. This research has made extensive use of the SIMBAD database, operated at CDS, Strasbourg, France, the FASTWIND stellar atmosphere code developed by Joachim Puls, the National Institute of Standards and Technology (NIST) atomic spectra database, the Vienna Atomic Line Database (VALD), and SciPy \url{ https://scipy.org/citing-scipy/}. We appreciate comments on the draft manuscript by Abel Schootemeijer and Jorick Vink.

\section*{Data Availability}\label{data_availability}

Table~\ref{table:targets} (physical properties of BLOeM OB stars) and Table~\ref{census} (catalogue of spectroscopically confirmed O stars in the SMC) are available in electronic form at the CDS via anonymous ftp to cdsarc.u-strasbg.fr (130.79.128.5) or via \url{http://cdsweb.u-strasbg.fr/cgi-bin/qcat?J/A+A/}.

Online material at \url{10.5281/zenodo.15526149} includes spectral fits for each star (model in red, observations in blue).

%\bibliographystyle{mnras}
%\bibliography{biblio}

%%%%%%%%%%%%%%%%%%%%%%%%%%%%%%%%%%%%%%%%%%%%%%%%%%%%%%%%%%%%%%%%%%%%%%%%%%%%%%%%%%%%%%%%%%%%%%%%%%%%%%%%%%%%%%%%%%%%%%%

%%%%%%%%%%%%%%%%%%%%%%%%%%%%%%%%%%%%%%%%%%%%%%%%%%%%%%%%%%%%%%%%%%%%%%%%%%%%%%%%%%%%%%%%%%%%%%%%%%%%%%%%%%%%%%%%%%%%%%%

\appendix

\section{Physical properties of BLOeM OB stars}

Table~\ref{table:targets} presents physical parameters of BLOeM OB stars, excluding systems which are problematic for spectroscopic analysis (OBe, SB2, sgB[e], strong nebulosity). Binarity has been investigated by \citet{BLOeM_O} for O stars, \citet{BLOeM_OeBe} for OBe stars, \citet{BLOeM_B} for early non-supergiant B stars, \citet{BLOeM_Bsuper} for early supergiant B stars, and \citet{BLOeM_BAF} for cooler supergiants.

\begin{landscape}
\begin{table}
\caption{Pipeline-derived physical parameters of BLOeM OB stars. Notes include known spectroscopic binaries (or line profile variability, lpv) and discrepancies in spectroscopic fits. Evolutionary masses, $M_{\rm evol}$, and ages, $\tau$, are from \citet{Brott+2011} with the exception of two luminous supergiants involving \citet{Hastings+2021} which are indicated with H2021.}
\label{table:targets} 
\begin{tabular}{l@{\hspace{1.5mm}}l@{\hspace{-1.5mm}}r@{\hspace{1.5mm}}r@{\hspace{1.5mm}}c@{\hspace{1.5mm}}c@{\hspace{1.5mm}}c@{\hspace{1.5mm}}r@{\hspace{1.5mm}}c@{\hspace{1.5mm}}c@{\hspace{1.5mm}}r@{\hspace{1.5mm}}r@{\hspace{1.5mm}}r@{\hspace{1.5mm}}r@{\hspace{1.5mm}}c@{\hspace{3mm}}c@{\hspace{3mm}}l}
\hline\hline
%\hline
BLOeM & Spect. & $\varv_{\rm rad}$ & $\sigma (\varv_{\rm rad})$ & $T_{\rm eff}$ & $\log g$ & $\log g_{c}$ & $R_{\rm eff}$ & $\log L$ & $Y$ & $\varv_{\rm e} \sin i$ & $M_{\rm spec}$ & $M_{\rm evol}$ & $\tau$ & $m_{\rm Ks}$ & $A_{\rm V}$ & Notes\\ 
       & Type & km\,s$^{-1}$  & km\,s$^{-1}$     & kK & cm s$^{-2}$ & cm s$^{-2}$ & $R_{\odot}$ & $L_{\odot}$  & $\cdots$  & km s$^{-1}$ & $M_{\odot}$ & $M_{\odot}$ & Myr & mag & mag\\
\hline
	   1-001 &B9\,Iab& 154 &    1.7 &  13.5$^{  +0.4}_{  -1.9}$ &  2.31$^{ +0.14}_{ -0.14}$ &  2.31$^{ +0.14}_{ -0.14}$ &  35.3$^{  +11.7}_{  -6.9}$  &  4.57$^{ +0.19}_{ -0.23}$ &  0.15$^{ +0.03}_{ -0.00}$ &  23$^{   +9}_{   -8}$ &   9.4$^{     +9.5}_{   -3.5}$      &  12.5$^{+1.1}_{-2.6}$ & 15.4$^{+3.6}_{-3.8}$ & 13.635 & 0.32 & He\,{\sc i} too strong, post-MS\\[2pt]
           1-002 &B0\,IV: & 206 &   21.0 &  33.4$^{  +1.6}_{  -3.1}$ &  4.10$^{ +0.17}_{ -0.51}$ &  4.10$^{ +0.17}_{ -0.51}$ &   8.1$^{   +3.5}_{  -1.9}$ &  4.86$^{ +0.27}_{ -0.28}$ &  0.24$^{ +0.10}_{ -0.05}$ &  77$^{  +19}_{  -21}$ &  30.0$^{    +42.7}_{  -17.1}$  & 17.0$^{+2.4}_{-2.7}$ & 5.7$^{+2.2}_{-2.1}$ & 15.839 & 0.39 &SB1\\[2pt] % 5.6+0.6-0.7 Myr 18.8+0.8-0.8 
           1-003 &B2.5\,II:& 135 &    4.0 &  22.5$^{  +3.5}_{  -3.9}$ &  3.30$^{ +0.34}_{ -0.51}$ &  3.40$^{ +0.34}_{ -0.51}$ &  11.6$^{   +5.7}_{  -3.8}$ &  4.49$^{ +0.28}_{ -0.30}$ &  0.18$^{ +0.15}_{ -0.03}$ & 201$^{  +25}_{  -24}$ &  12.2$^{    +33.1}_{   -7.3}$ & 10.6$^{+2.4}_{-1.5}$ & 13.3$^{+6.1}_{-4.1}$ & 15.536 & 0.32 &       \\
\hline
\end{tabular}
\end{table}
\end{landscape}

\begin{landscape}
\begin{table}
\contcaption{}
%\label{table:targets} 
\begin{tabular}{l@{\hspace{1.5mm}}l@{\hspace{-1.5mm}}r@{\hspace{1.5mm}}r@{\hspace{1.5mm}}c@{\hspace{1.5mm}}c@{\hspace{1.5mm}}c@{\hspace{1.5mm}}r@{\hspace{1.5mm}}c@{\hspace{1.5mm}}c@{\hspace{1.5mm}}r@{\hspace{1.5mm}}r@{\hspace{1.5mm}}r@{\hspace{1.5mm}}r@{\hspace{1.5mm}}c@{\hspace{3mm}}c@{\hspace{3mm}}l}
\hline\hline
%\hline
BLOeM & Spect. & $\varv_{\rm rad}$ & $\sigma (\varv_{\rm rad})$ & $T_{\rm eff}$ & $\log g$ & $\log g_{c}$ & $R_{\rm eff}$ & $\log L$ & $Y$ & $\varv_{\rm e} \sin i$ & $M_{\rm spec}$ & $M_{\rm evol}$ & $\tau$ & $m_{\rm Ks}$ & $A_{\rm V}$ & Notes\\ 
       & Type & km\,s$^{-1}$  & km\,s$^{-1}$     & kK & cm s$^{-2}$ & cm s$^{-2}$ & $R_{\odot}$ & $L_{\odot}$  & $\cdots$  & km s$^{-1}$ & $M_{\odot}$ & $M_{\odot}$ & Myr & mag & mag\\
\hline
           1-037 &B1.5\,III& 167 &   66.5 &  29.9$^{  +2.0}_{  -7.8}$ &  4.10$^{ +0.29}_{ -0.74}$ &  4.13$^{ +0.29}_{ -0.74}$ &   7.5$^{   +4.3}_{  -2.0}$ &  4.61$^{ +0.27}_{ -0.36}$ &  0.15$^{ +0.10}_{ -0.00}$ & 201$^{  +25}_{  -25}$ &  27.3$^{    +71.1}_{  -17.1}$ &  12.7$^{+2.5}_{-2.5}$  & 7.8$^{+6.1}_{-4.1}$ & 16.198 & 0.36 & SB2 \\[2pt]
           1-038 &B2.5\,Ib& 186 &    1.6 &  20.0$^{  +1.9}_{  -1.9}$ &  2.88$^{ +0.38}_{ -0.19}$ &  2.89$^{ +0.38}_{ -0.19}$ &   22.6$^{   +8.1}_{  -5.5}$ &  4.87$^{ +0.23}_{ -0.23}$ &  0.20$^{ +0.08}_{ -0.05}$ &  35$^{  +57}_{  -32}$ &    14.3$^{    +34.3}_{   -6.3}$   &  12.6$^{+2.7}_{-1.8}$ & 12.5$^{+3.8}_{-2.4}$ & 14.177 & 0.31 &\\[2pt]
           1-040 &O9.7\,III:n e & 33 & 16.2 & $\cdots$ & $\cdots$ & $\cdots$ & $\cdots$ & $\cdots$ & $\cdots$ & $\cdots$ & $\cdots$ & $\cdots$ & 13.299 & $\cdots$ &SB1, Oe \\[2pt]
           1-041 &O9.7:\,V:n+O9.7:& 166 & 3.8 &  29.9$^{  +2.7}_{  -2.3}$ &  3.50$^{ +0.38}_{ -0.33}$ &  3.66$^{ +0.40}_{ -0.33}$ &  12.4$^{   +5.5}_{  -3.3}$ &  5.04$^{ +0.28}_{ -0.28}$  &  0.31$^{ +0.13}_{ -0.11}$ & 349$^{ +139}_{  -35}$ &  25.7$^{    +75.1}_{  -13.4}$ &  17.2$^{+3.4}_{-2.3}$ & 6.8$^{+1.9}_{-1.6}$ & 15.050& 0.36 &\\
\hline
\end{tabular}
\end{table}
\end{landscape}

\begin{landscape}
\begin{table}
\contcaption{}
%\label{table:targets} 
\begin{tabular}{l@{\hspace{1.5mm}}l@{\hspace{-1.5mm}}r@{\hspace{1.5mm}}r@{\hspace{1.5mm}}c@{\hspace{1.5mm}}c@{\hspace{1.5mm}}c@{\hspace{1.5mm}}r@{\hspace{1.5mm}}c@{\hspace{1.5mm}}c@{\hspace{1.5mm}}r@{\hspace{1.5mm}}r@{\hspace{1.5mm}}r@{\hspace{1.5mm}}r@{\hspace{1.5mm}}c@{\hspace{3mm}}c@{\hspace{3mm}}l}
\hline\hline
%\hline
BLOeM & Spect. & $\varv_{\rm rad}$ & $\sigma (\varv_{\rm rad})$ & $T_{\rm eff}$ & $\log g$ & $\log g_{c}$ & $R_{\rm eff}$ & $\log L$ & $Y$ & $\varv_{\rm e} \sin i$ & $M_{\rm spec}$ & $M_{\rm evol}$ & $\tau$ & $m_{\rm Ks}$ & $A_{\rm V}$ & Notes\\ 
       & Type & km\,s$^{-1}$  & km\,s$^{-1}$     & kK & cm s$^{-2}$ & cm s$^{-2}$ & $R_{\odot}$ & $L_{\odot}$  & $\cdots$  & km s$^{-1}$ & $M_{\odot}$ & $M_{\odot}$ & Myr & mag & mag\\
\hline
           1-076 &O9\,III:     & 213 &    1.3 &  33.7$^{  +1.5}_{  -1.5}$ &  3.69$^{ +0.16}_{ -0.16}$ &  3.69$^{ +0.16}_{ -0.16}$ &  13.7$^{   +5.7}_{  -3.2}$ &  5.34$^{ +0.27}_{ -0.27}$ &  0.34$^{ +0.06}_{ -0.06}$ &  53$^{  +76}_{  -18}$ &  33.9$^{    +44.8}_{  -14.0}$ &  21.3$^{+5.8}_{-3.1}$  & 4.8$^{+0.9}_{-0.7}$ & 14.609 & 0.38 &\\[2pt]
           1-078 &O9.7\,III:  & 182 &   31.1 &  31.8$^{  +2.7}_{  -1.5}$ &  3.69$^{ +0.33}_{ -0.19}$ &  3.69$^{ +0.33}_{ -0.19}$ &  11.6$^{   +4.9}_{  -3.0}$ &  5.09$^{ +0.27}_{ -0.26}$ &  0.30$^{ +0.06}_{ -0.08}$ &  51$^{  +75}_{  -21}$ &  24.4$^{    +56.3}_{  -10.9}$ &  18.7$^{+3.9}_{-2.7}$ & 5.7$^{+1.4}_{-1.4}$ & 15.085& 0.40 &SB1\\[2pt]
           1-079 &O9.2\,III:(n) pe& -38 & 17.1 & $\cdots$ & $\cdots$ & $\cdots$ & $\cdots$ & $\cdots$ & $\cdots$ & $\cdots$ & $\cdots$ & $\cdots$ & $\cdots$ & 13.314 & $\cdots$ &Oe, neb, 2MASS \\[2pt]
           1-080 &O8:V:+B+B& 167 &   24.7 &  33.4$^{  +2.7}_{  -2.7}$ &  4.10$^{ +0.34}_{ -0.51}$ &  4.11$^{ +0.34}_{ -0.51}$ &   13.7$^{   +6.1}_{  -3.6}$ &  5.32$^{ +0.28}_{ -0.28}$ &  0.24$^{ +0.09}_{ -0.06}$ & 202$^{  +25}_{  -24}$ &  88.6$^{   +222.2}_{  -51.2}$ & 22.1$^{+4.8}_{-4.0}$ & 4.7$^{+1.4}_{-1.2}$ & 14.731& 0.37 &neb\\
\hline
\end{tabular}
\end{table}
\end{landscape}

\begin{landscape}
\begin{table}
\contcaption{}
%\label{table:targets} 
\begin{tabular}{l@{\hspace{1.5mm}}l@{\hspace{-1.5mm}}r@{\hspace{1.5mm}}r@{\hspace{1.5mm}}c@{\hspace{1.5mm}}c@{\hspace{1.5mm}}c@{\hspace{1.5mm}}r@{\hspace{1.5mm}}c@{\hspace{1.5mm}}c@{\hspace{1.5mm}}r@{\hspace{1.5mm}}r@{\hspace{1.5mm}}r@{\hspace{1.5mm}}r@{\hspace{1.5mm}}c@{\hspace{3mm}}c@{\hspace{3mm}}l}
\hline\hline
%\hline
BLOeM & Spect. & $\varv_{\rm rad}$ & $\sigma (\varv_{\rm rad})$ & $T_{\rm eff}$ & $\log g$ & $\log g_{c}$ & $R_{\rm eff}$ & $\log L$ & $Y$ & $\varv_{\rm e} \sin i$ & $M_{\rm spec}$ & $M_{\rm evol}$ & $\tau$ & $m_{\rm Ks}$ & $A_{\rm V}$ & Notes\\ 
       & Type & km\,s$^{-1}$  & km\,s$^{-1}$     & kK & cm s$^{-2}$ & cm s$^{-2}$ & $R_{\odot}$ & $L_{\odot}$  & $\cdots$  & km s$^{-1}$ & $M_{\odot}$ & $M_{\odot}$ & Myr & mag & mag\\
\hline
           1-116 &B1\,II & 184 & $\cdots$ &   29.9$^{  +2.7}_{  -7.4}$ &  3.87$^{ +0.40}_{ -0.63}$ &  3.90$^{ +0.40}_{ -0.63}$ &  10.8$^{   +6.0}_{  -3.1}$ &  4.93$^{ +0.27}_{ -0.35}$ &  0.21$^{ +0.14}_{ -0.06}$ & 195$^{  +28}_{  -30}$ &  33.9$^{   +118.3}_{  -20.7}$ &  15.4$^{+3.1}_{-3.0}$ & 7.0$^{+3.9}_{-2.2}$ & 15.345& 0.37 &\\[2pt]
           2-001 &O9.2\,V:& 90 & $\cdots$ &  33.7$^{  +3.4}_{  -1.5}$ &  3.91$^{ +0.59}_{ -0.22}$ &  3.92$^{ +0.59}_{ -0.22}$ &    5.7$^{   +2.5}_{  -1.6}$ &  4.58$^{ +0.28}_{ -0.27}$ &  0.30$^{ +0.09}_{ -0.09}$ & 110$^{  +22}_{  -21}$ &   9.9$^{    +48.9}_{   -4.7}$ &   16.7$^{+2.5}_{-2.2}$  & 4.0$^{+1.3}_{-4.0}$ & 16.563& 0.40 &\\[2pt]
           2-002 &B2\,II   & 91 &  36.4 &  23.3$^{  +8.6}_{  -3.9}$ &  3.76$^{ +0.69}_{ -0.69}$ &  3.82$^{ +0.69}_{ -0.69}$ &    7.3$^{   +4.8}_{  -4.4}$ &  4.15$^{ +0.40}_{ -0.31}$ &  0.15$^{ +0.17}_{ -0.00}$ & 204$^{  +28}_{  -27}$ &  12.8$^{   +109.7}_{   -8.6}$ &  9.6$^{+2.4}_{-1.9}$  & 10.7$^{+6.2}_{-10.7}$ & 16.441& 0.25 &lpv/SB1\\
\hline
\end{tabular}
\end{table}
\end{landscape}

\begin{landscape}
\begin{table}
\contcaption{}
%\label{table:targets} 
\begin{tabular}{l@{\hspace{1.5mm}}l@{\hspace{-1.5mm}}r@{\hspace{1.5mm}}r@{\hspace{1.5mm}}c@{\hspace{1.5mm}}c@{\hspace{1.5mm}}c@{\hspace{1.5mm}}r@{\hspace{1.5mm}}c@{\hspace{1.5mm}}c@{\hspace{1.5mm}}r@{\hspace{1.5mm}}r@{\hspace{1.5mm}}r@{\hspace{1.5mm}}r@{\hspace{1.5mm}}c@{\hspace{3mm}}c@{\hspace{3mm}}l}
\hline\hline
%\hline
BLOeM & Spect. & $\varv_{\rm rad}$ & $\sigma (\varv_{\rm rad})$ & $T_{\rm eff}$ & $\log g$ & $\log g_{c}$ & $R_{\rm eff}$ & $\log L$ & $Y$ & $\varv_{\rm e} \sin i$ & $M_{\rm spec}$ & $M_{\rm evol}$ & $\tau$ & $m_{\rm Ks}$ & $A_{\rm V}$ & Notes\\ 
       & Type & km\,s$^{-1}$  & km\,s$^{-1}$     & kK & cm s$^{-2}$ & cm s$^{-2}$ & $R_{\odot}$ & $L_{\odot}$  & $\cdots$  & km s$^{-1}$ & $M_{\odot}$ & $M_{\odot}$ & Myr & mag & mag\\
\hline
           2-037 &B2.5\,Ib      & 124 &    3.3 &  18.8$^{  +1.5}_{  -1.5}$ &  2.50$^{ +0.24}_{ -0.29}$ &  2.52$^{ +0.25}_{ -0.29}$ &  36.6$^{   +7.9}_{  -6.5}$ &  5.18$^{ +0.15}_{ -0.15}$ &  0.31$^{ +0.10}_{ -0.15}$ &  55$^{  +76}_{  -13}$ &  16.0$^{    +17.9}_{   -7.2}$ &  19.8$^{+3.4}_{-1.9}$ & 7.4$^{+1.4}_{-1.0}$ & 13.169& 0.47 &2MASS. Post-MS \\[2pt]
           2-038 &B1.5\,II       & 132 &    5.4 &  25.3$^{  +3.8}_{  -3.0}$ &  3.50$^{ +0.29}_{ -0.33}$ &  3.55$^{ +0.29}_{ -0.33}$ &  13.8$^{   +4.5}_{  -3.8}$ &  4.84$^{ +0.21}_{ -0.20}$ &  0.18$^{ +0.11}_{ -0.03}$ & 201$^{  +25}_{  -24}$ &  24.8$^{    +41.1}_{  -13.0}$  &  15.7$^{+2.5}_{-2.0}$  & 9.1$^{+1.9}_{-2.3}$ & 15.071& 0.61 &\\[2pt]
           2-039 &B1\,II-Ib      & 143 &    2.1 &  23.8$^{  +1.1}_{  -1.1}$ &  3.12$^{ +0.14}_{ -0.19}$ &  3.13$^{ +0.15}_{ -0.19}$ &  23.5$^{   +7.5}_{  -4.8}$ &  5.20$^{ +0.22}_{ -0.22}$ &  0.20$^{ +0.07}_{ -0.03}$ &  55$^{  +77}_{  -13}$ &  26.8$^{    +26.5}_{  -11.0}$  &  17.8$^{+5.1}_{-2.5}$  & 7.6$^{+2.2}_{-1.5}$ & 13.856& 0.39 &Si\,{\sc iv} $\lambda$4089 too weak\\
\hline
\end{tabular}
\end{table}
\end{landscape}

\begin{landscape}
\begin{table}
\contcaption{}
%\label{table:targets} 
\begin{tabular}{l@{\hspace{1.5mm}}l@{\hspace{-1.5mm}}r@{\hspace{1.5mm}}r@{\hspace{1.5mm}}c@{\hspace{1.5mm}}c@{\hspace{1.5mm}}c@{\hspace{1.5mm}}r@{\hspace{1.5mm}}c@{\hspace{1.5mm}}c@{\hspace{1.5mm}}r@{\hspace{1.5mm}}r@{\hspace{1.5mm}}r@{\hspace{1.5mm}}r@{\hspace{1.5mm}}c@{\hspace{3mm}}c@{\hspace{3mm}}l}
\hline\hline
%\hline
BLOeM & Spect. & $\varv_{\rm rad}$ & $\sigma (\varv_{\rm rad})$ & $T_{\rm eff}$ & $\log g$ & $\log g_{c}$ & $R_{\rm eff}$ & $\log L$ & $Y$ & $\varv_{\rm e} \sin i$ & $M_{\rm spec}$ & $M_{\rm evol}$ & $\tau$ & $m_{\rm Ks}$ & $A_{\rm V}$ & Notes\\ 
       & Type & km\,s$^{-1}$  & km\,s$^{-1}$     & kK & cm s$^{-2}$ & cm s$^{-2}$ & $R_{\odot}$ & $L_{\odot}$  & $\cdots$  & km s$^{-1}$ & $M_{\odot}$ & $M_{\odot}$ & Myr & mag & mag\\
\hline
           2-077 &B1\,II        & 140 &    7.5 &  29.9$^{  +1.2}_{  -2.7}$ &  3.93$^{ +0.23}_{ -0.51}$ &  3.93$^{ +0.23}_{ -0.51}$ &   9.7$^{   +3.9}_{  -2.2}$ &  4.83$^{ +0.25}_{ -0.26}$ &  0.18$^{ +0.09}_{ -0.03}$ &  78$^{  +19}_{  -20}$ &  29.4$^{    +47.8}_{  -16.7}$  & 14.4$^{+2.9}_{-2.0}$ & 7.9$^{+2.2}_{-1.4}$ & 15.572& 0.40 &lpv/SB1\\[2pt]
           2-078 &B0\,II        & 150 &    1.7 &  31.9$^{  +1.2}_{  -1.6}$ &  3.70$^{ +0.17}_{ -0.17}$ &  3.70$^{ +0.17}_{ -0.17}$ &  13.5$^{   +5.1}_{  -3.0}$ &  5.23$^{ +0.25}_{ -0.25}$ &  0.30$^{ +0.07}_{ -0.06}$ &  31$^{  +13}_{  -20}$ &  33.2$^{    +42.0}_{  -13.6}$ & 20.0$^{+4.5}_{-3.5}$ & 6.0$^{+0.9}_{-1.0}$ & 14.731& 0.43 &\\[2pt]
           2-079 &O4\,V:+early B& 151 & 1.6 &    37.6$^{  +3.1}_{  -1.5}$ &  4.50$^{ +0.00}_{ -0.29}$ &  4.50$^{ +0.00}_{ -0.29}$ &   8.6$^{   +3.7}_{  -2.2}$ &  5.12$^{ +0.28}_{ -0.27}$ &  0.47$^{ +0.08}_{ -0.06}$ &  89$^{  +24}_{  -19}$ &  84.9$^{    +62.1}_{  -42.1}$ & 23.1$^{+4.5}_{-3.1}$  & 3.4$^{+0.8}_{-2.1}$ & 15.503& 0.38 &\\[2pt]
           2-081 &B2\,II e & 165 & 4.4& $\cdots$ & $\cdots$ & $\cdots$ & $\cdots$ & $\cdots$ & $\cdots$ & $\cdots$ & $\cdots$ & $\cdots$ & $\cdots$ & 14.160 & $\cdots$ &Be \\
\hline
\end{tabular}
\end{table}
\end{landscape}

\begin{landscape}
\begin{table}
\contcaption{}
%\label{table:targets} 
\begin{tabular}{l@{\hspace{1.5mm}}l@{\hspace{-1.5mm}}r@{\hspace{1.5mm}}r@{\hspace{1.5mm}}c@{\hspace{1.5mm}}c@{\hspace{1.5mm}}c@{\hspace{1.5mm}}r@{\hspace{1.5mm}}c@{\hspace{1.5mm}}c@{\hspace{1.5mm}}r@{\hspace{1.5mm}}r@{\hspace{1.5mm}}r@{\hspace{1.5mm}}r@{\hspace{1.5mm}}c@{\hspace{3mm}}c@{\hspace{3mm}}l}
\hline\hline
%\hline
BLOeM & Spect. & $\varv_{\rm rad}$ & $\sigma (\varv_{\rm rad})$ & $T_{\rm eff}$ & $\log g$ & $\log g_{c}$ & $R_{\rm eff}$ & $\log L$ & $Y$ & $\varv_{\rm e} \sin i$ & $M_{\rm spec}$ & $M_{\rm evol}$ & $\tau$ & $m_{\rm Ks}$ & $A_{\rm V}$ & Notes\\ 
       & Type & km\,s$^{-1}$  & km\,s$^{-1}$     & kK & cm s$^{-2}$ & cm s$^{-2}$ & $R_{\odot}$ & $L_{\odot}$  & $\cdots$  & km s$^{-1}$ & $M_{\odot}$ & $M_{\odot}$ & Myr & mag & mag\\
\hline
           2-115 &B5\,Ib           & 154 &    1.6 &  14.6$^{  +0.8}_{  -1.1}$ &  2.50$^{ +0.14}_{ -0.33}$ &  2.52$^{ +0.16}_{ -0.33}$ &  35.1$^{   +9.6}_{  -6.6}$ &  4.71$^{ +0.18}_{ -0.19}$ &  0.15$^{ +0.06}_{ -0.00}$ &  55$^{  +77}_{  -13}$ &  14.7$^{    +13.7}_{   -7.1}$ & 12.7$^{+2.7}_{-1.2}$ & 12.5$^{+3.0}_{-2.4}$ & 13.673& 0.45 & Post-MS\\[2pt]
           2-116 &sgB[e] & 100 & 8.3& $\cdots$ & $\cdots$ & $\cdots$ & $\cdots$ & $\cdots$ & $\cdots$ & $\cdots$ & $\cdots$ & $\cdots$ & $\cdots$ & 11.109 & $\cdots$ &B[e], 2MASS \\[2pt]
           3-001 &B2.5\,Ib      & 129 &    1.1 &  18.8$^{  +1.9}_{  -1.9}$ &  2.69$^{ +0.14}_{ -0.33}$ &  2.70$^{ +0.15}_{ -0.33}$ &    38.3$^{   +9.9}_{  -8.1}$ &  5.22$^{ +0.17}_{ -0.17}$ &  0.28$^{ +0.15}_{ -0.07}$ &  55$^{  +77}_{  -14}$ &  26.8$^{    +22.9}_{  -13.2}$  & 19.9$^{+4.2}_{-1.8}$ & 7.2$^{+1.3}_{-1.3}$ & 13.198 & 0.59 & 2MASS. Post-MS\\[2pt]
           3-002 &O6\,V:nn    & 97 &   13.8 & 38.2$^{  +6.0}_{  -3.4}$ &  3.53$^{ +0.65}_{ -0.32}$ &  3.84$^{ +0.65}_{ -0.33}$ &   6.7$^{   +2.2}_{  -1.9}$ &  4.94$^{ +0.23}_{ -0.19}$ &  0.47$^{ +0.08}_{ -0.18}$ & 402$^{  +30}_{  -36}$ &  11.3$^{    +53.4}_{   -6.0}$ &  22.2$^{+3.6}_{-3.1}$ & 3.5$^{+1.1}_{-3.3}$ & 15.853 & 0.54 &\\
\hline
\end{tabular}
\end{table}
\end{landscape}

\begin{landscape}
\begin{table}
\contcaption{}
%\label{table:targets} 
\begin{tabular}{l@{\hspace{1.5mm}}l@{\hspace{-1.5mm}}r@{\hspace{1.5mm}}r@{\hspace{1.5mm}}c@{\hspace{1.5mm}}c@{\hspace{1.5mm}}c@{\hspace{1.5mm}}r@{\hspace{1.5mm}}c@{\hspace{1.5mm}}c@{\hspace{1.5mm}}r@{\hspace{1.5mm}}r@{\hspace{1.5mm}}r@{\hspace{1.5mm}}r@{\hspace{1.5mm}}c@{\hspace{3mm}}c@{\hspace{3mm}}l}
\hline\hline
%\hline
BLOeM & Spect. & $\varv_{\rm rad}$ & $\sigma (\varv_{\rm rad})$ & $T_{\rm eff}$ & $\log g$ & $\log g_{c}$ & $R_{\rm eff}$ & $\log L$ & $Y$ & $\varv_{\rm e} \sin i$ & $M_{\rm spec}$ & $M_{\rm evol}$ & $\tau$ & $m_{\rm Ks}$ & $A_{\rm V}$ & Notes\\ 
       & Type & km\,s$^{-1}$  & km\,s$^{-1}$     & kK & cm s$^{-2}$ & cm s$^{-2}$ & $R_{\odot}$ & $L_{\odot}$  & $\cdots$  & km s$^{-1}$ & $M_{\odot}$ & $M_{\odot}$ & Myr & mag & mag\\
\hline
           3-036 &B0.7\,III-II      & 142 &    1.5 &   28.3$^{  +2.7}_{  -1.2}$ &  3.53$^{ +0.46}_{ -0.17}$ &  3.53$^{ +0.46}_{ -0.17}$ &  11.2$^{   +4.5}_{  -2.9}$ &  4.86$^{ +0.26}_{ -0.25}$ &  0.32$^{ +0.08}_{ -0.09}$ &  25$^{  +12}_{  -20}$ &  15.4$^{    +49.3}_{   -6.7}$ & 15.0$^{+3.0}_{-2.0}$ & 7.9$^{+1.9}_{-1.8}$ & 15.293 & 0.39 &neb\\[2pt]
           3-037 &B3\,Ia            & 166 &    3.3 &  16.9$^{  +0.4}_{  -1.5}$ &  2.31$^{ +0.14}_{ -0.33}$ &  2.32$^{ +0.15}_{ -0.33}$ &   69.1$^{  +18.0}_{ -11.7}$ &  5.55$^{ +0.17}_{ -0.19}$ &  0.18$^{ +0.09}_{ -0.03}$ &  55$^{  +77}_{  -13}$ &  36.6$^{    +32.1}_{  -17.2}$ & 24.1$^{+9.7}_{-1.3}$ & 5.2$^{+1.1}_{-1.0}$ & 11.957 & 0.89 &lpv/SB1, 2MASS. Post-MS\\[2pt]
           3-039 &B3\,Ib            & 143 &    3.1 &  15.0$^{  +1.5}_{  -0.8}$ &  2.31$^{ +0.33}_{ -0.14}$ &  2.33$^{ +0.35}_{ -0.14}$ &  39.5$^{  +10.8}_{  -8.8}$ &  4.86$^{ +0.20}_{ -0.18}$ &  0.18$^{ +0.06}_{ -0.03}$ &  55$^{  +77}_{  -13}$ &  12.3$^{    +22.1}_{   -4.8}$ & 15.2$^{+2.3}_{-2.3}$ & 10.4$^{+2.5}_{-1.9}$ & 13.225 & 0.28 &lpv/SB1, Post-MS\\
\hline
\end{tabular}
\end{table}
\end{landscape}

\begin{landscape}
\begin{table}
\contcaption{}
%\label{table:targets} 
\begin{tabular}{l@{\hspace{1.5mm}}l@{\hspace{-1.5mm}}r@{\hspace{1.5mm}}r@{\hspace{1.5mm}}c@{\hspace{1.5mm}}c@{\hspace{1.5mm}}c@{\hspace{1.5mm}}r@{\hspace{1.5mm}}c@{\hspace{1.5mm}}c@{\hspace{1.5mm}}r@{\hspace{1.5mm}}r@{\hspace{1.5mm}}r@{\hspace{1.5mm}}r@{\hspace{1.5mm}}c@{\hspace{3mm}}c@{\hspace{3mm}}l}
\hline\hline
%\hline
BLOeM & Spect. & $\varv_{\rm rad}$ & $\sigma (\varv_{\rm rad})$ & $T_{\rm eff}$ & $\log g$ & $\log g_{c}$ & $R_{\rm eff}$ & $\log L$ & $Y$ & $\varv_{\rm e} \sin i$ & $M_{\rm spec}$ & $M_{\rm evol}$ & $\tau$ & $m_{\rm Ks}$ & $A_{\rm V}$ & Notes\\ 
       & Type & km\,s$^{-1}$  & km\,s$^{-1}$     & kK & cm s$^{-2}$ & cm s$^{-2}$ & $R_{\odot}$ & $L_{\odot}$  & $\cdots$  & km s$^{-1}$ & $M_{\odot}$ & $M_{\odot}$ & Myr & mag & mag\\
\hline
           3-075 &O6.5\,V:nnn& 139 &    4.5 &  35.5$^{  +1.5}_{  -2.6}$ &  4.34$^{ +0.11}_{ -0.70}$ &  4.35$^{ +0.11}_{ -0.70}$ &  15.4$^{   +6.3}_{  -3.6}$ &  5.53$^{ +0.26}_{ -0.27}$ &  0.25$^{ +0.11}_{ -0.05}$ & 294$^{  +30}_{  -33}$ & 195.6$^{   +210.4}_{ -119.7}$ & 24.9$^{+7.5}_{-3.6}$ & 4.2$^{+0.9}_{-0.7}$ & 14.373 & 0.41 &\\[2pt]
           3-076 &O8\,Vn         &  54 &   10.1 &  33.4$^{  +2.3}_{  -1.6}$ &  3.30$^{ +0.23}_{ -0.17}$ &  3.67$^{ +0.29}_{ -0.18}$ &    6.9$^{   +2.8}_{  -1.7}$ &  4.73$^{ +0.27}_{ -0.26}$ &  0.44$^{ +0.11}_{ -0.11}$ & 353$^{ +136}_{  -29}$ &   8.0$^{    +16.1}_{   -3.4}$ & 17.6$^{+2.3}_{-1.7}$ & 5.2$^{+1.1}_{-3.6}$ & 16.104 & 0.34 &\\[2pt]
           3-077 &B1\,II e & 214 & 7.5& $\cdots$ & $\cdots$ & $\cdots$ & $\cdots$ & $\cdots$ & $\cdots$ & $\cdots$ & $\cdots$ & $\cdots$ & $\cdots$ & 14.088 & $\cdots$ &Be \\[2pt]
           3-078 &O8\,III((f))     & 150 &    1.5 &  35.6$^{  +1.5}_{  -1.5}$ &  4.36$^{ +0.10}_{ -0.19}$ &  4.36$^{ +0.10}_{ -0.19}$ &  15.9$^{   +6.4}_{  -3.7}$ &  5.56$^{ +0.26}_{ -0.26}$ &  0.25$^{ +0.06}_{ -0.05}$ &  78$^{  +96}_{  -20}$ & 209.0$^{   +210.8}_{  -89.6}$  &  25.2$^{+9.3}_{-3.4}$ & 4.0$^{+0.6}_{-0.7}$ & 14.261 & 0.51 &\\
\hline
\end{tabular}
\end{table}
\end{landscape}

\begin{landscape}
\begin{table}
\contcaption{}
%\label{table:targets} 
\begin{tabular}{l@{\hspace{1.5mm}}l@{\hspace{-1.5mm}}r@{\hspace{1.5mm}}r@{\hspace{1.5mm}}c@{\hspace{1.5mm}}c@{\hspace{1.5mm}}c@{\hspace{1.5mm}}r@{\hspace{1.5mm}}c@{\hspace{1.5mm}}c@{\hspace{1.5mm}}r@{\hspace{1.5mm}}r@{\hspace{1.5mm}}r@{\hspace{1.5mm}}r@{\hspace{1.5mm}}c@{\hspace{3mm}}c@{\hspace{3mm}}l}
\hline\hline
%\hline
BLOeM & Spect. & $\varv_{\rm rad}$ & $\sigma (\varv_{\rm rad})$ & $T_{\rm eff}$ & $\log g$ & $\log g_{c}$ & $R_{\rm eff}$ & $\log L$ & $Y$ & $\varv_{\rm e} \sin i$ & $M_{\rm spec}$ & $M_{\rm evol}$ & $\tau$ & $m_{\rm Ks}$ & $A_{\rm V}$ & Notes\\ 
       & Type & km\,s$^{-1}$  & km\,s$^{-1}$     & kK & cm s$^{-2}$ & cm s$^{-2}$ & $R_{\odot}$ & $L_{\odot}$  & $\cdots$  & km s$^{-1}$ & $M_{\odot}$ & $M_{\odot}$ & Myr & mag & mag\\
\hline
           3-113 &B2\,II:             & 157 &    1.3 &  20.1$^{  +2.7}_{  -1.6}$ &  3.13$^{ +0.46}_{ -0.17}$ &  3.15$^{ +0.46}_{ -0.17}$ &  15.5$^{   +5.1}_{  -4.2}$ &  4.55$^{ +0.23}_{ -0.20}$ &  0.21$^{ +0.11}_{ -0.06}$ &  78$^{  +97}_{  -19}$ &  12.3$^{    +35.1}_{   -5.5}$ & 11.4$^{+1.7}_{-1.6}$ & 15.0$^{+3.7}_{-3.3}$ & 14.944 & 0.37 &\\[2pt]
           3-115 &B1\,III             & 148 &    2.4 &  23.7$^{  +1.2}_{  -1.2}$ &  3.30$^{ +0.23}_{ -0.17}$ &  3.32$^{ +0.23}_{ -0.17}$ &  16.6$^{   +6.1}_{  -3.7}$ &  4.89$^{ +0.24}_{ -0.24}$ &  0.18$^{ +0.08}_{ -0.03}$ & 113$^{  +20}_{  -19}$ &  21.2$^{    +31.7}_{   -8.7}$ &  13.9$^{+2.8}_{-1.9}$ & 10.7$^{+2.7}_{-1.8}$ & 14.677 & 0.47 &lpv/SB1, Si\,{\sc iv} $\lambda$4089 too weak\\[2pt]
           3-116 &B3\,II              & 151 &    2.1 &  17.9$^{  +2.6}_{  -0.8}$ &  2.88$^{ +0.33}_{ -0.14}$ &  2.91$^{ +0.35}_{ -0.14}$ &  17.2$^{   +5.3}_{  -4.7}$ &  4.43$^{ +0.23}_{ -0.18}$ &  0.15$^{ +0.03}_{ -0.00}$ &  78$^{  +98}_{  -19}$ &   8.7$^{    +17.1}_{   -3.8}$  &  10.2$^{+1.5}_{-1.2}$ & 18.2$^{+4.1}_{-3.7}$ & 14.849 & 0.45 &\\
\hline
\end{tabular}
\end{table}
\end{landscape}

\begin{landscape}
\begin{table}
\contcaption{}
%\label{table:targets} 
\begin{tabular}
{l@{\hspace{1.5mm}}l@{\hspace{-1.5mm}}r@{\hspace{1.5mm}}r@{\hspace{1.5mm}}c@{\hspace{1.5mm}}c@{\hspace{1.5mm}}c@{\hspace{1.5mm}}r@{\hspace{1.5mm}}c@{\hspace{1.5mm}}c@{\hspace{1.5mm}}r@{\hspace{1.5mm}}r@{\hspace{1.5mm}}r@{\hspace{1.5mm}}r@{\hspace{1.5mm}}c@{\hspace{3mm}}c@{\hspace{3mm}}l}
\hline\hline
%\hline
BLOeM & Spect. & $\varv_{\rm rad}$ & $\sigma (\varv_{\rm rad})$ & $T_{\rm eff}$ & $\log g$ & $\log g_{c}$ & $R_{\rm eff}$ & $\log L$ & $Y$ & $\varv_{\rm e} \sin i$ & $M_{\rm spec}$ & $M_{\rm evol}$ & $\tau$ & $m_{\rm Ks}$ & $A_{\rm V}$ & Notes\\ 
       & Type & km\,s$^{-1}$  & km\,s$^{-1}$     & kK & cm s$^{-2}$ & cm s$^{-2}$ & $R_{\odot}$ & $L_{\odot}$  & $\cdots$  & km s$^{-1}$ & $M_{\odot}$ & $M_{\odot}$ & Myr & mag & mag\\
\hline
           4-036 &B1.5\,II        & 142 &   15.5 & 23.7$^{  +3.1}_{  -1.6}$ &  3.30$^{ +0.29}_{ -0.23}$ &  3.33$^{ +0.29}_{ -0.23}$ &  14.7$^{   +5.9}_{  -4.3}$ &  4.79$^{ +0.27}_{ -0.25}$ &  0.18$^{ +0.10}_{ -0.03}$ & 113$^{  +20}_{  -19}$ &  16.8$^{    +32.5}_{   -8.2}$  & 13.2$^{+3.2}_{-1.6}$ & 10.1$^{+3.1}_{-2.1}$ & 14.980 & 0.36 &SB1\\[2pt]
           4-037 &O9.7\,IV      & 104 & 49.3& $\cdots$ & $\cdots$ & $\cdots$ & $\cdots$ & $\cdots$ & $\cdots$ & $\cdots$ & $\cdots$ & $\cdots$ & $\cdots$ & 15.795 & $\cdots$ &neb \\[2pt]
           4-038 &B1.5\,III-II    & 157 &    3.5 &   23.7$^{  +1.2}_{  -2.0}$ &  3.47$^{ +0.23}_{ -0.34}$ &  3.50$^{ +0.23}_{ -0.34}$ &    8.9$^{   +3.5}_{  -2.0}$ &  4.35$^{ +0.25}_{ -0.26}$ &  0.15$^{ +0.14}_{ -0.00}$ & 112$^{  +20}_{  -20}$ &   9.2$^{    +14.5}_{   -4.6}$  &   9.7$^{+1.8}_{-1.1}$ & 16.4$^{+3.6}_{-3.3}$ & 16.087 & 0.36 &\\[2pt]
           4-039 &O6.5f?pe & 148 & 10.7& $\cdots$ & $\cdots$ & $\cdots$ & $\cdots$ & $\cdots$ & $\cdots$ & $\cdots$ & $\cdots$ & $\cdots$ & $\cdots$ & 15.078 & $\cdots$ &Oe \\[2pt]
           4-040 &O7\,V(n) & 121 & 7.4& $\cdots$ & $\cdots$ & $\cdots$ & $\cdots$ & $\cdots$ & $\cdots$ & $\cdots$ & $\cdots$ & $\cdots$ & $\cdots$ & 15.431 & $\cdots$ &neb \\
\hline
\end{tabular}
\end{table}
\end{landscape}

\begin{landscape}
\begin{table}
\contcaption{}
%\label{table:targets} 
\begin{tabular}{l@{\hspace{1.5mm}}l@{\hspace{-1.5mm}}r@{\hspace{1.5mm}}r@{\hspace{1.5mm}}c@{\hspace{1.5mm}}c@{\hspace{1.5mm}}c@{\hspace{1.5mm}}r@{\hspace{1.5mm}}c@{\hspace{1.5mm}}c@{\hspace{1.5mm}}r@{\hspace{1.5mm}}r@{\hspace{1.5mm}}r@{\hspace{1.5mm}}r@{\hspace{1.5mm}}c@{\hspace{3mm}}c@{\hspace{3mm}}l}
\hline\hline
%\hline
BLOeM & Spect. & $\varv_{\rm rad}$ & $\sigma (\varv_{\rm rad})$ & $T_{\rm eff}$ & $\log g$ & $\log g_{c}$ & $R_{\rm eff}$ & $\log L$ & $Y$ & $\varv_{\rm e} \sin i$ & $M_{\rm spec}$ & $M_{\rm evol}$ & $\tau$ & $m_{\rm Ks}$ & $A_{\rm V}$ & Notes\\ 
       & Type & km\,s$^{-1}$  & km\,s$^{-1}$     & kK & cm s$^{-2}$ & cm s$^{-2}$ & $R_{\odot}$ & $L_{\odot}$  & $\cdots$  & km s$^{-1}$ & $M_{\odot}$ & $M_{\odot}$ & Myr & mag & mag\\
\hline
           4-072 &B9\,Ia            & 167 &    2.9 &  12.0$^{  +0.4}_{  -0.4}$ &  1.70$^{ +0.17}_{ -0.17}$ &  1.71$^{ +0.17}_{ -0.17}$ &  70.6$^{  +17.7}_{ -12.1}$ &  4.96$^{ +0.18}_{ -0.18}$ &  0.18$^{ +0.05}_{ -0.03}$ &  22$^{  +12}_{   -9}$ &   9.3$^{     +8.5}_{   -3.5}$ & 16.7$^{+2.5}_{-2.6}$ & 9.8$^{+1.6}_{-2.1}$ & 12.324 & 0.48 &SB?, 2MASS. Post-MS\\[2pt]
           4-073 &O9.2\,V         & 220 &   27.2 &  35.5$^{  +1.9}_{  -3.4}$ &  4.50$^{ +0.00}_{ -0.70}$ &  4.51$^{ +0.00}_{ -0.70}$ &   8.5$^{   +3.7}_{  -2.0}$ &  5.02$^{ +0.27}_{ -0.28}$ &  0.30$^{ +0.13}_{ -0.13}$ & 153$^{  +24}_{  -24}$ &  84.3$^{    +61.7}_{  -51.7}$ & 20.0$^{+3.1}_{-3.1}$ & 4.5$^{+1.8}_{-1.9}$ & 15.626 & 0.42 &SB1, neb\\[2pt]
           4-074 &O9\,V            & 165 &    1.1 &    35.5$^{  +1.5}_{  -1.5}$ &  4.07$^{ +0.16}_{ -0.22}$ &  4.07$^{ +0.16}_{ -0.22}$ &   8.0$^{   +3.4}_{  -1.9}$ &  4.96$^{ +0.27}_{ -0.27}$ &  0.36$^{ +0.05}_{ -0.07}$ &   0$^{  +25}_{   -0}$ &  27.3$^{    +36.9}_{  -12.2}$  & 19.0$^{+3.0}_{-1.9}$ & 4.5$^{+1.1}_{-1.9}$ & 15.750 & 0.38 &\\
\hline
\end{tabular}
\end{table}
\end{landscape}

\begin{landscape}
\begin{table}
\contcaption{}
%\label{table:targets} 
\begin{tabular}{l@{\hspace{1.5mm}}l@{\hspace{-1.5mm}}r@{\hspace{1.5mm}}r@{\hspace{1.5mm}}c@{\hspace{1.5mm}}c@{\hspace{1.5mm}}c@{\hspace{1.5mm}}r@{\hspace{1.5mm}}c@{\hspace{1.5mm}}c@{\hspace{1.5mm}}r@{\hspace{1.5mm}}r@{\hspace{1.5mm}}r@{\hspace{1.5mm}}r@{\hspace{1.5mm}}c@{\hspace{3mm}}c@{\hspace{3mm}}l}
\hline\hline
%\hline
BLOeM & Spect. & $\varv_{\rm rad}$ & $\sigma (\varv_{\rm rad})$ & $T_{\rm eff}$ & $\log g$ & $\log g_{c}$ & $R_{\rm eff}$ & $\log L$ & $Y$ & $\varv_{\rm e} \sin i$ & $M_{\rm spec}$ & $M_{\rm evol}$ & $\tau$ & $m_{\rm Ks}$ & $A_{\rm V}$ & Notes\\ 
       & Type & km\,s$^{-1}$  & km\,s$^{-1}$     & kK & cm s$^{-2}$ & cm s$^{-2}$ & $R_{\odot}$ & $L_{\odot}$  & $\cdots$  & km s$^{-1}$ & $M_{\odot}$ & $M_{\odot}$ & Myr & mag & mag\\
\hline
           4-112 &B5\,II            & 173 &    2.6 &  15.8$^{  +0.8}_{  -1.9}$ &  2.69$^{ +0.19}_{ -0.33}$ &  2.73$^{ +0.23}_{ -0.33}$ &  18.4$^{   +5.9}_{  -3.7}$ &  4.28$^{ +0.19}_{ -0.22}$ &  0.15$^{ +0.05}_{ -0.00}$ &  78$^{  +98}_{  -19}$ &   6.6$^{     +9.0}_{   -3.2}$  &  9.2$^{+1.6}_{-1.2}$ &  21.5$^{+6.1}_{-5.0}$ & 14.949 & 0.28 & Post-MS\\[2pt]
           4-113 &B2.5\,II pe    & 169 &   19.8 &  19.8$^{  +4.6}_{  -2.7}$ &  3.50$^{ +1.00}_{ -0.81}$ &  3.61$^{ +1.00}_{ -0.81}$ &  11.9$^{   +6.3}_{  -5.0}$ &  4.29$^{ +0.33}_{ -0.29}$ &  0.30$^{ +0.17}_{ -0.13}$ & 279$^{  +37}_{  -37}$ &  21.1$^{   +329.4}_{  -13.9}$ & 9.4$^{+1.9}_{-1.5}$ & 16.0$^{+8.8}_{-6.7}$ & 15.695 & 0.15 &SB1? H$\gamma$ poor fit\\[2pt]
           4-114 &B5\,II            & 211 &    3.2 &  15.9$^{  +0.8}_{  -0.8}$ &  2.67$^{ +0.51}_{ -0.17}$ &  2.75$^{ +0.52}_{ -0.17}$ &   18.6$^{   +5.1}_{  -3.5}$ &  4.29$^{ +0.19}_{ -0.19}$ &  0.16$^{ +0.09}_{ -0.01}$ & 113$^{  +20}_{  -19}$ &   7.2$^{    +21.5}_{   -2.8}$  &  9.6$^{+1.5}_{-1.2}$ & 20.8$^{+5.5}_{-4.6}$ & 14.916 & 0.28 &SB? Post-MS\\
\hline
\end{tabular}
\end{table}
\end{landscape}

\begin{landscape}
\begin{table}
\contcaption{}
%\label{table:targets} 
\begin{tabular}{l@{\hspace{1.5mm}}l@{\hspace{-1.5mm}}r@{\hspace{1.5mm}}r@{\hspace{1.5mm}}c@{\hspace{1.5mm}}c@{\hspace{1.5mm}}c@{\hspace{1.5mm}}r@{\hspace{1.5mm}}c@{\hspace{1.5mm}}c@{\hspace{1.5mm}}r@{\hspace{1.5mm}}r@{\hspace{1.5mm}}r@{\hspace{1.5mm}}r@{\hspace{1.5mm}}c@{\hspace{3mm}}c@{\hspace{3mm}}l}
\hline\hline
%\hline
BLOeM & Spect. & $\varv_{\rm rad}$ & $\sigma (\varv_{\rm rad})$ & $T_{\rm eff}$ & $\log g$ & $\log g_{c}$ & $R_{\rm eff}$ & $\log L$ & $Y$ & $\varv_{\rm e} \sin i$ & $M_{\rm spec}$ & $M_{\rm evol}$ & $\tau$ & $m_{\rm Ks}$ & $A_{\rm V}$ & Notes\\ 
       & Type & km\,s$^{-1}$  & km\,s$^{-1}$     & kK & cm s$^{-2}$ & cm s$^{-2}$ & $R_{\odot}$ & $L_{\odot}$  & $\cdots$  & km s$^{-1}$ & $M_{\odot}$ & $M_{\odot}$ & Myr & mag & mag\\
\hline
           5-035 &B1.5\,II: e      & 172 &    7.1 &    23.3$^{  +5.4}_{  -5.4}$ &  3.12$^{ +1.05}_{ -0.81}$ &  3.30$^{ +1.05}_{ -0.81}$ &  19.2$^{  +10.3}_{  -7.9}$ &  4.99$^{ +0.30}_{ -0.31}$ &  0.52$^{ +0.03}_{ -0.17}$ & 303$^{  +30}_{  -30}$ &  26.9$^{   +474.4}_{  -17.6}$ & 15.5$^{+3.7}_{-2.6}$ & 8.7$^{+2.8}_{-2.9}$ & 14.060 & 1.36 &SB?, 2MASS\\[2pt]
           5-036 &B9\,Ib            & 347 &    1.5 &  $\cdots$ &  $\cdots$ &  $\cdots$ &   $\cdots$ &  $\cdots$ &  $\cdots$ & $\cdots$ & $\cdots$ &  $\cdots$ & & 14.425 & &Poor fits\\[2pt]
           5-037 &B0.5\,V:+early B& 118 & 18.3 &  23.5$^{  +7.1}_{  -3.8}$ &  3.69$^{ +0.48}_{ -0.48}$ &  3.72$^{ +0.48}_{ -0.48}$ &  11.1$^{   +6.6}_{  -5.7}$ &  4.53$^{ +0.36}_{ -0.30}$ &  0.17$^{ +0.15}_{ -0.02}$ & 153$^{  +24}_{  -24}$ &  23.6$^{   +108.2}_{  -15.0}$ & 12.5$^{+2.5}_{-2.9}$ & 9.7$^{+5.9}_{-6.0}$ & 15.640 & 0.27 &SB2\\[2pt]
           5-038 &B1.5\,III-II          & 128 &    6.2 &  23.7$^{  +1.2}_{  -2.3}$ &  3.70$^{ +0.17}_{ -0.34}$ &  3.71$^{ +0.17}_{ -0.34}$ &  14.2$^{   +5.8}_{  -3.3}$ &  4.75$^{ +0.25}_{ -0.27}$ &  0.18$^{ +0.13}_{ -0.03}$ & 113$^{  +20}_{  -19}$ &  37.8$^{    +51.2}_{  -19.2}$  & 12.6$^{+2.2}_{-2.2}$ & 12.9$^{+3.0}_{-3.0}$ & 15.090 & 0.31 &SB1\\
\hline
\end{tabular}
\end{table}
\end{landscape}

\begin{landscape}
\begin{table}
\contcaption{}
%\label{table:targets} 
\begin{tabular}{l@{\hspace{1.5mm}}l@{\hspace{-1.5mm}}r@{\hspace{1.5mm}}r@{\hspace{1.5mm}}c@{\hspace{1.5mm}}c@{\hspace{1.5mm}}c@{\hspace{1.5mm}}r@{\hspace{1.5mm}}c@{\hspace{1.5mm}}c@{\hspace{1.5mm}}r@{\hspace{1.5mm}}r@{\hspace{1.5mm}}r@{\hspace{1.5mm}}r@{\hspace{1.5mm}}c@{\hspace{3mm}}c@{\hspace{3mm}}l}
\hline\hline
%\hline
BLOeM & Spect. & $\varv_{\rm rad}$ & $\sigma (\varv_{\rm rad})$ & $T_{\rm eff}$ & $\log g$ & $\log g_{c}$ & $R_{\rm eff}$ & $\log L$ & $Y$ & $\varv_{\rm e} \sin i$ & $M_{\rm spec}$ & $M_{\rm evol}$ & $\tau$ & $m_{\rm Ks}$ & $A_{\rm V}$ & Notes\\ 
       & Type & km\,s$^{-1}$  & km\,s$^{-1}$     & kK & cm s$^{-2}$ & cm s$^{-2}$ & $R_{\odot}$ & $L_{\odot}$  & $\cdots$  & km s$^{-1}$ & $M_{\odot}$ & $M_{\odot}$ & Myr & mag & mag\\
\hline
           5-073 &B2\,II:            & 139 &    4.4 &  24.9$^{  +3.8}_{  -2.7}$ &  3.69$^{ +0.48}_{ -0.48}$ &  3.72$^{ +0.48}_{ -0.48}$ &  10.3$^{   +4.5}_{  -3.3}$ &  4.56$^{ +0.28}_{ -0.26}$ &  0.18$^{ +0.13}_{ -0.03}$ & 153$^{  +24}_{  -24}$ &  20.1$^{    +73.0}_{  -11.8}$ &  12.6$^{+2.0}_{-2.3}$ & 10.8$^{+4.3}_{-3.7}$ & 15.719 & 0.35 & \\[2pt]
           5-074 &B0.5\,IV         & 165 &    3.6 &  29.9$^{  +1.9}_{  -2.7}$ &  4.12$^{ +0.29}_{ -0.38}$ &  4.12$^{ +0.29}_{ -0.38}$ &   8.6$^{   +3.6}_{  -2.1}$ &  4.72$^{ +0.26}_{ -0.27}$ &  0.24$^{ +0.09}_{ -0.08}$ &  53$^{  +20}_{  -23}$ &  35.4$^{    +71.6}_{  -18.7}$ & 14.1$^{+2.3}_{-2.1}$ & 7.9$^{+2.7}_{-2.3}$ & 15.905 & 0.34 &SB1\\[2pt]
           5-075 &B2\,II:            & 144 &   18.3 &  23.3$^{  +5.2}_{  -3.7}$ &  3.31$^{ +0.52}_{ -0.48}$ &  3.62$^{ +0.52}_{ -0.48}$ &  11.3$^{   +5.5}_{  -4.5}$ &  4.53$^{ +0.30}_{ -0.27}$ &  0.24$^{ +0.14}_{ -0.09}$ & 410$^{  +26}_{  -30}$ &  19.4$^{    +85.7}_{  -11.8}$ & 12.7$^{+2.1}_{-2.4}$ & 10.3$^{+5.6}_{-3.9}$ & 15.503 & 0.17 &SB2\\
\hline
\end{tabular}
\end{table}
\end{landscape}

\begin{landscape}
\begin{table}
\contcaption{}
%\label{table:targets} 
\begin{tabular}{l@{\hspace{1.5mm}}l@{\hspace{-1.5mm}}r@{\hspace{1.5mm}}r@{\hspace{1.5mm}}c@{\hspace{1.5mm}}c@{\hspace{1.5mm}}c@{\hspace{1.5mm}}r@{\hspace{1.5mm}}c@{\hspace{1.5mm}}c@{\hspace{1.5mm}}r@{\hspace{1.5mm}}r@{\hspace{1.5mm}}r@{\hspace{1.5mm}}r@{\hspace{1.5mm}}c@{\hspace{3mm}}c@{\hspace{3mm}}l}
\hline\hline
%\hline
BLOeM & Spect. & $\varv_{\rm rad}$ & $\sigma (\varv_{\rm rad})$ & $T_{\rm eff}$ & $\log g$ & $\log g_{c}$ & $R_{\rm eff}$ & $\log L$ & $Y$ & $\varv_{\rm e} \sin i$ & $M_{\rm spec}$ & $M_{\rm evol}$ & $\tau$ & $m_{\rm Ks}$ & $A_{\rm V}$ & Notes\\ 
       & Type & km\,s$^{-1}$  & km\,s$^{-1}$     & kK & cm s$^{-2}$ & cm s$^{-2}$ & $R_{\odot}$ & $L_{\odot}$  & $\cdots$  & km s$^{-1}$ & $M_{\odot}$ & $M_{\odot}$ & Myr & mag & mag\\
\hline
           5-111 &B2\,III:             & 131 &   28.9 &    22.9$^{  +7.8}_{  -3.7}$ &  3.28$^{ +0.66}_{ -0.51}$ &  3.81$^{ +0.66}_{ -0.51}$ &   6.8$^{   +4.3}_{  -3.8}$ &  4.06$^{ +0.39}_{ -0.31}$ &  0.20$^{ +0.18}_{ -0.05}$ & 461$^{  +29}_{  -39}$ &  10.8$^{    +84.8}_{   -7.0}$  & 9.4$^{+2.2}_{-1.7}$ & 10.8$^{+8.2}_{-8.4}$ & 16.498 & 0.17 &SB2\\[2pt]
           5-113 &B1\,II               & 140 &    1.2 & 23.8$^{  +0.8}_{  -1.1}$ &  3.12$^{ +0.29}_{ -0.14}$ &  3.12$^{ +0.29}_{ -0.14}$ &  20.8$^{   +8.0}_{  -4.6}$ &  5.10$^{ +0.25}_{ -0.25}$ &  0.24$^{ +0.06}_{ -0.09}$ &  36$^{  +52}_{  -34}$ &  20.9$^{    +39.3}_{   -8.2}$  & 15.2$^{+3.4}_{-2.9}$ & 9.9$^{+2.5}_{-2.4}$ & 14.219 & 0.30 &\\[2pt]
           5-114 &B1.5\,II            & 129 &    5.6 &    22.6$^{  +1.9}_{  -3.4}$ &  3.31$^{ +0.33}_{ -0.33}$ &  3.37$^{ +0.33}_{ -0.33}$ &  17.6$^{   +6.4}_{  -4.1}$ &  4.86$^{ +0.21}_{ -0.25}$ &  0.20$^{ +0.10}_{ -0.05}$ & 201$^{  +24}_{  -24}$ &  26.7$^{    +55.8}_{  -13.5}$ & 13.7$^{+3.0}_{-1.6}$ & 10.7$^{+2.9}_{-1.9}$ & 14.425 & 0.39 &lpv/SB1\\
           5-115&O9.5:\,V: pe          & -17 & 13.8 & $\cdots$ & $\cdots$ & $\cdots$ & $\cdots$ & $\cdots$ & $\cdots$ & $\cdots$ & $\cdots$ & $\cdots$ & $\cdots$ & 13.604 & $\cdots$ &SB2?, Oe \\
\hline
\end{tabular}
\end{table}
\end{landscape}

\begin{landscape}
\begin{table}
\contcaption{}
%\label{table:targets} 
\begin{tabular}{l@{\hspace{1.5mm}}l@{\hspace{-1.5mm}}r@{\hspace{1.5mm}}r@{\hspace{1.5mm}}c@{\hspace{1.5mm}}c@{\hspace{1.5mm}}c@{\hspace{1.5mm}}r@{\hspace{1.5mm}}c@{\hspace{1.5mm}}c@{\hspace{1.5mm}}r@{\hspace{1.5mm}}r@{\hspace{1.5mm}}r@{\hspace{1.5mm}}r@{\hspace{1.5mm}}c@{\hspace{3mm}}c@{\hspace{3mm}}l}
\hline\hline
%\hline
BLOeM & Spect. & $\varv_{\rm rad}$ & $\sigma (\varv_{\rm rad})$ & $T_{\rm eff}$ & $\log g$ & $\log g_{c}$ & $R_{\rm eff}$ & $\log L$ & $Y$ & $\varv_{\rm e} \sin i$ & $M_{\rm spec}$ & $M_{\rm evol}$ & $\tau$ & $m_{\rm Ks}$ & $A_{\rm V}$ & Notes\\ 
       & Type & km\,s$^{-1}$  & km\,s$^{-1}$     & kK & cm s$^{-2}$ & cm s$^{-2}$ & $R_{\odot}$ & $L_{\odot}$  & $\cdots$  & km s$^{-1}$ & $M_{\odot}$ & $M_{\odot}$ & Myr & mag & mag\\
\hline
           6-038&B2.5\,IV:           & 291 & 132.0& $\cdots$ & $\cdots$ & $\cdots$ & $\cdots$ & $\cdots$ & $\cdots$ & $\cdots$ & $\cdots$ & $\cdots$ & $\cdots$ & 16.658 & $\cdots$ &SB2, neb \\[2pt]
           6-039&B2\,III:              & 120 & 46.9& $\cdots$ & $\cdots$ & $\cdots$ & $\cdots$ & $\cdots$ & $\cdots$ & $\cdots$ & $\cdots$ & $\cdots$ & $\cdots$ & 16.687 & $\cdots$ &neb \\[2pt]
           6-040 &B1.5\,III:          & 193 &    1.4 &  29.9$^{  +2.3}_{  -7.0}$ &  4.10$^{ +0.34}_{ -0.51}$ &  4.13$^{ +0.34}_{ -0.51}$ &   5.8$^{   +2.9}_{  -1.5}$ &  4.38$^{ +0.25}_{ -0.33}$ &  0.25$^{ +0.10}_{ -0.10}$ & 201$^{  +26}_{  -25}$ &  16.4$^{    +45.5}_{   -9.5}$ & 11.7$^{+2.3}_{-2.2}$ & 7.5$^{+6.0}_{-6.2}$ & 16.616 & 0.36 &neb\\[2pt]
           6-041 &B1.5\,III:          & 174 &   10.1 &    22.9$^{  +8.2}_{  -5.1}$ &  3.53$^{ +0.74}_{ -0.69}$ &  3.68$^{ +0.74}_{ -0.69}$ &   6.5$^{   +4.5}_{  -3.8}$ &  4.01$^{ +0.40}_{ -0.34}$ &  0.15$^{ +0.16}_{ -0.00}$ & 252$^{  +31}_{  -29}$ &   7.3$^{    +75.0}_{   -4.9}$   & 8.5$^{+2.2}_{-1.8}$ & 10.9$^{+9.7}_{-10.9}$ & 16.686 & 0.24 &\\
\hline
\end{tabular}
\end{table}
\end{landscape}

\begin{landscape}
\begin{table}
\contcaption{}
%\label{table:targets} 
\begin{tabular}{l@{\hspace{1.5mm}}l@{\hspace{-1.5mm}}r@{\hspace{1.5mm}}r@{\hspace{1.5mm}}c@{\hspace{1.5mm}}c@{\hspace{1.5mm}}c@{\hspace{1.5mm}}r@{\hspace{1.5mm}}c@{\hspace{1.5mm}}c@{\hspace{1.5mm}}r@{\hspace{1.5mm}}r@{\hspace{1.5mm}}r@{\hspace{1.5mm}}r@{\hspace{1.5mm}}c@{\hspace{3mm}}c@{\hspace{3mm}}l}
\hline\hline
%\hline
BLOeM & Spect. & $\varv_{\rm rad}$ & $\sigma (\varv_{\rm rad})$ & $T_{\rm eff}$ & $\log g$ & $\log g_{c}$ & $R_{\rm eff}$ & $\log L$ & $Y$ & $\varv_{\rm e} \sin i$ & $M_{\rm spec}$ & $M_{\rm evol}$ & $\tau$ & $m_{\rm Ks}$ & $A_{\rm V}$ & Notes\\ 
       & Type & km\,s$^{-1}$  & km\,s$^{-1}$     & kK & cm s$^{-2}$ & cm s$^{-2}$ & $R_{\odot}$ & $L_{\odot}$  & $\cdots$  & km s$^{-1}$ & $M_{\odot}$ & $M_{\odot}$ & Myr & mag & mag\\
\hline
           6-074 &B1.5\,III          & 183 &    8.1 &  23.8$^{  +4.9}_{  -4.6}$ &  4.07$^{ +0.38}_{ -0.81}$ &  4.11$^{ +0.38}_{ -0.81}$ &   7.1$^{   +4.1}_{  -2.8}$ &  4.16$^{ +0.33}_{ -0.33}$ &  0.15$^{ +0.13}_{ -0.00}$ & 248$^{  +31}_{  -32}$ &  24.0$^{    +83.6}_{  -15.7}$  & 9.1$^{+2.2}_{-1.5}$ & 12.4$^{+8.4}_{-9.1}$ & 16.453 & 0.18 &SB1, neb\\[2pt]
           6-075 &B0.7:\,V          & 195 &   20.1 & 31.5$^{  +3.1}_{  -5.5}$ &  4.33$^{ +0.17}_{ -0.34}$ &  4.34$^{ +0.17}_{ -0.34}$ &   7.3$^{   +3.7}_{  -2.1}$ &  4.68$^{ +0.28}_{ -0.32}$ &  0.18$^{ +0.08}_{ -0.03}$ & 202$^{  +25}_{  -25}$ &  43.4$^{    +71.5}_{  -23.1}$  & 14.5$^{+3.0}_{-2.6}$ & 6.1$^{+3.9}_{-4.0}$ & 16.045 & 0.31 &SB1\\[2pt]
           6-076 &B0\,Ib              & 192 &    2.9 &  26.8$^{  +2.3}_{  -1.2}$ &  2.90$^{ +0.29}_{ -0.17}$ &  2.92$^{ +0.29}_{ -0.17}$ &   29.5$^{   +6.7}_{  -5.7}$ &  5.61$^{ +0.17}_{ -0.16}$ &  0.30$^{ +0.05}_{ -0.13}$ &  78$^{  +98}_{  -19}$ &  26.1$^{    +35.4}_{  -10.2}$ &  32.4$^{+5.1}_{-5.3}$ & 4.4$^{+0.8}_{-0.5}$ & 13.192 & 0.42 &\\
\hline
\end{tabular}
\end{table}
\end{landscape}

\begin{landscape}
\begin{table}
\contcaption{}
%\label{table:targets} 
\begin{tabular}{l@{\hspace{1.5mm}}l@{\hspace{-1.5mm}}r@{\hspace{1.5mm}}r@{\hspace{1.5mm}}c@{\hspace{1.5mm}}c@{\hspace{1.5mm}}c@{\hspace{1.5mm}}r@{\hspace{1.5mm}}c@{\hspace{1.5mm}}c@{\hspace{1.5mm}}r@{\hspace{1.5mm}}r@{\hspace{1.5mm}}r@{\hspace{1.5mm}}r@{\hspace{1.5mm}}c@{\hspace{3mm}}c@{\hspace{3mm}}l}
\hline\hline
%\hline
BLOeM & Spect. & $\varv_{\rm rad}$ & $\sigma (\varv_{\rm rad})$ & $T_{\rm eff}$ & $\log g$ & $\log g_{c}$ & $R_{\rm eff}$ & $\log L$ & $Y$ & $\varv_{\rm e} \sin i$ & $M_{\rm spec}$ & $M_{\rm evol}$ & $\tau$ & $m_{\rm Ks}$ & $A_{\rm V}$ & Notes\\ 
       & Type & km\,s$^{-1}$  & km\,s$^{-1}$     & kK & cm s$^{-2}$ & cm s$^{-2}$ & $R_{\odot}$ & $L_{\odot}$  & $\cdots$  & km s$^{-1}$ & $M_{\odot}$ & $M_{\odot}$ & Myr & mag & mag\\
\hline
           6-111 &B0.5\,Ib            & 175 &    3.0 &    26.8$^{  +1.2}_{  -1.2}$ &  2.90$^{ +0.17}_{ -0.17}$ &  2.92$^{ +0.18}_{ -0.17}$ &  30.5$^{   +5.7}_{  -4.4}$ &  5.63$^{ +0.14}_{ -0.14}$ &  0.40$^{ +0.06}_{ -0.08}$ &  78$^{  +98}_{  -19}$ &  27.8$^{    +21.9}_{   -9.8}$ & 32.8$^{+5.5}_{-4.5}$ & 4.4$^{+0.7}_{-0.5}$ & 13.125 & 0.48 &2MASS\\[2pt]
           6-112 &B1\,III-II           & 184 &    2.9 & 23.7$^{  +7.8}_{  -2.0}$ &  3.81$^{ +0.63}_{ -0.34}$ &  3.82$^{ +0.63}_{ -0.34}$ &   7.7$^{   +4.3}_{  -4.1}$ &  4.22$^{ +0.37}_{ -0.25}$ &  0.15$^{ +0.15}_{ -0.00}$ &  55$^{  +19}_{  -19}$ &  14.2$^{    +91.9}_{   -8.9}$ & 10.5$^{+2.1}_{-2.0}$ & 11.5$^{+5.4}_{-8.9}$ & 16.414 & 0.34 &Si\,{\sc iv} $\lambda$4089 too weak\\[2pt]
           6-113&Onn pe             &  -9 &    4.4& $\cdots$ & $\cdots$ & $\cdots$ & $\cdots$ & $\cdots$ & $\cdots$ & $\cdots$ & $\cdots$ & $\cdots$ & $\cdots$ & 14.508 & $\cdots$ &Oe \\
\hline
\end{tabular}
\end{table}
\end{landscape}

% ------

\begin{landscape}
\begin{table}
\contcaption{}
%\label{table:targets} 
\begin{tabular}{l@{\hspace{1.5mm}}l@{\hspace{-1.5mm}}r@{\hspace{1.5mm}}r@{\hspace{1.5mm}}c@{\hspace{1.5mm}}c@{\hspace{1.5mm}}c@{\hspace{1.5mm}}r@{\hspace{1.5mm}}c@{\hspace{1.5mm}}c@{\hspace{1.5mm}}r@{\hspace{1.5mm}}r@{\hspace{1.5mm}}r@{\hspace{1.5mm}}r@{\hspace{1.5mm}}c@{\hspace{3mm}}c@{\hspace{3mm}}l}
\hline\hline
%\hline
BLOeM & Spect. & $\varv_{\rm rad}$ & $\sigma (\varv_{\rm rad})$ & $T_{\rm eff}$ & $\log g$ & $\log g_{c}$ & $R_{\rm eff}$ & $\log L$ & $Y$ & $\varv_{\rm e} \sin i$ & $M_{\rm spec}$ & $M_{\rm evol}$ & $\tau$ & $m_{\rm Ks}$ & $A_{\rm V}$ & Notes\\ 
       & Type & km\,s$^{-1}$  & km\,s$^{-1}$     & kK & cm s$^{-2}$ & cm s$^{-2}$ & $R_{\odot}$ & $L_{\odot}$  & $\cdots$  & km s$^{-1}$ & $M_{\odot}$ & $M_{\odot}$ & Myr & mag & mag\\
\hline
           7-033 &B0.7\,III         & 170 &   24.4 &  28.0$^{  +3.0}_{  -3.4}$ &  3.69$^{ +0.38}_{ -0.24}$ &  3.71$^{ +0.38}_{ -0.24}$ &  10.1$^{   +3.9}_{  -2.6}$ &  4.75$^{ +0.24}_{ -0.25}$ &  0.24$^{ +0.09}_{ -0.09}$ & 112$^{  +20}_{  -19}$ &  18.8$^{    +47.2}_{   -8.9}$ & 14.5$^{+2.4}_{-2.0}$ & 8.4$^{+3.0}_{-2.4}$ & 15.499 & 0.45 &SB1\\[2pt]
           7-034 &B0\,V             & 139 &   14.9 &  29.9$^{  +1.9}_{  -2.7}$ &  3.50$^{ +0.29}_{ -0.33}$ &  3.59$^{ +0.29}_{ -0.33}$ &     7.8$^{   +2.7}_{  -1.7}$ &  4.64$^{ +0.23}_{ -0.24}$ &  0.36$^{ +0.13}_{ -0.10}$ & 201$^{  +24}_{  -25}$ &   8.6$^{    +15.1}_{   -4.3}$   & 14.2$^{+2.2}_{-1.6}$ & 7.6$^{+2.6}_{-2.4}$ & 16.021 & 0.43 &SB1\\[2pt]
           7-035 &B1\,II:             & 175 &   43.0 &  25.3$^{  +3.8}_{  -3.4}$ &  3.50$^{ +0.33}_{ -0.48}$ &  3.56$^{ +0.33}_{ -0.48}$ &   12.0$^{   +5.5}_{  -3.8}$ &  4.73$^{ +0.28}_{ -0.28}$ &  0.18$^{ +0.15}_{ -0.03}$ & 201$^{  +24}_{  -24}$ &  19.2$^{    +48.0}_{  -11.2}$  & 13.0$^{+3.3}_{-1.7}$ & 9.4$^{+3.8}_{-2.9}$ & 15.124 & 0.34 &SB2\\
\hline
\end{tabular}
\end{table}
\end{landscape}

\begin{landscape}
\begin{table}
\contcaption{}
%\label{table:targets} 
\begin{tabular}{l@{\hspace{1.5mm}}l@{\hspace{-1.5mm}}r@{\hspace{1.5mm}}r@{\hspace{1.5mm}}c@{\hspace{1.5mm}}c@{\hspace{1.5mm}}c@{\hspace{1.5mm}}r@{\hspace{1.5mm}}c@{\hspace{1.5mm}}c@{\hspace{1.5mm}}r@{\hspace{1.5mm}}r@{\hspace{1.5mm}}r@{\hspace{1.5mm}}r@{\hspace{1.5mm}}c@{\hspace{3mm}}c@{\hspace{3mm}}l}
\hline\hline
%\hline
BLOeM & Spect. & $\varv_{\rm rad}$ & $\sigma (\varv_{\rm rad})$ & $T_{\rm eff}$ & $\log g$ & $\log g_{c}$ & $R_{\rm eff}$ & $\log L$ & $Y$ & $\varv_{\rm e} \sin i$ & $M_{\rm spec}$ & $M_{\rm evol}$ & $\tau$ & $m_{\rm Ks}$ & $A_{\rm V}$ & Notes\\ 
       & Type & km\,s$^{-1}$  & km\,s$^{-1}$     & kK & cm s$^{-2}$ & cm s$^{-2}$ & $R_{\odot}$ & $L_{\odot}$  & $\cdots$  & km s$^{-1}$ & $M_{\odot}$ & $M_{\odot}$ & Myr & mag & mag\\
\hline
           7-070 &B1.5\,III:            & 134 &    9.2 &  25.7$^{  +5.3}_{  -2.7}$ &  4.12$^{ +0.33}_{ -0.52}$ &  4.14$^{ +0.33}_{ -0.52}$ &   6.5$^{   +3.0}_{  -2.5}$ &  4.21$^{ +0.30}_{ -0.26}$ &  0.15$^{ +0.15}_{ -0.00}$ & 154$^{  +25}_{  -25}$ &  20.9$^{    +52.8}_{  -12.8}$ & 10.5$^{+2.1}_{-1.6}$ & 11.1$^{+4.3}_{-8.4}$ & 16.595 & 0.37 &SB1\\[2pt]
           7-071 &B0\,II:+B0         & 143 &   54.9 & 29.9$^{  +2.7}_{  -2.0}$ &  3.59$^{ +0.40}_{ -0.34}$ &  3.59$^{ +0.40}_{ -0.34}$ &  17.4$^{   +6.3}_{  -4.2}$ &  5.34$^{ +0.24}_{ -0.23}$ &  0.24$^{ +0.06}_{ -0.08}$ &  78$^{  +19}_{  -20}$ &  43.0$^{   +109.2}_{  -22.1}$ & 22.9$^{+4.6}_{-4.3}$ & 5.2$^{+1.4}_{-0.8}$ & 14.207 & 0.47 &SB2\\[2pt]
           7-072 &O8\,Vnn            &  52 &    9.8 &  31.8$^{  +2.3}_{  -2.3}$ &  3.12$^{ +0.29}_{ -0.29}$ &  3.60$^{ +0.29}_{ -0.29}$ &   9.9$^{   +4.2}_{  -2.5}$ &  4.96$^{ +0.27}_{ -0.27}$ &  0.40$^{ +0.14}_{ -0.17}$ & 413$^{  +25}_{  -29}$ &  13.6$^{    +27.6}_{   -6.7}$  & 17.5$^{+3.0}_{-2.1}$ & 6.2$^{+2.1}_{-1.8}$ & 15.321 & 0.32 &\\
\hline
\end{tabular}
\end{table}
\end{landscape}

\begin{landscape}
\begin{table}
\contcaption{}
%\label{table:targets} 
\begin{tabular}{l@{\hspace{1.5mm}}l@{\hspace{-1.5mm}}r@{\hspace{1.5mm}}r@{\hspace{1.5mm}}c@{\hspace{1.5mm}}c@{\hspace{1.5mm}}c@{\hspace{1.5mm}}r@{\hspace{1.5mm}}c@{\hspace{1.5mm}}c@{\hspace{1.5mm}}r@{\hspace{1.5mm}}r@{\hspace{1.5mm}}r@{\hspace{1.5mm}}r@{\hspace{1.5mm}}c@{\hspace{3mm}}c@{\hspace{3mm}}l}
\hline\hline
%\hline
BLOeM & Spect. & $\varv_{\rm rad}$ & $\sigma (\varv_{\rm rad})$ & $T_{\rm eff}$ & $\log g$ & $\log g_{c}$ & $R_{\rm eff}$ & $\log L$ & $Y$ & $\varv_{\rm e} \sin i$ & $M_{\rm spec}$ & $M_{\rm evol}$ & $\tau$ & $m_{\rm Ks}$ & $A_{\rm V}$ & Notes\\ 
       & Type & km\,s$^{-1}$  & km\,s$^{-1}$     & kK & cm s$^{-2}$ & cm s$^{-2}$ & $R_{\odot}$ & $L_{\odot}$  & $\cdots$  & km s$^{-1}$ & $M_{\odot}$ & $M_{\odot}$ & Myr & mag & mag\\
\hline
           7-107 &B1.5\,III-II             & 157 &    1.8 &  23.7$^{  +1.2}_{  -2.3}$ &  3.87$^{ +0.23}_{ -0.29}$ &  3.88$^{ +0.23}_{ -0.29}$ &   7.1$^{   +2.8}_{  -1.6}$ &  4.15$^{ +0.25}_{ -0.26}$ &  0.17$^{ +0.10}_{ -0.02}$ &  56$^{  +17}_{  -15}$ &  13.8$^{    +22.2}_{   -6.7}$  & 8.8$^{+1.4}_{-1.1}$ & 18.1$^{+5.6}_{-4.0}$ & 16.544 & 0.32 &Si\,{\sc iv} $\lambda$4089 too weak\\[2pt]
           7-108 &B2.5\,Ia                & 198 &    6.0 &  18.9$^{  +3.0}_{  -1.5}$ &  2.50$^{ +0.33}_{ -0.33}$ &  2.66$^{ +0.34}_{ -0.34}$ &   23.7$^{   +7.7}_{  -6.8}$ &  4.81$^{ +0.23}_{ -0.19}$ &  0.26$^{ +0.16}_{ -0.08}$ & 153$^{  +24}_{  -24}$ &   9.4$^{    +18.2}_{   -5.0}$  & 13.9$^{+2.1}_{-1.7}$ & 11.1$^{+3.1}_{-1.7}$ & 13.992 & 0.35 &SB1 \\[2pt]
           7-109 &B2\,III:                  & 176 &   10.6 &  21.3$^{  +9.4}_{  -3.1}$ &  3.30$^{ +0.74}_{ -0.51}$ &  3.59$^{ +0.74}_{ -0.52}$ &      7.3$^{   +5.0}_{  -5.0}$ &  3.99$^{ +0.43}_{ -0.27}$ &  0.15$^{ +0.18}_{ -0.00}$ & 306$^{  +33}_{  -31}$ &   7.4$^{    +75.7}_{   -5.0}$  & 8.8$^{+2.1}_{-1.5}$ & 12.4$^{+7.4}_{-12.4}$ & 16.596 & 0.33 &\\
\hline
\end{tabular}
\end{table}
\end{landscape}

\begin{landscape}
\begin{table}
\contcaption{}
%\label{table:targets} 
\begin{tabular}{l@{\hspace{1.5mm}}l@{\hspace{-1.5mm}}r@{\hspace{1.5mm}}r@{\hspace{1.5mm}}c@{\hspace{1.5mm}}c@{\hspace{1.5mm}}c@{\hspace{1.5mm}}r@{\hspace{1.5mm}}c@{\hspace{1.5mm}}c@{\hspace{1.5mm}}r@{\hspace{1.5mm}}r@{\hspace{1.5mm}}r@{\hspace{1.5mm}}r@{\hspace{1.5mm}}c@{\hspace{3mm}}c@{\hspace{3mm}}l}
\hline\hline
%\hline
BLOeM & Spect. & $\varv_{\rm rad}$ & $\sigma (\varv_{\rm rad})$ & $T_{\rm eff}$ & $\log g$ & $\log g_{c}$ & $R_{\rm eff}$ & $\log L$ & $Y$ & $\varv_{\rm e} \sin i$ & $M_{\rm spec}$ & $M_{\rm evol}$ & $\tau$ & $m_{\rm Ks}$ & $A_{\rm V}$ & Notes\\ 
       & Type & km\,s$^{-1}$  & km\,s$^{-1}$     & kK & cm s$^{-2}$ & cm s$^{-2}$ & $R_{\odot}$ & $L_{\odot}$  & $\cdots$  & km s$^{-1}$ & $M_{\odot}$ & $M_{\odot}$ & Myr & mag & mag\\
\hline
           8-027 &B1\,III:                    & 165 &    3.9 &    29.9$^{  +2.3}_{  -4.2}$ &  4.12$^{ +0.29}_{ -0.43}$ &  4.12$^{ +0.29}_{ -0.43}$ &   6 6.9$^{   +3.1}_{  -1.8}$ &  4.54$^{ +0.26}_{ -0.29}$ &  0.25$^{ +0.10}_{ -0.09}$ &  93$^{  +24}_{  -20}$ &  23.5$^{    +49.7}_{  -12.9}$ & 12.9$^{+2.3}_{-2.0}$ & 7.9$^{+4.3}_{-4.0}$ & 16.331 & 0.38 &\\
           8-028 &O9.7\,II-Ib(n)          & 142 &    6.1 &    26.5$^{  +1.5}_{  -1.5}$ &  2.88$^{ +0.19}_{ -0.29}$ &  3.09$^{ +0.19}_{ -0.29}$ &  20.4$^{   +8.0}_{  -4.7}$ &  5.27$^{ +0.25}_{ -0.25}$ &  0.34$^{ +0.21}_{ -0.09}$ & 250$^{  +30}_{  -30}$ &  18.3$^{    +25.6}_{   -8.8}$ & 19.4$^{+4.9}_{-3.2}$ & 7.0$^{+1.7}_{-1.5}$ & 13.987 & 0.36 &SB1\\[2pt]
           8-029 &B1\,IV:                    & 167 &    6.6 &  29.9$^{  +1.9}_{  -1.1}$ &  4.36$^{ +0.10}_{ -0.29}$ &  4.36$^{ +0.10}_{ -0.29}$ &   6.5$^{   +3.0}_{  -1.7}$ &  4.49$^{ +0.29}_{ -0.28}$ &  0.15$^{ +0.03}_{ -0.00}$ & 121$^{  +28}_{  -24}$ &  36.0$^{    +40.7}_{  -17.8}$ & 13.5$^{+1.7}_{-1.6}$ & 7.4$^{+1.8}_{-3.7}$ & 16.580 & 0.57 &SB1\\
\hline
\end{tabular}
\end{table}
\end{landscape}

\begin{landscape}
\begin{table}
\contcaption{}
%\label{table:targets} 
\begin{tabular}{l@{\hspace{1.5mm}}l@{\hspace{-1.5mm}}r@{\hspace{1.5mm}}r@{\hspace{1.5mm}}c@{\hspace{1.5mm}}c@{\hspace{1.5mm}}c@{\hspace{1.5mm}}r@{\hspace{1.5mm}}c@{\hspace{1.5mm}}c@{\hspace{1.5mm}}r@{\hspace{1.5mm}}r@{\hspace{1.5mm}}r@{\hspace{1.5mm}}r@{\hspace{1.5mm}}c@{\hspace{3mm}}c@{\hspace{3mm}}l}
\hline\hline
%\hline
BLOeM & Spect. & $\varv_{\rm rad}$ & $\sigma (\varv_{\rm rad})$ & $T_{\rm eff}$ & $\log g$ & $\log g_{c}$ & $R_{\rm eff}$ & $\log L$ & $Y$ & $\varv_{\rm e} \sin i$ & $M_{\rm spec}$ & $M_{\rm evol}$ & $\tau$ & $m_{\rm Ks}$ & $A_{\rm V}$ & Notes\\ 
       & Type & km\,s$^{-1}$  & km\,s$^{-1}$     & kK & cm s$^{-2}$ & cm s$^{-2}$ & $R_{\odot}$ & $L_{\odot}$  & $\cdots$  & km s$^{-1}$ & $M_{\odot}$ & $M_{\odot}$ & Myr & mag & mag\\
\hline
           8-063 &B3\,Ib                       & 147 &    1.8 &  17.9$^{  +1.1}_{  -0.8}$ &  2.69$^{ +0.29}_{ -0.14}$ &  2.70$^{ +0.29}_{ -0.14}$ &  30.2$^{   +8.2}_{  -5.9}$ &  4.92$^{ +0.19}_{ -0.19}$ &  0.18$^{ +0.06}_{ -0.03}$ &  36$^{  +57}_{  -34}$ &  16.5$^{    +24.3}_{   -6.1}$ & 16.6$^{+1.8}_{-3.1}$ & 9.8$^{+2.2}_{-1.9}$ & 13.664 & 0.33 & Post-MS\\[2pt]
           8-064 &B3\,II e                     & 171 &    9.2 &  16.8$^{  +4.9}_{  -1.9}$ &  2.88$^{ +0.92}_{ -0.70}$ &  3.15$^{ +0.92}_{ -0.70}$ &  14.2$^{   +6.4}_{  -6.6}$ &  4.16$^{ +0.31}_{ -0.20}$ &  0.25$^{ +0.19}_{ -0.10}$ & 251$^{  +30}_{  -29}$ &  10.3$^{   +117.4}_{   -6.7}$ & 9.2$^{+1.5}_{-1.1}$ & 19.9$^{+7.5}_{-6.8}$ & 15.174 & 0.38 &\\[2pt]
           8-065 &B1\,III-II                & 155 & $\cdots$ &  23.7$^{  +1.2}_{  -2.0}$ &  3.70$^{ +0.23}_{ -0.51}$ &  3.70$^{ +0.23}_{ -0.51}$ &  10.9$^{   +4.4}_{  -2.5}$ &  4.53$^{ +0.25}_{ -0.26}$ &  0.18$^{ +0.16}_{ -0.03}$ &  52$^{  +75}_{  -20}$ &  21.9$^{    +35.4}_{  -12.4}$ & 10.8$^{+1.9}_{-1.5}$ & 14.5$^{+4.1}_{-2.4}$ & 15.666 & 0.29 &Si\,{\sc iv} $\lambda$4089 too weak\\
\hline
\end{tabular}
\end{table}
\end{landscape}

\begin{landscape}
\begin{table}
\contcaption{}
%\label{table:targets} 
\begin{tabular}{l@{\hspace{1.5mm}}l@{\hspace{-1.5mm}}r@{\hspace{1.5mm}}r@{\hspace{1.5mm}}c@{\hspace{1.5mm}}c@{\hspace{1.5mm}}c@{\hspace{1.5mm}}r@{\hspace{1.5mm}}c@{\hspace{1.5mm}}c@{\hspace{1.5mm}}r@{\hspace{1.5mm}}r@{\hspace{1.5mm}}r@{\hspace{1.5mm}}r@{\hspace{1.5mm}}c@{\hspace{3mm}}c@{\hspace{3mm}}l}
\hline\hline
%\hline
BLOeM & Spect. & $\varv_{\rm rad}$ & $\sigma (\varv_{\rm rad})$ & $T_{\rm eff}$ & $\log g$ & $\log g_{c}$ & $R_{\rm eff}$ & $\log L$ & $Y$ & $\varv_{\rm e} \sin i$ & $M_{\rm spec}$ & $M_{\rm evol}$ & $\tau$ & $m_{\rm Ks}$ & $A_{\rm V}$ & Notes\\ 
       & Type & km\,s$^{-1}$  & km\,s$^{-1}$     & kK & cm s$^{-2}$ & cm s$^{-2}$ & $R_{\odot}$ & $L_{\odot}$  & $\cdots$  & km s$^{-1}$ & $M_{\odot}$ & $M_{\odot}$ & Myr & mag & mag\\
\hline
           8-109 &B0\,III:pe        & 16 & 10.5& $\cdots$ & $\cdots$ & $\cdots$ & $\cdots$ & $\cdots$ & $\cdots$ & $\cdots$ & $\cdots$ & $\cdots$ & $\cdots$ & 13.364 & $\cdots$ &Be \\[2pt]
           8-110 &B1\,III:+B1\,III:& 248 &   12.3 &  15.8$^{  +4.9}_{  -0.8}$ &  3.88$^{ +0.57}_{ -0.57}$ &  4.03$^{ +0.57}_{ -0.57}$ &   9.9$^{   +6.0}_{  -5.2}$ &  3.74$^{ +0.38}_{ -0.29}$ &  0.15$^{ +0.08}_{ -0.00}$ & 459$^{  +39}_{  -39}$ &  37.9$^{   +226.7}_{  -24.7}$ & 6.7$^{+1.4}_{-1.1}$ & 23.9$^{+15.4}_{-13.8}$ & 16.399 & 0.00 &SB2. H\,{\sc i}, He\,{\sc i} poor fits\\[2pt]
           8-111 &B2\,III              & 141 &   19.7 &  22.6$^{  +2.7}_{  -3.4}$ &  3.93$^{ +0.52}_{ -0.67}$ &  3.98$^{ +0.52}_{ -0.67}$ &   5.7$^{   +2.4}_{  -1.6}$ &  3.89$^{ +0.24}_{ -0.26}$ &  0.15$^{ +0.14}_{ -0.00}$ & 204$^{  +27}_{  -26}$ &  11.3$^{    +44.5}_{   -6.9}$  & 8.1$^{+1.2}_{-1.2}$ & 18.3$^{+10.5}_{-10.0}$ & 16.927 & 0.23 &SB1\\
\hline
\end{tabular}
\end{table}
\end{landscape}

\clearpage

\section{Pipeline versus literature results}

Tables~\ref{O-lit} and \ref{B-lit} compare pipeline-derived physical parameters of BLOeM O and B stars, respectively, with representative literature results.

\begin{table*}
\caption{Comparison of pipeline-derived physical parameters of BLOeM O stars with representative literature results. Previous analyses  involve {\sc fastwind} \citep{Puls+2005, Rivero-Gonzalez+2012} or {\sc cmfgen} \citep{Hillier-Miller1998}.
% or {\sc PoWR} \citep{Grafener+2002, Sander+2015}. 
}
% Mass-loss rates are presented as $\dot{M}/\sqrt{f_{v}}$ to reflect differences in adopted or derived wind clumping factors.
\label{O-lit}
\begin{center}
\begin{tabular}
{l@{\hspace{2mm}}l@{\hspace{2mm}}l@{\hspace{2mm}}l@{\hspace{3mm}}l@{\hspace{3mm}}l@{\hspace{3mm}}c@{\hspace{2mm}}c@{\hspace{2mm}}l}

    \hline\hline
    BLOeM    & Alias    &  Spect.  & $T_{\rm eff}$    & $\log g$     & $\log L$   & $\varv_{\rm e} \sin i$ & Fitting & Ref\\
              &          & Type    &     kK          & cm\,s$^{-2}$ & $L_{\odot}$ & km\,s$^{-1}$ & Tool \\ [2pt]
\hline
2-016   & AzV 80 & O6\,III:nn(f)p     & 38.0 & 3.70 & 5.71 & 350       & {\sc cmfgen} (He) & MBH24 \\ [2pt]
         &          &           & 35.4$^{+1.9}_{-3.1}$ & 3.30$^{+0.17}_{-0.34}$ & 5.65$^{+0.23}_{-0.24}$ & 357$^{+131}_{-30}$ & Pipeline & This work \\ [4pt] 
%
%3-042 AzV26 O6\,I(f)+O7.5 MKE06
%
7-069    & AzV 243 & O6.5\,V     & 42.6$^{+0.8}_{-0.6}$     & 3.94$^{+0.09}_{-0.07}$  &
           5.68$\pm$0.07 & 59$^{+8}_{-6}$     & {\sc fastwind} (He) & MKE06 \\ [2pt] %
          &          &           & 39.6$\pm$1.5 & 3.90$\pm$0.10 & 5.59$\pm$0.10 & 60      & {\sc cmfgen} (He) & BLM13   \\ [2pt]
          &          &           & 39.7$^{+2.2}_{-3.4}$     & 4.07$^{+0.22}_{-0.49}$  & 5.57$^{+0.26}_{-0.27}$ & 154$^{+25}_{-25}$ & Pipeline & This work \\ [4pt] 
4-057     & NGC346 ELS 46 & O6.5\,Vnn & 39.7$^{+1.7}_{-1.8}$ & 4.17$^{+0.23}_{-0.29}$ &
           4.81$\pm$0.10 & 340$^{+45}_{-27}$ & {\sc fastwind} (He) & MKE06 \\ [2pt] %
           &         &          & 39.0$\pm$1.5 & 4.15$\pm$0.10 & 4.81$\pm$0.10   & 300 
          & {\sc cmfgen} (He) & BLM13 \\ [2pt]
          &         &               &  $35.5^{+3.7}_{-1.5}$ & $3.31^{+0.33}_{-0.19}$ & $4.80^{+0.27}_{-0.26}$ & 471$^{+20}_{-30}$ & Pipeline    & This work \\ [4pt] 
4-049    & AzV 226 & O7\,IIIn((f)) & 35.9$^{+1.3}_{-1.0}$ & 3.54$^{+0.13}_{-0.08}$ & 5.20$\pm$0.09 & 313$^{+27}_{-23}$       & {\sc fastwind} (He) & MKE06 \\ [2pt]
          &          &           & 33.7$^{+1.5}_{-1.5}$     & 3.10$^{+0.16}_{-0.16}$ & 5.17$^{+0.25}_{-0.25}$ & 354$^{+137}_{-26}$ & Pipeline & This work \\ [4pt] 
2-020   & AzV 83 & O7\,Iaf$^{+}$     & 32.8 & 3.25 & 5.54 & 70:       & {\sc cmfgen} (He) & HLH03 \\ [2pt]
         &          &           & 35.7$^{+1.5}_{-3.1}$ & 3.31$^{+0.14}_{-0.29}$ & 5.61$^{+0.21}_{-0.23}$ & 77$^{+97}_{-19}$ & Pipeline & This work \\ [4pt] 
4-058     & Sk~80 & O7\,Iaf$^{+}$ & $34.1^{+0.6}_{-0.6}$ & $3.35^{+0.17}_{-0.12}$
 & $6.02^{+0.06}_{-0.06}$   & 74$^{+15}_{-9}$ & {\sc fastwind} (He) & MKE06 \\ [2pt]
          &         &               & $33.5\pm1.0$          & $3.16\pm0.10$         &
          $5.89\pm0.10$          & 75             & {\sc cmfgen} (He) & BMH21 \\ [2pt]
          &         &               &  $35.7^{+1.5}_{-1.9}$ & $3.50^{+0.14}_{-0.14}$ & $6.12^{+0.15}_{-0.16}$ & 78$^{+98}_{-19}$ & Pipeline    & This work \\ [4pt] 
 1-012    & AzV 267 & O7.5\,Vn     & 35.7$\pm$1.5 & $4.00\pm0.20$ & 4.90$\pm$0.10 & 220      & {\sc cmfgen} (He) & BLM13  \\ [2pt]
          &          &           & 33.7$^{+3.0}_{-2.7}$     & 3.69$^{+0.67}_{-0.48}$  & 4.93$^{+0.28}_{-0.28}$ & 303$^{+29}_{-28}$ & Pipeline & This work \\ [4pt] 
1-027   & AzV 296 & O7.5\,V((f))n  & 35.0 & 3.5 & 5.30 & $\cdots$       & {\sc cmfgen} (He) & MKB04 \\ [2pt]
         &          &           & 33.7$^{+1.5}_{-3.0}$ & 3.53$^{+0.27}_{-0.54}$ & 5.16$^{+0.26}_{-0.27}$ & 354$^{+134}_{-31}$ & Pipeline & This work \\ [4pt] 
2-035   & AzV 95 & O7.5\,III((f))  & 38.0$\pm$0.10 & 3.70$\pm$0.10 & 5.46$\pm$0.10 & 55       & {\sc cmfgen} (He) & BMH21 \\ [2pt]
         &          &           & 35.6$^{+1.5}_{-1.5}$ & 3.50$^{+0.14}_{-0.14}$ & 5.50$^{+0.21}_{-0.21}$ & 77$^{+97}_{-20}$ & Pipeline & This work \\ [4pt] %
7-072    & AzV 251 & O8\,Vnn     & 36.0 & 3.90 & 5.01 & 500      & {\sc cmfgen} (He) & MBH24   \\ [2pt]
          &          &           & 31.8$^{+2.3}_{-2.3}$     & 3.12$^{+0.29}_{-0.29}$  & 4.96$^{+0.27}_{-0.27}$ & 413$^{+25}_{-29}$ & Pipeline & This work \\ [4pt] 
3-078    & AzV 47 & O8\,III((f))   & 35.0$\pm$1.0 & 3.75$\pm$0.10& 5.44$\pm$0.10 & 60       & {\sc cmfgen} (He) & BMH21  \\ [2pt]
           &          &           & 35.6$^{+1.5}_{-1.5}$     & 4.36$^{+0.11}_{-0.19}$ & 5.56$^{+0.26}_{-0.26}$ & 78$^{+96}_{-20}$ & Pipeline & This work \\ [4pt] 
7-001     & NGC330 ELS 13 & O8.5\,III((f)) & $34.5^{+0.8}_{-0.9}$ & $3.40^{+0.14}_{-0.15}$
 & $5.40^{+0.07}_{-0.07}$  & 73$^{+9}_{-11}$ & {\sc fastwind} (He) & MKE06 \\ [2pt]
        &                 &         & $33.7^{+1.1}_{-1.5}$ & $3.50^{+0.14}_{-0.14}$ & $5.35^{+0.24}_{-0.24}$ &  54$^{+76}_{-16}$ & Pipeline    & This work \\ [4pt] 
4-074     & NGC346 ELS 31 & O9\,V & $39.5^{+1.4}_{-1.2}$ & $3.99^{+0.18}_{-0.24}$
 & $4.99^{+0.08}_{-0.08}$  & 18$^{+10}_{-9}$ & {\sc fastwind} (He) & MKE06 \\ [2pt]
        &                 &         & 37.2$\pm$1.5         & $4.00 \pm 0.10$        &
        $4.95\pm0.10$               & 25          & {\sc cmfgen} (He)   & BLM13 \\ [2pt] % new
        &                 &         & $35.5^{+1.5}_{-1.5}$ & $4.07^{+0.16}_{-0.22}$ & $4.96^{+0.27}_{-0.27}$ &  0$^{+25}_{-0}$ & Pipeline    & This work \\ [4pt] 
4-073     & NGC346 ELS 25 & O9.2\,V & $36.2^{+1.2}_{-0.8}$ & $4.07^{+0.24}_{-0.08}$
 & $4.90^{+0.08}_{-0.08}$  & 138$^{+17}_{-14}$ & {\sc fastwind} (He) & MKE06 \\ [2pt]
        &                 &         & $35.5^{+1.9}_{-3.4}$ & $4.50^{+0}_{-0.70}$ & $45.02^{+0.07}_{-0.06}$ &  202$^{+26}_{-26}$ & Pipeline    & This work \\ [4pt] 
4-026    & NGC346 ELS 18 & O9.5\,IIIpe  & 32.7$^{+1.1}_{-1.3}$  & 3.33$^{+0.15}_{-0.14}$ & 5.10$\pm$0.09 & 138$^{+38}_{-30}$ & {\sc fastwind} (He) & MKE06  \\ [2pt]
           &          &           & 29.9$^{+3.1}_{-1.2}$    & 3.21$^{+1.05}_{-0.24}$  & 5.14$^{+0.20}_{-0.18}$ & 353$^{+42}_{-43}$ & Pipeline & This work \\ [4pt] 
           2-007   & AzV 70 & O9.5\,II-I     & 28.5 & 3.1 & 5.68 & 100       & {\sc cmfgen} (He) & ECF04 \\ [2pt]
         &          &           & 29.9$^{+1.5}_{-1.1}$ & 3.31$^{+0.14}_{-0.14}$ & 5.90$^{+0.15}_{-0.15}$ & 113$^{+20}_{-19}$ & Pipeline & This work \\ [4pt] 
1-066   & AzV 327 & O9.7\,II-Ib  & 30.8                 & 3.2                   &
         5.60                    & 150              & {\sc fastwind} (He) & MZM09 \\ [2pt]
         &          &           & 30.0$\pm$1.0          & 3.12$\pm$0.10         & 5.54$\pm$0.10           & 95               & {\sc cmfgen} (He) & BMH21 \\ [2pt]
         &          &           & 29.9$^{+1.1}_{-1.1}$ & 3.31$^{+0.14}_{-0.14}$ & 5.47$^{+0.25}_{-0.25}$ & 55$^{+77}_{-14}$ & Pipeline & This work \\ [4pt] 
\hline
\end{tabular}
\end{center}
%\tablebib{  
{\footnotesize
BLM13 \citet{Bouret+2013};
BMH21 \citet{Bouret+2021}; 
ECF04 \citet{Evans+2004-cmfgen};
HLH03 \citet{Hillier+2003}; \\ 
%{\bf HLD08} \citet{Hunter+2008b};
MBK04 \citet{Massey+2004}; 
MKE06 \citet{Mokiem+2006-SMC}; 
% {\bf MKE07} \citet{Mokiem+2007-LMC}; 
MZM09 \citet{Massey+2009}; 
MBH24 \citet{Martins+2024}  
% {\bf RHH18} \citet{Ramachandran+2018b}; 
%{\bf RPM12} \citet{Rivero-Gonzalez+2012}; 
%{\bf RSK17} \citet{Ramirez-Agudelo+2017}; 
%{\bf TLP04} \citet{Trundle+2004}; 
%{\bf TL05} \citet{Trundle+2005};
% {\bf UKG17} \citet{Urbaneja+2017}.
}
\end{table*}

\begin{table*}
\caption{Comparison of pipeline-derived physical parameters of BLOeM B-type stars with representative literature results. Previous analyses  involve {\sc fastwind} \citep{Puls+2005, Rivero-Gonzalez+2012}, {\sc cmfgen} \citep{Hillier-Miller1998} or {\sc tlusty} \citep{HubenyLanz1995}.
% or {\sc PoWR} \citep{Grafener+2002, Sander+2015}. 
}
% Mass-loss rates are presented as $\dot{M}/\sqrt{f_{v}}$ to reflect differences in adopted or derived wind clumping factors.
\label{B-lit}
\begin{center}
\begin{tabular}
{l@{\hspace{2mm}}l@{\hspace{2mm}}l@{\hspace{2mm}}l@{\hspace{3mm}}l@{\hspace{3mm}}l@{\hspace{3mm}}c@{\hspace{2mm}}c@{\hspace{2mm}}l}

    \hline\hline
    BLOeM    & Alias    &  Spect.  & $T_{\rm eff}$    & $\log g$     & $\log L$   & $\varv_{\rm e} \sin i$ & Fitting & Ref\\
              &          & Type    &     kK          & cm\,s$^{-2}$ & $L_{\odot}$ & km\,s$^{-1}$ & Tool \\ [2pt]
\hline
 4-013   &  NGC346 ELS 43& B0\,V & 33.0$\pm$1.0 & 4.25$\pm$0.20
 & 4.71  & 10$\pm$5 & {\sc tlusty} (Si) & HDS07 \\ [2pt]
         &               &       & 31.8$^{+3.0}_{-2.7}$ & $4.12^{+0.33}_{-0.52}$ & $4.77^{+0.26}_{-0.26}$ & 36$^{+18}_{-21}$ & Pipeline & This work \\ [4pt] 
%          &               &         & $33.4^{+1.2}_{-0.9}$ & $4.10^{+0.13}_{-0.12}$ & $4.82^{+0.07}_{-0.06}$ & 68$^{+24}_{-27}$ & Pipeline    & This work \\ [4pt] 
          %
 4-014   &  NGC346 ELS 26 & B0\,III & $31.0\pm1.5$ & $3.65\pm0.10$               &
           $4.93\pm0.10$         & 60      & {\sc cmfgen} (He) & BLM13 \\ [2pt]
        &                 &        & $32.6^{+0.4}_{-1.2}$ & $3.76^{+0.05}_{-0.17}$
 & $4.93^{+0.09}_{-0.09}$  & 67$^{+9}_{-5}$ & {\sc fastwind} (He) & MKE06 \\ [2pt]
          &               &       & 31.9$^{+1.2}_{-2.7}$ & $3.70^{+0.17}_{-0.34}$ & $4.95^{+0.21}_{-0.22}$ & 75$^{+20}_{-23}$ & Pipeline & This work \\ [4pt] 
%          &               &         & $31.6^{+0.9}_{-0.8}$ & $3.76^{+0.15}_{-0.13}$ & $4.96^{+0.06}_{-0.06}$ & 78$^{+20}_{-21}$ & Pipeline    & This work \\ [4pt] 
%
2-110   & AzV 148 & B0\,II & 31.0 & 3.60 & 5.16 & 35       & {\sc cmfgen} (He) & MBH24 \\ [2pt]
         &               &       & 29.9$^{+1.5}_{-1.1}$ & $3.50^{+0.14}_{-0.14}$ & $5.12^{+0.25}_{-0.25}$ & 0$^{+19}_{-0}$ & Pipeline & This work \\ [4pt] 
%         &          &           & 31.5$^{+0.9}_{-0.8}$ & 3.70$^{+0.10}_{-0.09}$ & 5.08$^{+0.06}_{-0.06}$ & 0$^{+27}_{-0}$ & Pipeline & This work \\ [4pt] %
%
6-080    & AzV 488 & B0\,Ia  & 27.5  & 2.9 & 5.74 & 80       & {\sc cmfgen} (Si) & ECF04  \\ [2pt]
%         &               &       & 25.2$^{+1.2}_{-1.2}$ & $2.69^{+0.14}_{-0.14}$ & $5.79^{+0.09}_{-0.09}$ & 56$^{+75}_{-14}$ & Pipeline &BCT24\\ [2pt] 
           &          &           & 26.8$^{+1.1}_{-1.5}$    & 2.88$^{+0.19}_{-0.14}$  & 5.98$^{+0.14}_{-0.15}$ & 78$^{+98}_{-19}$ & Pipeline & This work \\ [4pt] 
7-064   & AzV 235 & B0\,Ia     & 27.5 & 2.9 & 5.72 & 80       & {\sc cmfgen} (He) & ECF04 \\ [2pt]
%         &               &       & 28.3$^{+1.2}_{-1.6}$ & ($3.50^{+0.24}_{-0.14}$) & $6.01^{+0.09}_{-0.11}$ & 31$^{+14}_{-20}$ & Pipeline & BCT24 \\ [2pt] 
         &          &           & 26.8$^{+1.1}_{-1.5}$ & 2.88$^{+0.19}_{-0.14}$ & 5.93$^{+0.15}_{-0.16}$ & 78$^{+97}_{-19}$ & Pipeline & This work \\ [4pt] 
5-105    & AzV 420 & B0.7\,II  & 27.0$\pm$1.5  & 3.05$\pm$0.15 & 5.35 & 80       & {\sc fastwind} (Si) & TL05  \\ [2pt]
           &          &           & 26.8$^{+1.9}_{-1.5}$    & 3.12$^{+0.24}_{-0.19}$  & 5.41$^{+0.26}_{-0.25}$ & 55$^{+77}_{-13}$ & Pipeline & This work \\ [4pt] 
4-015    & AzV 202 & B1\,II-Ib  & 26.3$^{+0.8}_{-0.5}$  & 3.35$^{+0.10}_{-0.05}$ & 4.80$\pm$0.08 & 29$\pm$4       & {\sc fastwind} (Si) & MKE06  \\ [2pt]
           &          &           & 23.8$^{+1.1}_{-1.1}$    & 3.12$^{+0.14}_{-0.19}$  & 4.83$^{+0.24}_{-0.24}$ & 54$^{+76}_{-15}$ & Pipeline & This work \\ [4pt] 
4-020    & AzV 210 & B1\,Ib-Iab & 20.5$\pm$1.5 & 2.40$\pm$0.15 & 5.41 & 65       & {\sc fastwind} (Si) & TLP04 \\ [2pt]
%         &               &       & 23.7$^{+1.2}_{-2.0}$ & $2.69^{+0.14}_{-0.29}$ & $5.56^{+0.10}_{-0.14}$ & 55$^{+76}_{-13}$ & Pipeline & BCT24 \\ [2pt] 
         &          &           & 23.7$^{+1.2}_{-3.1}$     & 2.90$^{+0.17}_{-0.34}$ & 5.65$^{+0.15}_{-0.20}$ & 55$^{+77}_{-13}$ & Pipeline & This work \\ [4pt] 
8-008    & AzV 96 & B1\,Iab   & 22.0$\pm$1.5  & 2.55$\pm$0.15 & 5.39 & 90       & {\sc fastwind} (Si) & TL05  \\ [2pt]
%         &               &       & 21.3$^{+1.6}_{-1.2}$ & $2.31^{+0.29}_{-0.19}$ & $5.23^{+0.13}_{-0.11}$ & 78$^{+98}_{-19}$ & Pipeline & BCT24 \\ [2pt] 
           &          &           & 23.7$^{+1.2}_{-2.7}$    & 2.73$^{+0.23}_{-0.29}$  & 5.54$^{+0.14}_{-0.18}$ & 78$^{+98}_{-19}$ & Pipeline & This work \\ [4pt] 
%4-045     & AzV 224 & B1\,Iab  & \\ [2pt]
%         &               &       & 23.7$^{+1.2}_{-2.0}$ & $3.12^{+0.14}_{-0.33}$ & $4.83^{+0.10}_{-0.14}$ & 113$^{+20}_{-19}$ & Pipeline & BCT24 \\ [4pt] 
%
4-078    & AzV 242 & B1\,Ia     & 25.0$\pm$1.5 & 2.85$\pm$0.15 & 5.67 & 90       & {\sc fastwind} (Si) & TL05   \\ [2pt]
%         &               &       & 22.5$^{+0.8}_{-0.8}$ & $2.31^{+0.14}_{-0.14}$ & $5.28^{+0.08}_{-0.08}$ & 112$^{+20}_{-19}$ & Pipeline & BCT24 \\ [2pt] 
          &          &           & 23.8$^{+1.1}_{-1.1}$     & 2.69$^{+0.14}_{-0.14}$  & 5.79$^{+0.15}_{-0.15}$ & 78$^{+98}_{-19}$ & Pipeline & This work \\ [4pt] 
         1-009    & AzV 264 & B1\,Ia     & 22.5$\pm$1.5 & 2.55$\pm$0.15 & 5.44 & 85       & {\sc fastwind} (Si) & TL05   \\ [2pt]
%                  &               &       & 21.3$^{+0.8}_{-1.2}$ & 2.31$^{+0.14}_{-0.14}$ & $5.30^{+0.08}_{-0.11}$ & 78$^{+98}_{-19}$ & Pipeline & BCT24 \\ [2pt] 
          &          &           & 22.3$^{+2.3}_{-1.9}$     & 2.50$^{+0.33}_{-0.14}$  & 5.55$^{+0.18}_{-0.17}$ & 78$^{+98}_{-19}$ & Pipeline & This work \\ [4pt] 
%4-066    & AzV 234 & B2.5\,Ib \\ [2pt]
%
2-113    & AzV 151 & B2.5\,Ia   & 16.0$\pm$1.5  & 2.10$\pm$0.15 & 5.28 & 62       & {\sc fastwind} (Si) & TL05  \\ [2pt]
           &          &           & 17.7$^{+1.9}_{-1.5}$     & 2.31$^{+0.29}_{-0.14}$ & 5.45$^{+0.17}_{-0.16}$ & 55$^{+77}_{-13}$ & Pipeline & This work \\ [4pt] 

1-111   & AzV 362 & B3\,Ia     & 14.0$\pm$1.5 & 1.70$\pm$0.15 & 5.50 & 51       & {\sc fastwind} (Si) & TLP04 \\ [2pt]
%         &               &       & 15.9$^{+0.8}_{-0.8}$ & $1.64^{+0.19}_{-0.10}$ & $4.86^{+0.10}_{-0.10}$ & 5$^{+8}_{-5}$ & Pipeline & BCT24 \\ [2pt] 
         &          &           & 14.9$^{+1.5}_{-0.4}$ & 1.64$^{+0.38}_{-0.10}$ & 5.62$^{+0.17}_{-0.14}$ & 55$^{+77}_{-13}$ & Pipeline & This work \\ [4pt] 
      \hline
\end{tabular}
\end{center}
%\tablebib{  
{\footnotesize
 BLM13 \citet{Bouret+2013};
ECF04 \citet{Evans+2004-cmfgen}; 
% {\bf HLH03} \citet{Hillier+2003}; 
HDS07 \citet{Hunter+2007};
% {\bf MBK04} \citet{Massey+2004}; 
MKE06 \citet{Mokiem+2006-SMC}; \\ 
% {\bf MKE07} \citet{Mokiem+2007-LMC}; 
% {\bf MZM09} \citet{Massey+2009}; 
MBH24 \citet{Martins+2024};  
% {\bf RHH18} \citet{Ramachandran+2018b}; 
%{\bf RPM12} \citet{Rivero-Gonzalez+2012}; 
%{\bf RSK17} \citet{Ramirez-Agudelo+2017}; 
TLP04 \citet{Trundle+2004}; 
TL05 \citet{Trundle+2005};
% {\bf UKG17} \citet{Urbaneja+2017}.
}
\end{table*}

\clearpage

\section{Pipeline versus RIOTS4 results}

%\onecolumn

Table~\ref{RIOTS4} presents temperatures and luminosities of BLOeM targets (this work) in common  with RIOTS4 \citep{Castro+2018}.

\begin{table*}
\caption{Comparison of pipeline-derived physical parameters for BLOeM (this work) targets in common with
the RIOTS4 study of \citet{Castro+2018}, sorted by spectral type. [M2002] catalogue numbers \citep{Massey2002}
used in the RIOTS4 survey are included.
% or {\sc PoWR} \citep{Grafener+2002, Sander+2015}. 
}
% Mass-loss rates are presented as $\dot{M}/\sqrt{f_{v}}$ to reflect differences in adopted or derived wind clumping factors.
\label{RIOTS4}
\begin{center}
\begin{tabular}
{l@{\hspace{2mm}}r@{\hspace{2mm}}l@{\hspace{2mm}}l@{\hspace{3mm}}l@{\hspace{2mm}}c@{\hspace{3mm}}l@{\hspace{4mm}}l@{\hspace{1mm}}c}

    \hline\hline
    BLOeM    & M2002    &  Spect.  & \multicolumn{2}{c}{$\log T_{\rm eff}$/K}  & $\Delta \log T_{\rm eff}$ & \multicolumn{2}{c}{$\log L/L_{\odot}$} & $\Delta \log L/L_{\odot}$\\
    &     & Type    &     RIOTS4         & BLOeM & & RIOTS4 & BLOeM\\ [2pt]
\hline
6-105 & 77368 & O6\,V:n       & 4.57$\pm$0.03 & 4.58$^{+0.04}_{-0.03}$ & +0.01 & 5.31$\pm$0.18 & 5.45$^{+0.27}_{-0.25}$ & +0.14\\ [2pt]
4-049 & 46035 & O7\,IIIn((f)) & 4.54$\pm$0.02 & 4.53$\pm$0.02          & --0.01 & 5.04$\pm$0.21 & 5.17$\pm$0.25          & +0.13 \\ [2pt]
3-014 &  7782 & O8\,Vn        & 4.53$\pm$0.02 & 4.53$^{+0.03}_{-0.02}$ & --0.00 & 5.08$\pm$0.31 & 5.14$^{+0.25}_{-0.23}$ & +0.06 \\ [2pt]
8-020 & 21877 & O8\,V         & 4.32$\pm$0.02 & 4.55$^{+0.04}_{-0.02}$ & +0.23 & 5.28$\pm$0.30 & 4.89$^{+0.27}_{-0.26}$ &--0.39 \\ [2pt]
2-005 & 15742 & O8.5\,II:(n)  & 4.48$\pm$0.02 & 4.48$\pm$0.02          & --0.00 & 5.27$\pm$0.21 & 5.34$\pm$0.22          & +0.07 \\ [2pt]
4-074 & 47478 & O9\,V         & 4.57$\pm$0.03 & 4.55$\pm$0.02          & --0.02 & 4.71$\pm$0.18 & 4.96$\pm$0.27          & +0.25 \\ [2pt]
2-008 & 16230 & O9\,II:       & 4.48$\pm$0.01 & 4.48$\pm$0.02          & --0.00 & 5.40$\pm$0.32 & 5.47$\pm$0.25          & +0.07 \\ [2pt]
6-025 & 75210 & O9.2\,V       & 4.54$\pm$0.02 & 4.55$\pm$0.02          & +0.01 & 5.08$\pm$0.19 & 5.16$\pm$0.18          & +0.08 \\ [2pt]
5-044 & 62416 & O9.5\,IV      & 4.49$\pm$0.02 & 4.53$\pm$0.02          & +0.04 & 4.96$\pm$0.32 & 4.95$\pm$0.26          &--0.01 \\ [2pt]
6-067 & 76371 & O9.7\,III     & 4.49$\pm$0.01 & 4.50$\pm$0.02          & +0.01 & 5.11$\pm$0.19 & 5.16$^{+0.22}_{-0.23}$ & +0.05 \\ [2pt]
6-005 & 73913 & O9.7\,II-Ib(n)& 4.41$\pm$0.02 & 4.45$\pm$0.03          & +0.04 & 5.17$\pm$0.19 & 5.31$^{+0.17}_{-0.18}$ & +0.14 \\ [2pt]
1-002 & 49825 & B0\,IV:       & 4.49$\pm$0.01 & 4.52$^{+0.02}_{-0.04}$ & +0.03 & 4.78$\pm$0.31 & 4.86$^{+0.27}_{-0.28}$ & +0.08 \\ [2pt]
6-035 & 75626 & B0\,IV        & 4.51$\pm$0.01 & 4.50$^{+0.03}_{-0.02}$ &--0.01 & 4.66$\pm$0.24 & 4.79$^{+0.27}_{-0.26}$ & +0.13 \\ [2pt]
7-071 & 48601 & B0\,II:+B0    & 4.45$\pm$0.01 & 4.48$\pm$0.03          & +0.03 & 5.15$\pm$0.25 & 5.34$^{+0.24}_{-0.23}$ & +0.19 \\ [2pt]
8-045 & 24096 & B0.2\,IV      & 4.48$\pm$0.02 & 4.48$^{+0.02}_{-0.04}$ &--0.00 & 5.15$\pm$0.31 & 4.82$^{+0.26}_{-0.27}$ &--0.33 \\ [2pt]
6-056 & 76253 & B0.5\,III     & 4.48$\pm$0.02 & 4.48$^{+0.02}_{-0.05}$ &--0.00 & 4.55$\pm$0.19 & 4.64$^{+0.27}_{-0.28}$ & +0.09 \\ [2pt]
3-028 &  8609 & B0.5\,II      & 4.45$\pm$0.01 & 4.48$^{+0.01}_{-0.05}$ & +0.03 & 5.06$\pm$0.31 & 5.20$^{+0.18}_{-0.20}$ & +0.14 \\ [2pt]
8-022 & 22178 & B0.5\,II      & 4.18$\pm$0.03 & 4.43$^{+0.02}_{-0.01}$ & +0.25 & 4.47$\pm$0.19 & 5.28$\pm$0.23          & +0.81 \\ [2pt]
6-111 & 77609 & B0.5\,Ib      & 4.38$\pm$0.01 & 4.43$\pm$0.02          & +0.05 & 5.40$\pm$0.30 & 5.63$\pm$0.14          & +0.23 \\ [2pt]
1-069 & 55952 & B0.7\,III     & 4.45$\pm$0.01 & 4.43$\pm$0.06          &--0.02 & 4.65$\pm$0.24 & 4.78$\pm$0.29          & +0.13 \\ [2pt]
2-047 & 20939 & B1\,Ib        & 4.30$\pm$0.08 & 4.38$\pm$0.04          & +0.08 & 4.54$\pm$0.25 & 4.83$\pm$0.21          & +0.29 \\ [2pt]
8-008 & 19728 & B1\,Iab       & 4.32$\pm$0.02 & 4.37$^{+0.03}_{-0.05}$ & +0.05 & 5.28$\pm$0.28 & 5.50$^{+0.14}_{-0.18}$ & +0.22 \\ [2pt]
% 8-043 23859
4-090 & 49450 & B1\,II        & 4.40$\pm$0.03 & 4.37$^{+0.12}_{-0.04}$ &--0.03 & 4.74$\pm$0.21 & 4.72$^{+0.36}_{-0.27}$ &--0.02 \\ [2pt]
7-051 & 46241 & B1\,II:e      & 4.18$\pm$0.09 & 4.36$^{+0.06}_{-0.12}$ & +0.18 & 4.14$\pm$0.31 & 4.51$^{+0.22}_{-0.31}$ & +0.37 \\ [2pt]
5-062 & 62981 & B1.5+early B+ & 4.46$\pm$0.02 & 4.48$^{+0.03}_{-0.13}$ & +0.02 & 4.76$\pm$0.24 & 4.78$^{+0.27}_{-0.35}$ & +0.02 \\ [2pt]
 \hline
\end{tabular}
\end{center}
\end{table*}

\clearpage

\section{Pipeline results from BLOeM versus XShootU}

Table~\ref{OB-BLOeM-XShootU} presents physical parameters of OB stars from BLOeM (this study) and XShootU datasets \citep{Bestenlehner+2025}.

%\begin{landscape}

\begin{table*}
\caption{Comparison of pipeline-derived physical parameters of OB stars from BLOeM (FLAMES/LR02, this work) and XShootU (XShooter) datasets \citep{Bestenlehner+2025}. Physical quantities shown in parentheses are not considered reliable.
% or {\sc PoWR} \citep{Grafener+2002, Sander+2015}. 
}
% Mass-loss rates are presented as $\dot{M}/\sqrt{f_{v}}$ to reflect differences in adopted or derived wind clumping factors.
\label{OB-BLOeM-XShootU}
\begin{center}
\begin{tabular}
{
l@{\hspace{2mm}}l@{\hspace{2mm}}
l@{\hspace{3mm}}l@{\hspace{3mm}}c@{\hspace{2mm}} % Teff
l@{\hspace{3mm}}l@{\hspace{3mm}}c@{\hspace{2mm}} % logg
l@{\hspace{3mm}}l@{\hspace{3mm}}c@{\hspace{2mm}} % logL
l@{\hspace{3mm}}l@{\hspace{3mm}}c} % vsini

    \hline\hline
    BLOeM    % & Alias    
    &  Spectral  & 
    \multicolumn{2}{c}{$T_{\rm eff}$/kK} & $\Delta T_{\rm eff}$ &
    \multicolumn{2}{c}{$\log g$/cm\,s$^{-2}$} & $\Delta \log g$ & 
    \multicolumn{2}{c}{$\log L/L_{\odot}$} & $\Delta \log L$ & 
    \multicolumn{2}{c}{$\varv_{\rm e} \sin i$} 
    & $\Delta \varv_{\rm e} \sin i$ \\
              % &          
              & Type    &     
            XShootU   & BLOeM &  kK & 
            XShootU   & BLOeM &  cm\,s$^{-2}$&
            XShootU   & BLOeM &  $L_{\odot}$&
            XShootU   & BLOeM & km\,s$^{-1}$ \\ [2pt]
\hline
2-016   %& AzV 80 
& O6\,III:nn(f)p               & 
35.4$^{+4.7}_{-1.6}$   & 35.4$^{+1.9}_{-3.1}$  & +0.0 & 
$3.31^{+0.52}_{-0.14}$ & 3.30$^{+0.17}_{-0.34}$ & --0.01 & 
$5.56^{+0.21}_{-0.09}$ & 5.65$^{+0.23}_{-0.24}$ & +0.09 & 
250$^{+30}_{-30}$      & 357$^{+131}_{-30}$     & +107 \\ [4pt] 
%
%3-042 AzV26 O6\,I(f)+O7.5 MKE06
%
7-069    %& AzV 243 
& O6.5\,V     & 
39.9$^{+1.8}_{-3.6}$   & 39.7$^{+2.2}_{-3.4}$   & --0.2 &
$3.69^{+0.19}_{-0.33}$ & 4.07$^{+0.22}_{-0.49}$ & +0.38 &
$5.49^{+0.09}_{-0.16}$ & 5.57$^{+0.26}_{-0.27}$  & +0.08 &
113$^{+20}_{-19}$ & 75$^{+20}_{-22}$ & -38 \\ [4pt] 
4-057     %& Cl* NGC 346 ELS 46 
& O6.5\,Vnn & 
40.1$^{+3.9}_{-3.1}$    & $35.5^{+3.7}_{-1.5}$   & --4.6 &
$4.31^{+0.19}_{-0.52}$  & $3.31^{+0.24}_{-0.23}$ & --1.00 &
$4.94^{+0.16}_{-0.14}$  & $4.80^{+0.27}_{-0.26}$ & --0.14 &
250$^{+30}_{-30}$       & 471$^{+20}_{-30}$      & +221  \\  [4pt] 
2-020   %& AzV 83 
& O7\,Iaf$^{+}$     & 
37.7$^{+1.6}_{-2.0}$     & 35.7$^{+1.5}_{-3.1}$   & --2.0    &
($4.07^{+0.38}_{-0.29}$) & 3.31$^{+0.14}_{-0.29}$ & (--0.76) & 
$5.68^{+0.09}_{-0.10}$   & 5.61$^{+0.21}_{-0.23}$ & --0.07   & 
0$^{+19}_{0}$            & 77$^{+97}_{-19}$       & +77 \\ [4pt] 
4-058     %& AzV 232 
& O7\,Iaf$^{+}$ &  
37.7$^{+1.6}_{-2.0}$    & $35.7^{+1.5}_{-1.9}$   & --2.0  &
$3.69^{+0.24}_{-0.14}$  & $3.50^{+0.14}_{-0.14}$ & --0.19 &
$6.17^{+0.09}_{-0.10}$  & $6.12^{+0.15}_{-0.16}$ & --0.05 &
54$^{+74}_{-18}$        &  78$^{+98}_{-18}$      & +24    \\ [4pt] 
 1-012    %& AzV 267 
 & O7.5\,Vn    & 
 35.4$^{+3.1}_{-1.6}$   & 33.7$^{+3.0}_{-2.7}$   & --1.7 &
 $3.69^{+0.52}_{-0.19}$ & 3.69$^{+0.67}_{-0.48}$ & +0.00 &
 $4.89^{+0.15}_{-0.09}$ & 4.93$^{+0.28}_{-0.28}$ & +0.04 &
 250$^{+30}_{-30}$      &  303$^{+28}_{-28}$    & +53  \\ [4pt] 
1-027   %& AzV 296 
& O7.5\,V((f))n   & 
33.4$^{+5.1}_{-3.3}$    & 33.7$^{+1.5}_{-3.0}$   & +0.2 &
$3.31^{+0.81}_{-0.35}$  & 3.53$^{+0.27}_{-0.54}$ & +0.22 &
$5.06^{+0.23}_{-0.16}$  & 5.16$^{+0.26}_{-0.27}$ & +0.10 &
251$^{+30}_{-29}$       & 354$^{+134}_{-31}$      & +103 \\ [4pt] 
2-035   %& AzV 95 
& O7.5\,III((f))  & 
37.7$^{+3.1}_{-2.0}$    & 35.6$^{+1.5}_{-1.5}$  & --2.1 & 
$3.69^{+0.29}_{-0.14}$  & 3.50$^{+0.14}_{-0.14}$ & +0.19 &
$5.56^{+0.14}_{-0.10}$  & 5.50$^{+0.21}_{-0.21}$ & --0.06 & 
53$^{+77}_{-17}$        &       77$^{+97}_{-19}$ & +24 \\ [4pt] %
3-078    %& AzV 47 
& O8\,III((f))    & 
35.4$^{+1.6}_{-1.2}$   & 35.6$^{+1.5}_{-1.5}$   & +0.2 &
$3.69^{+0.14}_{-0.14}$ & 4.36$^{+0.10}_{-0.19}$ & +0.67 &
$5.46^{+0.09}_{-0.08}$ & 5.56$^{+0.26}_{-0.26}$ & +0.10 &
113$^{+20}_{-19}$      &  78$^{+96}_{-20}$     & --35\\ [4pt] 
5-097    %& 2dFS 2266 
& O8\,II(f)   & 
35.4$^{+2.0}_{-1.6}$  & 37.5$^{+1.9}_{-1.5}$   & +2.1 &
$3.69^{+0.14}_{-0.14}$ & 4.31$^{+0.14}_{-0.24}$ & +0.62 &
$4.77^{+0.11}_{-0.09}$ & 4.94$^{+0.27}_{-0.27}$ & +0.17 &
33$^{+14}_{-20}$       &  22$^{+10}_{-22}$      & --11 \\ [4pt]
7-001     %& Cl* NGC 330 ELS 13 
& O8.5\,III((f)) & 
33.4$^{+1.6}_{-1.6}$   & $33.7^{+1.1}_{-1.5}$   & +0.3 &
$3.31^{+0.14}_{-0.14}$ & $3.50^{+0.14}_{-0.14}$ & +0.19 &
$5.27^{+0.09}_{-0.09}$ & $5.35^{+0.24}_{-0.24}$ & +0.08 &
78$^{+98}_{-19}$       &  78$^{+19}_{-20}$      & 0 \\ [4pt] 
4-074     %& Cl* NGC 346 ELS 31 
& O9\,V  & 
35.4$^{+3.5}_{-1.6}$ & $35.5^{+1.5}_{-1.5}$ & +0.1 & 
$3.69^{+0.33}_{-0.14}$ & $4.07^{+0.16}_{-0.22}$ & +0.38 & 
$4.86^{+0.16}_{-0.09}$ & $4.96^{+0.27}_{-0.27}$ & +0.10 &
0$^{+18}_{0}$ &  0$^{+25}_{0}$ & +0 \\ [4pt] 
4-073     %& Cl* NGC 346 ELS 25 
& O9.2\,V & 
35.4$^{+1.6}_{-2.3}$ & $35.5^{+1.9}_{-3.4}$  & +0.1 &
$4.12^{+0.14}_{-0.48}$ & $4.50^{+0}_{-0.70}$ & +0.38 &
$4.92^{+0.09}_{-0.12}$ & $5.02^{+0.27}_{-0.28}$ & +0.10 &
153$^{+24}_{-24}$ & 153$^{+24}_{-24}$ & +0 \\ [4pt] 
2-007   %& AzV 70 
& O9.5\,II-I     & 
28.3$^{+1.2}_{-1.2}$   & 29.9$^{+1.5}_{-1.1}$   & +1.6 &
$2.88^{+0.19}_{-0.14}$ & 3.31$^{+0.14}_{-0.14}$ & +0.43 &
$5.66^{+0.09}_{-0.09}$ & 5.90$^{+0.15}_{-0.15}$ & +0.24 & 
79$^{+93}_{-20}$       & 113$^{+20}_{-19}$      & +34 \\ [4pt] 
1-056    %& AzV 321 
& O9.5\,Ibn   & 
29.9$^{+3.1}_{-2.0}$   & 28.4$^{+1.1}_{-1.5}$ & --1.5 &
$3.12^{+0.33}_{-0.29}$ & 2.88$^{+0.19}_{-0.14}$ & --0.24 &
$5.14^{+0.17}_{-0.12}$ & 5.16$^{+0.26}_{-0.26}$ & +0.02  &
249$^{+30}_{-30}$      & 301$^{+28}_{-27}$ & +52 \\ [4pt]
4-076    %& AzV 238 
& O9.7\,III   & 
33.4$^{+1.6}_{-1.2}$   & 31.8$^{+1.5}_{-1.5}$  & --1.6 &
$3.69^{+0.14}_{-0.14}$ & 3.50$^{+0.14}_{-0.14}$& --0.19 &
$5.36^{+0.09}_{-0.08}$ & 5.36$^{+0.26}_{-0.26}$& +0.00 &
55$^{+77}_{-14}$       &  55$^{+76}_{-14}$ & +0 \\ [4pt]%
1-066   %& AzV 327 
& O9.7\,II-Ib  & 
29.9$^{+1.2}_{-1.2}$   & 29.9$^{+1.1}_{-1.1}$ & +0.0 &
$3.31^{+0.14}_{-0.14}$ & 3.31$^{+0.14}_{-0.14}$ & +0.00 &
$5.40^{+0.08}_{-0.08}$ & 5.47$^{+0.25}_{-0.25}$ & +0.07 &
55$^{+76}_{-14}$       & 55$^{+77}_{-14}$ & +0\\ [4pt] 
%\hline
 4-013   %&  Cl* NGC 346 ELS 43
 &  B0\,V  & 
 31.9$^{+1.2}_{-1.6}$ & $31.8^{+3.0}_{-2.7}$ & --0.1 &
 $4.12^{+0.14}_{-0.19}$ & $4.12^{+0.33}_{-0.52}$ & +0.00 &
 $4.70^{+0.08}_{-0.10}$ & $4.77^{+0.26}_{-0.26}$ & +0.07 & 
 19$^{+12}_{-19}$ & 36$^{+18}_{-21}$ & +17 \\ [4pt] 
 4-014   %&  Cl* NGC 346 ELS 26 
 & B0\,III & 
 31.5$^{+1.6}_{-2.3}$ & $31.9^{+1.2}_{-2.7}$ & +0.4 &
 $3.69^{+0.14}_{-0.29}$ & $3.70^{+0.17}_{-0.34}$ & +0.01 &
 $4.93^{+0.10}_{-0.13}$ & $4.95^{+0.21}_{-0.22}$  & +0.02 &
 55$^{+77}_{-14}$ & 75$^{+20}_{-23}$ & +20 \\ [4pt] 
2-110   %& AzV 148 
& B0\,II &  
29.9$^{+1.2}_{-1.2}$ & 29.9$^{+1.5}_{-1.1}$  & +0.0 &
$3.50^{+0.14}_{-0.14}$ & 3.50$^{+0.14}_{-0.14}$ & +0.00 &
$5.04^{+0.08}_{-0.08}$ & 5.12$^{+0.25}_{-0.25}$ & +0.08 &
5$^{+14}_{-5}$ & 0$^{+19}_{-0}$ & --5 \\ [4pt] %
6-080    %& AzV 488 
& B0\,Ia  & 
25.2$^{+1.2}_{-1.2}$ & 26.8$^{+1.1}_{-1.5}$  & +1.6 & 
$2.69^{+0.14}_{-0.14}$ & 2.88$^{+0.19}_{-0.14}$ & +0.19 &
$5.79^{+0.09}_{-0.09}$ & 5.98$^{+0.14}_{-0.15}$ & +0.19 &
56$^{+75}_{-14}$ &  78$^{+98}_{-18}$ & +22\\ [4pt] 
7-064   %& AzV 235 
& B0\,Ia     & 
28.3$^{+1.2}_{-1.6}$ &  26.8$^{+1.1}_{-1.5}$ & --1.5 &
($3.50^{+0.24}_{-0.14}$) & 2.88$^{+0.19}_{-0.14}$ & (--0.62) &
$6.01^{+0.09}_{-0.11}$ & 5.93$^{+0.15}_{-0.16}$ & --0.08 &
31$^{+14}_{-20}$ & 78$^{+97}_{-19}$ & +47  \\ [4pt] 
4-020    %& AzV 210 
& B1\,Ib-Iab & 
23.7$^{+1.2}_{-2.0}$ & 23.7$^{+1.2}_{-3.1}$ & +0.0 &
$2.69^{+0.14}_{-0.29}$ & 2.90$^{+0.17}_{-0.34}$ & +0.21 &
$5.56^{+0.10}_{-0.14}$ & 5.65$^{+0.15}_{-0.20}$  & +0.09 &
55$^{+76}_{-13}$ & 55$^{+77}_{-13}$ & +0\\ [4pt] 
8-008    %& AzV 96 
& B1\,Iab   & 
21.3$^{+1.6}_{-1.2}$ & 23.7$^{+1.2}_{-2.7}$ & +2.4 &
$2.31^{+0.29}_{-0.19}$ & 2.73$^{+0.23}_{-0.29}$  & +0.42 &
$5.23^{+0.13}_{-0.11}$ & 5.54$^{+0.13}_{-0.18}$ & +0.31 &
78$^{+98}_{-19}$ &  78$^{+98}_{-19}$ & +0 \\ [4pt] 
4-045     %& AzV 224 
& B1\,Iab  & 
23.7$^{+1.2}_{-2.0}$ & 23.7$^{+2.2}_{-3.0}$  & +0.0 &
$3.12^{+0.14}_{-0.33}$ & 2.88$^{+0.33}_{-0.29}$ & --0.24 &
$4.83^{+0.10}_{-0.14}$ & 4.88$^{+0.25}_{-0.27}$  & +0.05 &
113$^{+20}_{-19}$ &  201$^{+24}_{-24}$ & +88 \\ [4pt] 
4-078    %& AzV 242 
& B1\,Ia     & 
22.5$^{+0.8}_{-0.8}$ & 23.8$^{+1.1}_{-1.1}$  & +1.3 &
$2.31^{+0.14}_{-0.14}$ & 2.69$^{+0.14}_{-0.14}$  & +0.38 &
$5.28^{+0.08}_{-0.08}$ & 5.79$^{+0.15}_{-0.15}$ & +0.51 &
112$^{+20}_{-19}$ &  78$^{+98}_{-19}$ & --34 \\ [4pt] 
         1-009    %& AzV 264 
         & B1\,Ia     & 
         21.3$^{+0.8}_{-1.2}$ & 22.3$^{+2.3}_{-1.9}$ & +1.0 &
         2.31$^{+0.14}_{-0.14}$ & 2.49$^{+0.12}_{-0.11}$ & +0.18 &
         $5.30^{+0.08}_{-0.11}$ & 5.55$^{+0.18}_{-0.17}$ & +0.25 &
         78$^{+98}_{-19}$ & 78$^{+97}_{-19}$ & +0\\ [4pt] 
4-066    %& AzV 234 
& B2.5\,Ib & 
17.8$^{+0.8}_{-0.8}$ & 18.8$^{+0.8}_{-1.5}$ & +1.0 &
2.50$^{+0.14}_{-0.14}$ & 2.69$^{+0.14}_{-0.33}$ & +0.19 &
$5.06^{+0.09}_{-0.09}$ & 5.16$^{+0.18}_{-0.20}$ & +0.10 &
36$^{137}_{-34}$ &  36$^{+56}_{-34}$ & +0\\ [4pt] 
1-111   %& AzV 362 
& B3\,Ia     & 
15.9$^{+0.8}_{-0.8}$ & 14.9$^{+1.5}_{-0.4}$ & --1.0 &
$1.64^{+0.19}_{-0.10}$ & 1.64$^{+0.38}_{-0.10}$ & +0.00 &
$4.86^{+0.10}_{-0.10}$ & 5.62$^{+0.17}_{-0.14}$ & +0.76 &
5$^{+8}_{-5}$ & 55$^{+77}_{-13}$ & +50 \\ [4pt] 
1-062   %& AzV 324 
& B8\,Iab & 
12.7$^{+0.4}_{-0.4}$ & 13.5$^{+0.4}_{-0.8}$ & +0.8 &
1.88$^{+0.19}_{-0.33}$ & 2.10$^{+0.17}_{-0.17}$ & +0.22 &
$4.65^{+0.07}_{-0.07}$ & 4.77$^{+0.16}_{-0.17}$ & +0.12 &
55$^{+77}_{-13}$ &  35$^{+57}_{-32}$ & --20 \\ [4pt] 
      \hline
\end{tabular}
\end{center}
\end{table*}

%\end{landscape}

\clearpage

\section{Hertzsprung-Russell diagrams of single and binary systems}\label{HRD-extra}

\begin{figure*}
%\centering
  \includegraphics[width=1.5\columnwidth]{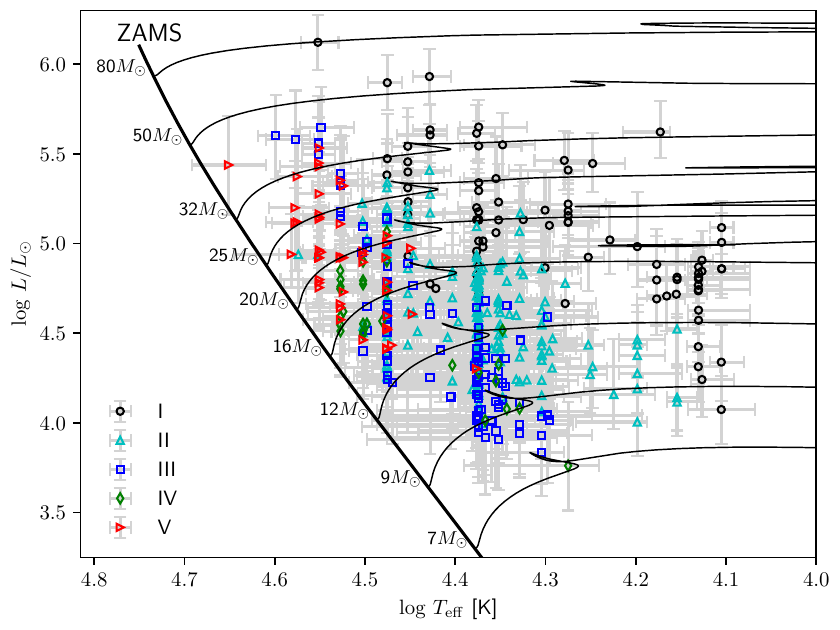}
  \includegraphics[width=1.5\columnwidth]{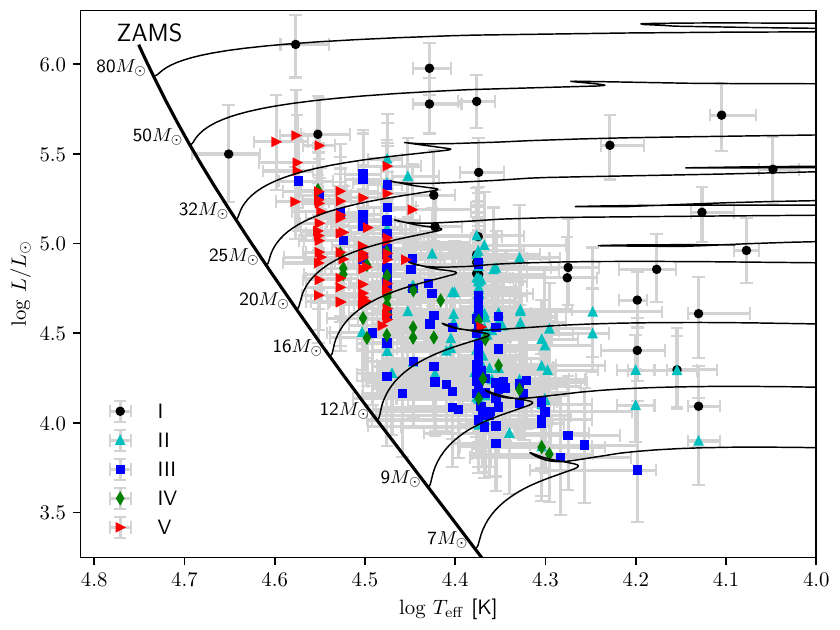}
  \caption{Hertzsprung-Russell diagram  of single (upper panel) and multiple (lower panel) OB stars (colour coded by luminosity class) on the basis of the initial 9 BLOeM epochs, together with evolutionary tracks for non-rotating SMC massive stars from \citet{Brott+2011}, with the exception of two luminous O supergiants drawn from \citet{Hastings+2021}.}
  \label{HRD2}
\end{figure*}

\begin{figure*}
%\centering
% \sidecaption
\includegraphics[width=1.75\columnwidth]{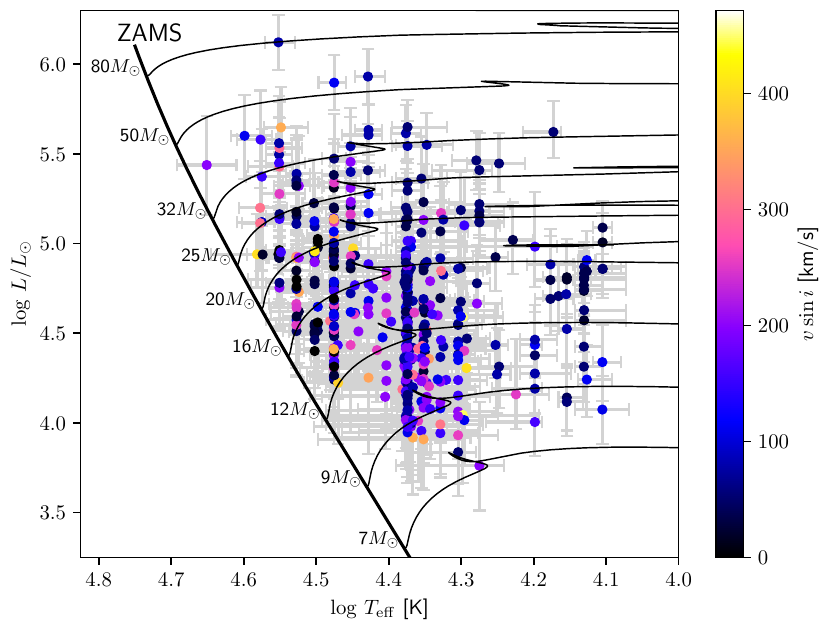}
 \includegraphics[width=1.75\columnwidth]{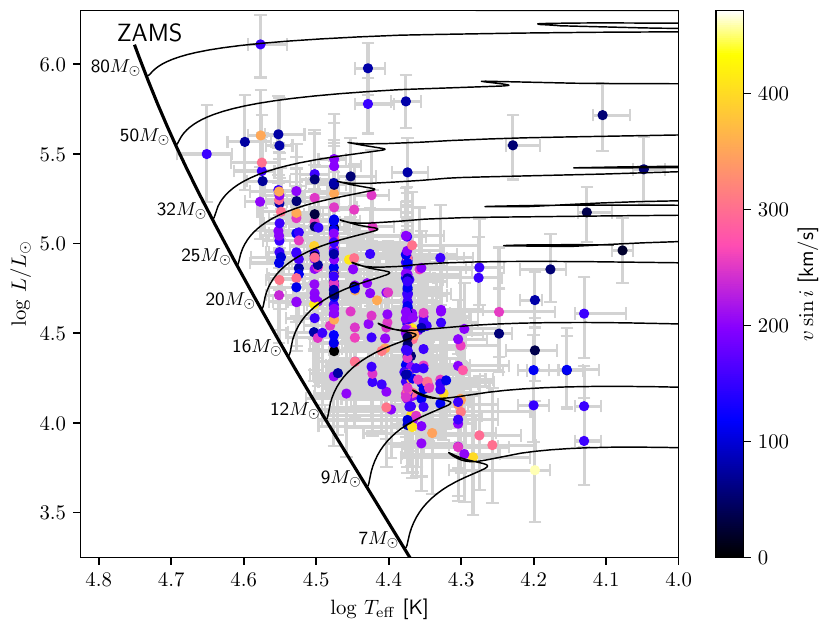}
  \caption{Hertzsprung-Russell diagram  of OB stars (colour coded by $\varv_{\rm e} \sin i$) for single (upper panel) and multiple (lower panel) systems on the basis of the initial 9 BLOeM epochs, together with evolutionary tracks for SMC massive stars from %\citet{Brott+2011}. 
  %Evolutionary masses of post-main sequence stars are determined from inspection of SMC tracks from 
  \citet{Schootemeijer+2019} for non-rotating stars ($\alpha_{\rm SC} = 10, \alpha_{\rm OV} = 0.33$).
  }
  \label{HRD_vsini_2}
\end{figure*}

Figure~\ref{HRD2} shows HR diagrams of single (upper) and binary (lower) BLOeM OB stars, colour coded by luminosity class, together with \citet{Brott+2011} non-rotating SMC tracks. Figure~\ref{HRD_vsini_2} shows HR diagrams for single (upper) and binary (lower) BLOeM OB stars, colour coded by $\varv_{\rm e} \sin i$, together with \citet{Schootemeijer+2019} non-rotating SMC tracks.

\clearpage

\section{Individual results from {\sc iacob-broad} and {\sc iacob-gbat} }\label{Sergio}

Table~\ref{iacob} presents physical parameters of a subset of OB stars from BLOeM (labels X-XX0) obtained with {\sc iacob-broad} \citep{SimonDiaz-Herrero2014} and {\sc iacob-gbat} \citep{SimonDiaz+2011}.

\begin{table*}
\caption{Physical parameters for subset of BLOeM OB stars obtained with {\sc iacob-broad} \citep{SimonDiaz-Herrero2014} and {\sc iacob-gbat}  \citep{SimonDiaz+2011}. Rotation velocities are obtained via Fourier Transform (FT) or Goodness of Fit (GOF) for He\,{\sc i} $\lambda$4387.  Helium abundances are provided by number, $y$ = N(He)/N(H) and by mass, $Y$ where $y$ = 0.085 ($Y$=0.25) is the baseline He content in the SMC adopted by \citet{Brott+2011}.}
% or {\sc PoWR} \citep{Grafener+2002, Sander+2015}. 
% Mass-loss rates are presented as $\dot{M}/\sqrt{f_{v}}$ to reflect differences in adopted or derived wind clumping factors.
\label{iacob}
\begin{tabular}{
l@{\hspace{3mm}}
l@{\hspace{3mm}}
r@{\hspace{0.5mm}}
l@{\hspace{3mm}}
r@{\hspace{0.5mm}}
l@{\hspace{3mm}}
c@{\hspace{3mm}}
r@{\hspace{5mm}}
r@{\hspace{0.5mm}}
l@{\hspace{3mm}}
r@{\hspace{0.5mm}}
l@{\hspace{3mm}}
r}
\hline
    BLOeM    &  Spect.  & \multicolumn{2}{c}{$T_{\rm eff}$}    & \multicolumn{2}{c}{$\log g$}       & \multicolumn{2}{c}{$\varv_{e} \sin i$ (km\,s$^{-1}$)} & \multicolumn{2}{c}{$y \times 10^{2}$} & \multicolumn{2}{c}{$Y$} & Note\\
    & Type    &  &   kK         & & cm\,s$^{-2}$ & FT & GOF & \\
    \hline
% BLOEM  SpT               Teff               logg            logL  vsini(FT)   vsini(GOF)        zeta    y
% BLOEM  SpT               Teff               logg            logL  vsini(FT)   vsini(GOF)           y
1-010 & B1.5\,III: &     &  ---         &    & ---                  &  158   &  110$^{+63}_{-96}$  &     & ---          &   & ---            &    \\ [2pt]
1-020 & B0\,III    &     & $30.8\pm1.2$ &    & $3.75\pm0.18$        &   77   &   70$^{+28}_{-36}$  & $<$ & $6.0^{+2.0}$ &$<$& $0.19^{+0.05}$ & SB1\\ [2pt]
1-030 & B1\,II     &     &  ---         &    & ---                  &   80   &   81$^{+77}_{-67}$  &     & ---          &   & ---            &    \\ [2pt]
1-060 & B1.5\,Ib   &     & ---          &    &---                   &  191   &  190$^{+19}_{-30}$  &     & ---          &   & ---            &    \\ [2pt]
1-070 & B1.5\,II   &     & ---          &    &---                   &   57   &   57$^{+19}_{-24}$  &     & ---          &   & ---            &    \\ [2pt]
1-080 & O8:V:+B+B  &     & $34.5\pm1.3$ & $>$& $4.30_{-0.28}$       &   86   &   63$^{+86}_{-49}$  & $<$ & $6.0^{+1.3}$ &$<$& $0.19^{+0.03}$ &    \\ [2pt]
1-100 & B1\,II     &     & ---          &    & ---                  &   98   &   79$^{+19}_{-23}$  &     & ---          &   & ---            & SB1\\ [2pt]
1-110 & B1\,Ib     &     & ---          &    & ---                  &   57   &   21$^{+15}_{- 7}$  &     & ---          &   & ---            &    \\ [5pt]
2-010 &B1.5\,III-II&     & ---          &    & ---                  &   47   &   14$^{+36}_{- 0}$  &     & ---          &   & ---            &    \\ [2pt]
2-020 &O7\,Iaf$^{+}$&    & $37.6\pm1.5$ &    & $3.58\pm0.17$        &   77   &   78$^{+30}_{-64}$  &     &$10.2\pm2.9$  &   & $0.29^{+0.05}_{-0.07}$ & SB1\\ [2pt]
2-030 & B2\,II     &     & ---          &    & ---                  &  169   & 148$^{+130}_{-135}$ &     & ---          &   & ---            & lpv/SB1\\[2pt]
2-040 & B2\,II     &     & ---          &    & ---                  &   53   &   28$^{+23}_{-15}$  &     & ---          &   & ---            & \\ [2pt]
2-060 & B1.5\,Ib   &     & ---          &    & ---                  &   74   &   64$^{+21}_{-26}$  &     & ---          &   & ---            & \\ [2pt]
2-070 & B1\,II e   &     & ---          &    & ---                  &  104   &   90$^{+29}_{-38}$  &     & ---          &   & ---            & SB?\\ [2pt]
2-090 & O7.5\,Vn   &     & $35.8\pm1.4$ &    & $3.99\pm0.24$        &  309   & 276$^{+151}_{-228}$ & $<$ & $6.0^{+3.2}$ &$<$& $0.19^{+0.08}$ & SB2\\ [2pt]
2-100 & B0\,V      &     & $32.7\pm0.8$ &    & $4.10\pm0.11$        & 145    &  130$^{+34}_{-54}$  &     & $7.7\pm1.0$  &   & $0.23^{+0.03}_{-0.02}$ & \\ [2pt]
2-110 & B0\,II     &     & $31.3\pm0.9$ &    & $3.59\pm0.13$        &  44    &   14$^{+19}_{- 0}$  &     & $8.6\pm2.2$  &   & $0.25^{+0.05}_{-0.05}$ & \\ [5pt]
3-010 & O9.7\,V:   &     & $35.4\pm1.0$ & $>$& $4.50_{-0.32}$       &  80    &   64$^{+83}_{-51}$  & $<$ & $6.0^{+1.0}$ &$<$& $0.19^{+0.03}$ & SB1\\ [2pt]
3-020 & B0\,III    &     & $31.0\pm1.2$ &    & $3.91\pm0.13$        &  72    &   28$^{+48}_{-14}$  & $<$ & $6.0^{+2.3}$ &$<$& $0.19^{+0.06}$ & SB2\\ [2pt]
3-030 & B1\,II     &     &  ---         &    & ---                  &  128   &  120$^{+17}_{-24}$  &     & ---          &   & ---            & \\ [2pt]
3-050 & B1.5\,III  &     & ---          &    & ---                  &  173   &  132$^{+53}_{-60}$  &     & ---          &   & ---            & SB1\\ [2pt]
3-060 & O6\,Vn:    &     & $37.7\pm1.1$ &    & $3.62\pm0.16$        &  294   & 285$^{+118}_{-271}$ &     & $12.5\pm3.1$ &   & $0.33^{+0.05}_{-0.06}$ & \\ [2pt]
3-070 & B1\,II     &     & ---          &    & ---                  &   38   &   14$^{+10}_{- 0}$  &     & ---          &   & ---           & SB1\\ [2pt]
3-080 & B1\,III-II &     & ---          &    & ---                  &   59   &   14$^{+20}_{- 0}$  &     & ---          &   & ---           & SB1\\ [2pt]
3-090 & B0.2\,Ia   &     & $28.0\pm1.1$ &    & $3.19\pm0.21$        &   75   &   43$^{+12}_{-17}$  & $<$ & $6.0^{+2.3}$ &$<$& $0.19^{+0.06}$& \\ [2pt]
3-100 & B3\,II     &     & ---          &    & ---                  &   61   &   50$^{+32}_{-37}$  &     & ---          &   & ---           & \\ [2pt]
3-110 & B8\,II-Ib  &     & ---          &    & ---                  &   46   &   14$^{+38}_{- 0}$  &     & ---          &   & ---           & Post-MS\\ [5pt]
4-020 & B1\,Iab-Ib &     & ---          &    & ---                  &   62   &   31$^{+19}_{-18}$  &     & ---          &   & ---           & \\ [2pt]
4-030 & B1\,Ia     &     & ---          &    & ---                  &   58   &   37$^{+13}_{-18}$  &     & ---          &   & ---           & lpv/SB1\\ [2pt]
4-050 & B1\,II:    &     & ---          &    & ---                  &   55   &   14$^{+42}_{- 0}$  &     & ---          &   & ---           & \\ [2pt]
4-060 & B8 II-Ib   &     & ---          &    & ---                  &   60   &   14$^{+44}_{- 0}$  &     & ---          &   & ---           & Post-MS\\ [2pt]
4-070 & B2\,II     &     & ---          &    & ---                  &  301   &  305$^{+54}_{-218}$ &     & ---          &   & ---           & lpv/SB1\\ [2pt]
4-080 &O9.7+O8-8.5+B&    &  $\cdots$    &    &   $\cdots$           &  227   &  21$^{+151}_{-7}$   &     & ---          &   & ---           & SB2\\ [2pt]
4-090 & B1\,II     &     &  ---         &    & ---                  &  134   &  116$^{+26}_{-30}$  &     & ---          &   & ---           & SB1\\ [2pt]
4-100 & B1\,III    &     &  ---         &    & ---                  &   73   &   28$^{+37}_{-14}$  &     & ---          &   & ---           & \\ [2pt]
4-110 & O7\,V:(n)  &     & $35.8\pm1.5$ &    & $3.90\pm0.28$        &  228   & 104$^{+206}_{-91}$  & $<$ & $6.0^{+2.3}$ &$<$& $0.19^{+0.06}$& SB1\\ [5pt]
5-010 & B3\,II     &     &  ---         &    &  ---                 &   44   &   14$^{+26}_{- 0}$  &     & ---          &   & ---           &\\ [2pt]
5-030 & B1.5\,III: &     & ---          &    & ---                  &  107   &  108$^{+59}_{-80}$  &     & ---          &   & ---           & SB2\\ [2pt]
5-040 & B1\,II     &     & ---          &    & ---                  &   66   &   16$^{+30}_{- 2}$  &     & ---          &   & ---           & SB1\\ [2pt]
5-050 &O9.7\,V:+early B& & $32.0\pm0.4$ &    & $3.74\pm0.08$        &  317   &  307$^{+24}_{-32}$  &     & $8.9\pm1.5$  &   & $0.26^{+0.03}_{-0.03}$ & SB2 \\ [2pt]
\hline
\end{tabular}
\end{table*}

\begin{table*}
\contcaption{}
% Mass-loss rates are presented as $\dot{M}/\sqrt{f_{v}}$ to reflect differences in adopted or derived wind clumping factors.
%\label{iacob}
\begin{tabular}{
l@{\hspace{3mm}}
l@{\hspace{3mm}}
r@{\hspace{0.5mm}}
l@{\hspace{3mm}}
r@{\hspace{0.5mm}}
l@{\hspace{3mm}}
c@{\hspace{3mm}}
r@{\hspace{5mm}}
r@{\hspace{0.5mm}}
l@{\hspace{3mm}}
r@{\hspace{0.5mm}}
l@{\hspace{3mm}}
r}
\hline
    BLOeM    &  Spect.  & \multicolumn{2}{c}{$T_{\rm eff}$}    & \multicolumn{2}{c}{$\log g$}       & \multicolumn{2}{c}{$\varv_{e} \sin i$ (km\,s$^{-1}$)} & \multicolumn{2}{c}{$y \times 10^{2}$} & \multicolumn{2}{c}{$Y$} & Note\\
    & Type    &  &   kK         & & cm\,s$^{-2}$ & FT & GOF & \\
    \hline
5-060 & B1.5\,II   &     & ---          &    & ---                  &   43   &   14$^{+15}_{- 0}$  &     & ---         &    & --- & \\ [2pt]
5-070 & B2\,III:   &     & ---          &    & ---                  &  293   & 260$^{+123}_{-222}$ &     & ---         &    & --- & \\ [2pt]
5-080 & B2\,III:   &     & ---          &    & ---                  &  344   &  344$^{+42}_{-87}$  &     & ---         &    & --- & SB2\\ [2pt]
5-090 & O9.5\,III  &     & $34.4\pm0.6$ &    & $3.64\pm0.06$        &   86   &   66$^{+20}_{-24}$  &     &$16.0\pm3.9$ &    &$0.39^{+0.05}_{-0.07}$ & \\ [2pt]
5-100 & B0\,V      &     & $31.9\pm1.3$ &    & $4.10\pm0.21$        &  122   &  122$^{+59}_{-78}$  & $<$ &$6.0^{+1.8}$ &$<$ &$0.19^{+0.05}$ & SB2\\ [2pt]
5-110 & B1\,III    &     & ---          &    & ---                  &  107   &   75$^{+63}_{-61}$  &     & ---         &    & --- & SB2\\ [5pt]
6-010 & B2\,IV:    &     & ---          &    & ---                  &  133   &  52$^{+212}_{-39}$  &     & ---         &    & --- & SB2\\ [2pt]
6-020 & B2.5\,III  &     &              &    & ---                  &  139   &  142$^{+33}_{-44}$  &     & ---         &    & --- & SB1\\ [2pt]
6-030 & B0\,IV:    &     & $32.9\pm1.6$ &    & $4.02\pm0.23$        &   41   &   14$^{+38}_{- 0}$  & $<$ &$8.0^{+2.1}$ &$<$ &$0.24^{+0.05}$ & \\ [2pt]
6-040 & B1.5\,III: &     & ---          &    & ---                  &  109   &  104$^{+54}_{-91}$  &     &  ---        &    & --- &  \\ [2pt]
6-050 & B1\,III    &     & ---          &    & ---                  &   49   &   14$^{+22}_{- 0}$  &     & ---         &    & --- & \\ [2pt]
6-060 & O9.7\,IV   &     & $35.0\pm1.1$ &    & $4.11\pm0.16$        &   47   &   14$^{+30}_{- 0}$  &     &$8.8\pm1.9$  &    &$0.26^{+0.04}_{-0.04}$ & \\ [2pt]
6-070 & B1:\,II    &     & ---          &    & ---                  &  123   &  121$^{+39}_{-49}$  &     & ---         &    & --- & SB1\\ [2pt]
6-080 & B0\,Ia     &     & $29.0\pm1.6$ &    & $3.43\pm0.28$        &   56   &   55$^{+10}_{- 7}$  & $<$ & $8.0^{+1.6}$&$<$ &$0.24^{+0.04}$ & lpv/SB1\\ [2pt]
6-090 & B2\,III    &     &  ---         &   & ---                   &  358   & 365$^{+112}_{-352}$ &     & ---         &    & --- & \\ [2pt]
6-100 & B1\,II     &     & ---          &   & ---                   &   35   &  14$^{+ 14}_{- 0}$  &     & ---         &    & --- & \\ [2pt]
6-110 & B1.5\,III: &     & ---          &   & ---                   &   34   &  14$^{+ 20}_{- 0}$  &     & ---         &    & --- & \\ [5pt]
7-010 & B1\,III    &     & ---          &   & ---                   &   56   &  14$^{+ 27}_{- 0}$  &     & ---         &    & --- & SB1\\ [2pt]
7-030 & B0.5:\,V   &     & $30.9\pm2.4$ & $>$&$4.10_{-0.28}$        &  101   &  87$^{+ 97}_{-73}$  & $<$ &$6.0^{+3.1}$ &$<$ &$0.19^{+0.08}$ & \\ [2pt]
7-040 &B1.5\,III-II&     & ---          &   & ---                   &  108   & 101$^{+ 31}_{-46}$  &     & ---         &    & --- & SB1\\ [2pt]
7-050 & B2\,III:   &     & ---          &   & ---                   &  201   & 173$^{+ 82}_{-159}$ &     & ---         &    & --- & \\ [2pt]
7-060 & B2\,III:   &     & ---          &   & ---                   &  180   & 208$^{+ 54}_{-84}$  &     & ---         &    & --- & SB2\\ [2pt]
7-070 & B1.5\,III: &     & ---          &   & ---                   &  117   &  93$^{+ 77}_{-79}$  &     & ---         &    & --- & SB1\\ [2pt]
7-080 & B1.5\,III: &     & ---          &   & ---                   &  119   &  96$^{+ 57}_{-82}$  &     & ---         &    & --- & \\ [2pt]
7-090 & B2\,III:   &     & ---          &   & ---                   &  162   & 112$^{+ 72}_{-98}$  &     & ---         &    & --- & \\ [2pt]
7-100 & B2\,II     &     & ---          &   & ---                   &  114   & 111$^{+ 61}_{-97}$  &     &  ---        &    & --- & SB1\\ [2pt]
7-110 & B1.5\,III: &     & ---          &   & ---                   &  157   & 134$^{+ 65}_{-121}$ &     & ---         &    & --  & SB1\\ [5pt]
8-020 & O8\,V      &     & $39.3\pm1.3$ & $>$&$4.30_{-0.20}$        &   73   &  63$^{+ 74}_{-49}$  & $<$ &8.0$^{+1.2}$ &$<$ &$0.24^{+0.03}$ & SB1\\ [2pt]
8-030 & O6.5\,Vn   &     & $38.2\pm0.8$ &    & $3.82\pm0.08$        &  290   & 293$^{+ 50}_{-122}$ &     &$13.0\pm2.3$ &    &$0.34^{+0.04}_{-0.04}$ & \\ [2pt]
8-040 & B2\,IV     &     &  ---         &    &     ---              &   95   &  58$^{+ 67}_{-45}$  &     & ---         &    & --- &  \\ [2pt]
8-050 & O9.7\,IV   &     & $35.7\pm1.3$ &    & $4.11\pm0.19$        &   42   &  14$^{+ 35}_{- 0}$  & $<$ &$8.0^{+2.2}$ &    &$0.24^{+0.05}_{-0.04}$ & \\ [2pt]
8-060 & B2\,II:    &     & ---          &    & ---                  &  115   &  90$^{+ 52}_{-76}$  &     & ---         &    & --- & \\ [2pt]
8-070 & B0.5\,IV   &     & $30.0\pm2.0$ & $>$& $4.20_{-0.33}$       &  149   & 126$^{+ 55}_{-92}$  & $<$ &$8.0^{+2.3}$ &    &$0.24^{+0.05}_{-0.04}$ & SB1\\ [2pt]
8-080 & B2\,III:   &     &---           &   & ---                   &  267   & 253$^{+ 79}_{-187}$ &     & ---         &    & --- & SB2\\ [2pt]
8-090 & B1\,II     &     & ---          &   & ---                   &  114   & 137$^{+ 41}_{-62}$  &     & ---         &    & --- & SB1\\ [2pt]
8-100 & B2\,II e   &     & ---          &   & ---                   &  245   & 238$^{+ 40}_{-94}$  &     & ---         &    & --- & \\ [2pt]
8-110 &B1\,III:+
       B1\,III:    &     &  ---         &   & ---                   &  ---  &  ----                &     & ---         &    & --- & SB2\\ [2pt]
\hline
\end{tabular}
\end{table*}

\clearpage

\section{Spectroscopic versus evolutionary models}

\begin{figure*}
%\sidecaption
\includegraphics[width=1.5\columnwidth]{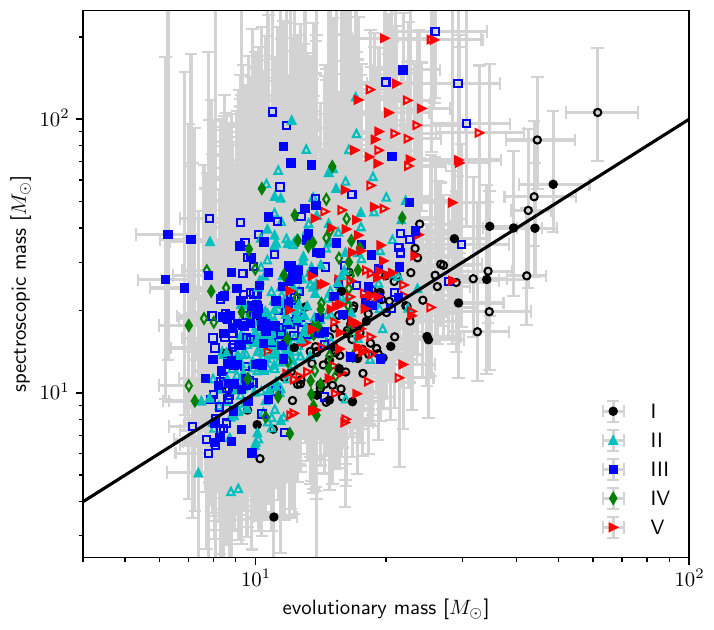}
  \caption{Comparison between (current) evolutionary masses and spectroscopic masses of BLOeM OB stars, based on \citet{Schootemeijer+2019} non-rotating models, colour coded by luminosity class (filled symbols are binaries).}
  \label{Msp_vs_Mev_Abel}
\end{figure*}

Figure~\ref{Msp_vs_Mev_Abel} compares spectroscopic and (current) evolutionary masses of OB stars from the BLOeM survey (filled symbols are known binaries) based on \citet{Schootemeijer+2019} non-rotating SMC metallicity models, which reveals a similar discrepancy to Fig.~\ref{Msp_vs_Mev} based on \citet{Brott+2011} rotating SMC metallicity models. 

\clearpage

\section{Updated catalogue of SMC O stars}

Table~\ref{census} presents an updated catalogue of spectroscopically confirmed O stars, listed in RA order, combining the unpublished compilation from I.D.~Howarth (priv. comm.) with BLOeM \citep{Shenar+2024}. Coordinates and ($G$-band) photometry  are from {\it Gaia} with the exception of the compact H\,{\sc ii} regions N81 and N88A. For N81 we adopt positions and {\it Hubble Space Telescope} WFPC2/F547M photometry from \citet{HeydariMalayeri+1999a} while for the central ionizing source in N88A \citep{HeydariMalayeri+1999b} we adopt positions and {\it HST} WFPC2/F547M photometry from \citet{Testor+2010}. Classifications exclusively from UV spectroscopy are excluded \citep[e.g.][]{SmithNeubigBruhweiler1997}. The BLOeM sample includes 159/449 O star systems in the SMC, of which 75 -- 1/6 of the current total -- have been newly spectroscopically classified as O stars via BLOeM.

%
%\begin{landscape}
\begin{table*}
\caption{Catalogue of spectroscopically confirmed O stars in the SMC including [M2002] SMC catalogue numbers from \citet{Massey2002}. Primary stellar catalogues
include Lin \citep{Lindsay1961}, Sk \citep{Sanduleak1968, Sanduleak1969}, AzV \citep{Azzopardi+1975}, Cl* NGC 346 MPG \citep{Massey+1989}, [MA93] \citep{MeyssonnierAzzopardi1993}, 2dFS \citep{Evans+2004-2dF}, Cl* NGC 330/346 ELS \citep{Evans+2006}, [BLK2010] \citep{Bonanos+2010} and [SBV2013] \citep{Sheets+2013}. Photometry (G-band) and coordinates are from {\it Gaia} with the exception of the compact clusters N81 \citep{HeydariMalayeri+1999a} and N88A \citep{HeydariMalayeri+1999b} for which {\it HST} WFPC2/F547M photometry is indicated in parentheses.}
\label{census}
\begin{center}
\begin{tabular}{c@{\hspace{2mm}}c@{\hspace{3mm}}c@{\hspace{3mm}}l@{\hspace{2mm}}l@{\hspace{2mm}}
l@{\hspace{2mm}}c@{\hspace{2mm}}r@{\hspace{3mm}}l}
\hline\hline
%\hline
RA & Dec                   & $G$   & Spectral & Ref & Name & BLOeM & M2002 & Note\\
\multicolumn{2}{c}{J2000}  & mag & Type   &     &      &       & \\
\hline
00 27 46.56  & --73 16 45.8 &  15.47 &   O9.5:                &   EHI04 & 2dFS 1                  & $\cdots$       & $\cdots$  &\\ 
 00 36 58.24  & --73 23 33.2 &  15.00 &   O7\,Ib(f) + early B  &  BCH25  & 2dFS 163                & $\cdots$       & $\cdots$ & EHI04: O8\,Ib(f)\\
 00 42 09.92  & --73 13 56.8 &  14.69 &   O8.5 V               &  EHI04  & 2dFS 404                & $\cdots$       & 1600     & \\
 00 43 36.69  & --73 02 26.7 &  15.77 &   O3--4                &  LOS16 & LHA 115-N9               & $\cdots$       & 3173     & \\      
 00 43 49.86  & --73 09 02.8 &  13.48 &   O9.5\,I              & LOS16 & AzV 2b, 2dFS 482          & $\cdots$       & 3459     & EHI04: B0\,(II)\\
 00 44 57.04  & --73 59 13.4 &  13.80 &   O9.5 Ib-II           & EHI04   & Sk 4, 2dFS 538          & $\cdots$       & 4919     & LOS16: B0\,III\\
 00 44 57.12 &  -73 00 47.0 &  14.57 &  O6 V                   & EHI04   & 2dFS 5001               & $\cdots$       & 4922     & \\  
 00 45 14.52 & --73 35 58.4  &  15.14 &  O8.5 V                & LOS16   & $\cdots$                & $\cdots$       &  5313    & \\
 00 45 18.20 & --73 15 23.0  &  13.76  & O9.2\,IV:              & BCH25  & AzV 6, 2dFS 5002        & $\cdots$       & 5391     & EHI04: O9\,III\\   
 00 45 25.82 & --73 23 00.7  &  15.08  & O8:\,V                & Cro25   & LHA 115-N 13A           & $\cdots$       & 5552     & \\
 00 45 30.72 & --73 03 29.6  &  15.39 &  O9 V                   & EHIO04 & 2dFS 5004                & $\cdots$      & 5655     & \\
 00 45 37.49 & --73 12 36.2  &  15.19 & O9                       & SBV13 &  [SBV2013] B4            & $\cdots$      & 5822     & \\
 00 45 40.16 & --73 14 40.9  & 15.44  & O6\,Vnn               & SBS24   & $\cdots$                  & 3-002         & 5880     & \\ % NEW
 00 46 02.12 & --73 06 27.3  &  13.16 & O9                       & AM77 & Sk 8, AzV 12              & $\cdots$      & 6406     & AVM75: B2 \\
 00 46 10.66 & --73 16 43.5  & 16.01  & O9.7\,IV:              & SBS24   &  $\cdots$                 & 3-004        &$\cdots$  & \\ % NEW
 00 46 20.00 & --73 06 31.7  &  15.47 & O9.5 III               & EHI04    & 2dFS 5006               & $\cdots$      & 6840     & \\ 
 00 46 22.57 & --73 23 17.5  &  15.09 & O9.5--B0\,III          & LOS16    &  $\cdots$                & $\cdots$     &  6908    & \\
 00 46 23.91 & --73 12 52.4  &  15.30 & O9.5 V                 & EHI04    &  2dFS 5008              & $\cdots$      &  6946    & \\
 00 46 32.62 & --73 06 05.6  &  13.82 & O3 V((fc))z+?          & BCH25     & Sk 9, AzV  14          & $\cdots$      & 7187     & MBK04: O5\,V\\
 00 46 34.85 & --73 21 35.7  &  14.76 & O9.5\,IIIe             & LOS16    & $\cdots$                & $\cdots$      & 7254     & \\
 00 46 40.01 & --73 31 17.7  & 15.03 &  O7--8 V                & EHI04    & 2dFS 610                & $\cdots$      & 7382     & \\
 00 46 40.42 & --73 21 53.3  & 15.66 & O7.5\,V:(n)             & SBS24    & $\cdots$                & 3-008         &$\cdots$  & \\
 00 46 42.16 & --73 24 55.5  &  13.12 & O6.5\,III(f)           & BCH25    & Sk 10, AzV 15           & $\cdots$      & 7437     & WLH00: O6.5\,II(f)\\  
 00 46 43.90 & --73 21 47.8 &   15.25 & O9.7\,V:               & SBS24    & $\cdots$                & 3-010         &$\cdots$  & \\ % NEW
 00 46 56.13 & --73 18 57.0  &  14.51 & O8\,Vn                 & SBS24    & $\cdots$                & 3-014         & 7782     & \\ % UPDATED
00 47 04.74 & --73 07 57.5  &  14.24 & O9.2\,V                 & SBS24    & 2dFS 5010               & 3-019         & 8003     & EHI04: O8.5\,V\\
 00 47 17.42 & --73 21 24.8  &  14.27 & O8\,V((f))             & EHI04   &  2dFS 5012               & $\cdots$      & 8344     & \\
 00 47 35.48 & --73 08 30.8 & 14.79 &  O9.5\,IV:n e?           & SBS24   &  $\cdots$                & 3-033         &$\cdots$  & \\ % NEW
 00 47 41.92 & --73 02 37.0 &  13.74 &  O9\,I                  & AVM75   & Sk 16, AzV  24           & $\cdots$      & 9079     & \\
 00 47 47.52 & --73 17 27.9  & 15.49 & O7                      & SBV13   &  [SBV2013] B23           & $\cdots$      & 9251     & \\
 00 47 50.05 & --73 08 21.1 &  12.42  & O6\,I(f) + O7.5        & SBS24   & Sk 18, AzV 26            & 3-042         & 9337     & MBK04: O6\,I(f)\\ %
 00 47 55.09 & --73 10 31.8 &  14.12  & O8.5\,V(n)             & SBS24   & $\cdots$                 & 3-046         & 9532     & \\ % NEW
 00 47 58.15 & --73 25 56.1 &  14.37 & O4\,I(n)                & SBS24   & $\cdots$                 & 3-049         & 9647     & \\ % NEW
 00 48 00.60  & --73 34 38.0 &  15.09 &  O7\,Vz                & LOS16   & $\cdots$                 &  $\cdots$     & 9732     & \\
 00 48 02.64 & --73 16 38.7  &  14.46  & O5.5\,V               & SBS24   & $\cdots$                 & 3-051         &$\cdots$  & \\ % NEW
 00 48 02.96 & --73 16 12.5   &   14.52 & O7\,V: + O7.5        & SBS24   & $\cdots$                 & 3-052         & 9845     & \\ % NEW
 00 48 09.26  & --73 14 16.0  & 14.40  & O7.5\,V(n)            & SBS24   &  Lin 106                 & 3-053         & $\cdots$ & Tes01: O8\,V \\ % 
 00 48 10.69  & --73 19 50.1  &  14.70  & O7\,V(n)             & SBS24   & [MA93] 210               & 3-054         & $\cdots$ &\\ % NEW
 00 48 18.92 & --73 27 15.1 &  14.60  & O6\,Vn:                & SBS24   & [SBV2013] B33            & 3-060         & 10505    & SBV13: O7--9 \\ % 
 00 48 24.15 & --73 06 49.1 &    14.05 & O9.7\,IIe             & SBS24   & Sk 22                    & 3-062         & 10756    & \\ % NEW
 00 48 25.53 & --73 08 47.9 &    14.15 & O9.2\,II(n)           & SBS24   &  $\cdots$                & 3-063         & 10818    & \\ % NEW
\hline
\end{tabular}
\end{center}
\footnotesize{  
AHZ09 \citet{Antoniou+2009};
AM77 \citet{Ardeberg-Maurice1977};
AVM75 \citet{Azzopardi+1975};
BCH25 \citet{Bestenlehner+2025}; \\ 
CG82 \citet{CramptonGreasley1982}; 
Cro25 Crowther (priv. comm.);
CNC01 \citet{Covino+2001};
DEH19 \citet{Dufton+2019}; \\
ECF04 \citet{Evans+2004-cmfgen};
EHI04 \citet{Evans+2004-2dF}; 
ELS06 \citet{Evans+2006};
EHO12 \citet{Evans+2012}; \\
FMG03 \citet{Foellmi+2003}; 
GCM87 \citet{Garmany+1987};
% {\bf GOL16} \citet{Golden-Marx+2016};
HS10 \citet{HeydariSelier10};
HHH05 \citet{Hilditch+2005}; \\
HT88 \citet{HutchingsThompson1988}; 
% mcn08{\bf Len97} \citet{Lennon1997};
% {\bf LOG13} \citet{Lamb+2013}; 
LOS16 \citet{Lamb+2016};
MCN08 \citet{McBride+2008};
MFH07 \citet{Martayan+2007}; \\
MSH04 \citet{Martins+2004}; 
MPG89 \citet{Massey+1989};
MWD00 \citet{Massey+2000};
MD01 \citet{MasseyDuffy2001}; \\
MBK04 \citet{Massey+2004}; 
MPP05 \citet{Massey+2005};
MZM09 \citet{Massey+2009};
MNM14 \citet{Massey+2014}; \\
MBS85 \citet{Moffat+1985}; 
MOP03 \citet{Morrell+2003}; 
% {\bf NMC86} \citet{Niemela+1986};
NB94 \citet{NiemelaBassino1994};
PSM12 \citet{Paul+2012}; \\
POH22 \citet{Pauli+2022}; 
RHO19 \citet{Ramachandran+2019};
% {\bf Pri87} \citet{Prinja1987};
RHH22 \citet{Rickard+2022}; 
RP23 \citet{RickardPauli2023}; \\
RSE12 \citet{Ritchie+2012}; 
SHT18 \citet{Shenar+2018}; 
SBS24 \citet{Shenar+2024};
SBV13 \citet{Sheets+2013}; \\
Tes01 \citet{Testor2001}; 
TL87 \citet{TestorLortet1987}; 
% {\bf TMN88} \citet{Thompson+1988};
Wal83 \citet{Walborn1983};  
WLH95 \citet{Walborn+1995}; \\
WLH00 \citet{Walborn+2000};  
WFC02 \citet{Walborn+2002-FUSE}; 
WMH04 \citet{Walborn+2004}; 
WHE10 \citet{Walborn+2010}; \\
Wil94 \citet{Wilcots1994}
}
\end{table*}
%\end{landscape}
%

\begin{table*}
\contcaption{}

\end{table*}

\bsp	% typesetting comment
\label{lastpage}
\end{document}